\documentclass[prx,letterpaper,nobalancelastpage,twocolumn,superscriptaddress,nofootinbib,longbibliography]{revtex4-2}

\usepackage{xr-hyper}
\usepackage{graphicx}
\usepackage{amsmath}
\usepackage{bbold}
\usepackage{amssymb}
\usepackage[english]{babel}
\usepackage{color}
\usepackage[version=4]{mhchem}
\usepackage[hidelinks]{hyperref}
\usepackage{dsfont}
\usepackage{placeins}
\usepackage{mathtools}
\usepackage[normalem]{ulem}
\usepackage{varwidth}
\usepackage{siunitx}
\usepackage{svg}
\usepackage{comment}
\usepackage{algorithm2e}

\newcommand{\Conv}{%
  \mathop{\scalebox{1.5}{\raisebox{-0.2ex}{$\circledast$}}
  }
}

\usepackage{soul}
\newcommand{\of}[1]{\left( #1 \right)}

\newcommand{\ket}[1]{| #1 \rangle}

\newcommand{\bra}[1]{\langle #1 |}

\newcommand{\Caltech}{California Institute of Technology, Pasadena, CA 91125, USA}
\newcommand{\MIT}{Center for Theoretical Physics, Massachusetts Institute of Technology, Cambridge, MA 02139, USA}
\newcommand{\Stanford}{Department of Electrical Engineering, Stanford University, Stanford, CA 94305, USA}

\usepackage{caption}

\DeclareCaptionLabelSeparator{bar}{ \textbf{\textbar}~}
\usepackage{enumitem}
\setlist{nolistsep}

\makeatletter

\begin{document}


\title{Experimental signatures of Hilbert-space ergodicity:\\Universal bitstring distributions and applications in noise learning}

\author{Adam L. Shaw}
\thanks{These authors contributed equally}
\affiliation{\Caltech}
\author{Daniel K. Mark}
\thanks{These authors contributed equally}
\affiliation{\MIT}
\author{Joonhee Choi}
\affiliation{\Stanford}
\author{Ran Finkelstein}
\affiliation{\Caltech}
\author{\\Pascal Scholl}
\affiliation{\Caltech}
\author{Soonwon Choi}
\thanks{soonwon@mit.edu}
\affiliation{\MIT}
\author{Manuel Endres}
\thanks{mendres@caltech.edu}
\affiliation{\Caltech}


\maketitle

\textbf{Systems reaching thermal equilibrium are ubiquitous. For classical systems, this phenomenon is typically understood statistically through ergodicity in phase space, but translating this to quantum systems is a long-standing problem of interest. Recently a strong notion of quantum ergodicity has been proposed, namely that isolated, global quantum states uniformly explore their available state space, dubbed \textit{Hilbert-space ergodicity}. Here we observe signatures of this process with an experimental Rydberg quantum simulator and various numerical models, before generalizing to the case of a local quantum system interacting with its environment. For a closed system, where the environment is a complementary subsystem, we predict and observe a smooth quantum-to-classical transition in that observables progress from large, quantum fluctuations to small, Gaussian fluctuations as the bath size grows. This transition exhibits universal properties on a quantitative level amongst a wide range of systems, including those at finite temperature, those with itinerant particles, and random circuits. For an open system, where the environment is uncontrolled, we predict the statistics of observables under largely arbitrary noise channels including those with correlated errors, allowing us to discriminate between candidate error models both for continuous Hamiltonian time evolution and for digital random circuits. This allows for computationally efficient experimental noise learning, and more broadly is a new avenue for quantitatively classifying the behavior of noisy quantum systems. Ultimately our results clarify the role of ergodicity in quantum dynamics, with fundamental and practical consequences.}

\section{Introduction}
The developing study of \textit{quantum thermalization}~\cite{Srednicki1994ChaosQuantum,Deutsch1991QuantumStatistical,Rigol2008ThermalizationIts,Abanin2019ColloquiumManybody,Nandkishore2015ManyBodyLocalization,DAlessio2016QuantumChaos,Ueda2020QuantumEquilibration,Deutsch2018EigenstateThermalization} has sought to create a consistent formulation for both statistical and quantum mechanics. In this framework, entanglement entropy between subsystems of a quantum state plays the role of thermal entropy in statistical mechanics. Considering a closed, bipartite quantum system composed of interacting subsystems $A$ and $B$ under a time-independent Hamiltonian, quantum thermalization suggests that at late times, local observables in $A$ will behave as if they were drawn from a thermal state. Here $B$ acts as a thermal bath, for which the system Hamiltonian and initial state define an effective temperature~\cite{Deutsch2018EigenstateThermalization}, behavior observed by numerous experiments~\cite{Kaufman2016QuantumThermalization,Choi2023PreparingRandom,Clos2016TimeResolvedObservation,Kranzl2023ExperimentalObservation,Kim2018DetailedBalance,Neill2016ErgodicDynamics}.

\begin{figure}[t!]
	\centering
	\includegraphics[width=90mm]{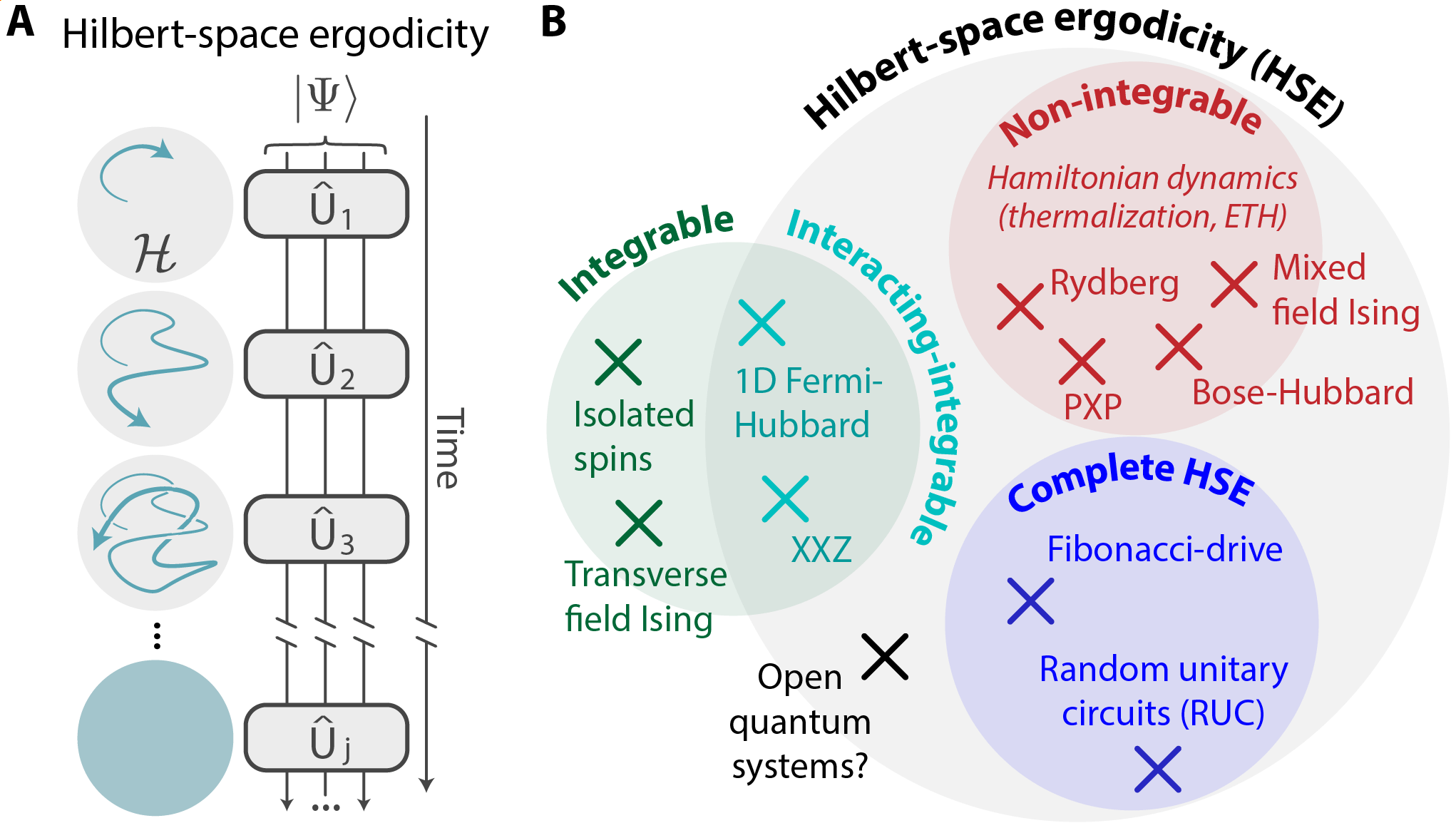}
	\caption{\textbf{Hilbert-space ergodicity.} \textbf{A.} Hilbert-space ergodicity (HSE) characterizes unitary quantum evolution under which a global quantum state, $|\Psi\rangle$, uniformly explores its Hilbert space, $\mathcal{H}$. \textbf{B.} While originally introduced~\cite{Pilatowsky-Cameo2023CompleteHilbertSpace} as \textit{complete} HSE, which characterizes certain symmetry-free dynamics, in recent theoretical work~\cite{Mark2024MaximumEntropy} HSE was extended to systems with symmetries and conservation laws. This expanded the framework and consequences of HSE to a host of new models of time-constant Hamiltonian systems, such as those exhibiting non-integrable dynamics traditionally understood via the eigenstate thermalization hypothesis (ETH), as well as certain interacting-integrable systems. In this work we show experimental evidence consistent with the predictions of HSE with a Rydberg quantum simulator, we generalize the theory to describe and then observe universal behavior of subsystems of closed systems displaying HSE, and finally we introduce a framework inspired by HSE for describing the behavior of open quantum systems.
 }
	\vspace{-0.5cm}
	\label{Fig0}
\end{figure}

In order to explain the ubiquity of this phenomenon, a set of interconnected theories have been studied, most notably the \textit{eigenstate thermalization hypothesis}~\cite{Deutsch2018EigenstateThermalization} (ETH). The ETH predicts that the above system-bath behavior occurs even on the level of individual many-body energy eigenstates, and accordingly it suggests quantum thermalization occurs for generic states because at late times, coherent contributions between energy eigenstates will be effectively dephased. Underlying the ETH is the assumption that energy eigenstates act as random vectors in certain ways in accordance with predictions from random matrix theory~\cite{Wang2022EigenstateThermalization,Deutsch2018EigenstateThermalization,Guhr1998RandommatrixTheories}, which is generally believed to hold when the Hamiltonian is non-integrable, meaning it has no exact solution.

 


Despite its success, the ETH restricts itself to the expectation values of \textit{local} observables. In contrast, newer theoretical perspectives have begun to shed light on the behavior of \textit{global} quantum systems. In particular, it was proposed that under certain quantum dynamics global quantum states homogeneously explore their available Hilbert space~\cite{Mark2024MaximumEntropy,Pilatowsky-Cameo2023CompleteHilbertSpace}. This phenomenon, dubbed \textit{Hilbert-space ergodicity} (HSE), is analogous to ergodicity in classical mechanics, whereby over time a system visits its state space uniformly (Fig.~\ref{Fig0}A). While HSE was originally~\cite{Pilatowsky-Cameo2023CompleteHilbertSpace} restricted only to certain symmetry-free models - such as random unitary circuits (RUCs) and aperiodic sequences of unitaries, for which it was delineated as \textit{complete} HSE - recent theoretical work~\cite{Mark2024MaximumEntropy} has extended it to systems with symmetries and conservation laws, including non-integrable thermalizing systems and certain interacting-integrable systems (Fig.~\ref{Fig0}B). 

Importantly, a system obeying HSE entails consequences beyond standard predictions of quantum thermalization and the ETH. For instance, consider a global projective observable which is not described by the ETH, like the probability to measure a specific basis state (e.g. a `bitstring' for a system of qubits). HSE predicts that \textit{relative} probability fluctuations over time precisely follow an exponential (or Porter-Thomas~\cite{Porter1956FluctuationsNuclear,Boixo2018CharacterizingQuantum}) distribution. While this is well-known for the case of random quantum circuits~\cite{Boixo2018CharacterizingQuantum,Mullane2020SamplingRandom}, it applies to all HSE systems including time-independent, thermalizing Hamiltonian dynamics. Given the power and generality of this framework, it is important to first test its existing conjectures before exploring further generalizations and potential practical applications.

Here we employ an experimental Rydberg quantum simulator~\cite{Bernien2017ProbingManybody,Browaeys2020ManybodyPhysics,Shaw2024BenchmarkingHighly} undergoing thermalizing dynamics to study both existing and new predictions of HSE. To start, we find temporal fluctuations of bitstring probabilities are consistent with the emergence of the Porter-Thomas distribution, a result also born out by numerics. Then, we extend the framework of HSE to the case of a system interacting with its environment to make concrete predictions beyond the scope of conventional quantum thermalization, both for the case of closed and open quantum systems.

For closed quantum systems, we consider the typical setting of quantum thermalization: when the environment is a complementary subsystem, which we call the intrinsic bath. For thermalization at infinite effective temperature, we find temporal fluctuations of projective observables follow a universal form as a direct consequence of HSE: the Erlang distribution~\cite{Leemis2008UnivariateDistribution}. Solely parameterized by the dimension of the bath, the Erlang distribution undergoes a smooth quantum-to-classical transition, progressing from the global Porter-Thomas distribution to a narrow Gaussian distribution as the intrinsic bath size increases. Using a precisely defined notion of effective bath dimension~\cite{Mark2023BenchmarkingQuantum}, we find consistent behavior for systems evolving at finite effective temperature or with conservation laws. Our results are exact and universal (i.e. independent of system details) for ergodic systems both globally and locally. This in contrast to studies of quantum thermalization which had typically either only bounded~\cite{Short2011EquilibrationQuantum,Reimann2008FoundationStatistical} or predicted the scaling of fluctuations with the total system dimension~\cite{Srednicki1999ApproachThermal,Nation2018OffdiagonalObservable,Nation2019ErgodicityProbes}; previous predictions have been made more exact for special cases like explicitly randomized dynamics~\cite{Bauer2020UniversalFluctuations} or the weakly non-integrable regime~\cite{Kiendl2017ManyParticleDephasing}.

For open quantum systems we focus our study on global observables. Such a setting is ubiquitous for noisy quantum experiments~\cite{Preskill2018QuantumComputing}. The coupling to the environment is generically thought to damp out fluctuations of observables, but we find the manner in which this occurs is highly dependent on the microscopic system-environment interaction for certain observables like bitstring probabilities. Based on our understanding of HSE principles, we devise a formalism to account for this behavior, and in particular we find that such fluctuations follow a hypoexponential distribution, a generalization of the Erlang distribution which captures the microscopic effects of the environment. We leverage the accuracy of this approach for practical applications by identifying classes of noise models with distinct signatures in the hypoexponential distribution in order to discriminate candidate models directly with many-body bitstring measurements. Our method can handle bespoke models with correlated noise and with heavy tails~\cite{Clader2021ImpactCorrelations}, both of which could be detrimental to quantum error correction~\cite{Cafaro2010QuantumStabilizer,Clader2021ImpactCorrelations,Carroll2024SubsystemSurface}. 

Overall, our work examines the role of ergodicity in open and closed quantum dynamics and studies the consequent emergence of universal statistical behavior. This leads to uncovering both fundamental and practical ramifications for modern quantum experiments.

\section{Hilbert-Space ergodicity}
\label{sec:background}
Hilbert-space ergodicity (HSE) makes universal predictions about the statistics of bitstring measurements readily available from quantum simulators. In this section, we first briefly recapitulate the theoretical concepts of HSE (see Ref.~\cite{Mark2024MaximumEntropy} and  Appendix~\ref{app:erlang_distribution} for further details); in subsequent sections we will discuss its original predictions for the distribution of global bitstring probabilities, and finally extend it to the broader settings of subsystems and open systems.

Central to HSE is the \textit{temporal ensemble}, the set of pure states, $|\Psi(t)\rangle=\exp(-i\hat{H}t/\hbar)|\Psi(t=0)\rangle$, evaluated at different times $t$ for evolution under a time-independent Hamiltonian, $\hat{H}$. Explicitly, if we discretize the time evolution with infinitesimal step size, $\delta$, and define the time $t_j=\delta j$ for $j\in\mathbb{Z}_+$, the temporal ensemble is the infinite set $\{|\Psi(t_j)\rangle\}_{j\in\mathbb{Z}_+}$. We also define the \textit{diagonal ensemble}~\cite{Reimann2008FoundationStatistical,Short2011EquilibrationQuantum}, which is the average over all states in the temporal ensemble,
\begin{align}
\hat{\rho}_d &= \lim_{T\rightarrow \infty}\frac{1}{T}\int_0^T dt\  |\Psi(t)\rangle\langle\Psi(t)| \\
&=\mathbb{E}_t[|\Psi(t)\rangle\langle\Psi(t)|] = \sum_n |c_n|^2|E_n\rangle\langle E_n|\,,
\label{eq:diagonal_ensemble_mt}
\end{align}
where $|E_n\rangle$ are energy eigenstates and $|c_n|^2 \equiv |\langle\Psi(0)|E_n\rangle|^2$ are their corresponding populations, set by the initial state $|\Psi(0)\rangle$.

In the case of time-independent Hamiltonian evolution considered here, the Schr\"odinger equation conserves the populations $|c_n|^2$, thus restricting the temporal ensemble to a subset of Hilbert space obeying energy conservation. On this manifold, the only remaining degrees of freedom are the complex phases $\text{arg}(\langle E_n|\Psi(t)\rangle)$. It was shown~\cite{Mark2024MaximumEntropy} that HSE equates the temporal ensemble with the so-called \textit{random phase ensemble}~\cite{Nakata2012PhaserandomStates,Mark2023BenchmarkingQuantum,Mark2024MaximumEntropy}. The only assumptions for this relation are the \textit{no-resonance conditions}, a commonly-held assumption~\cite{Srednicki1999ApproachThermal,Reimann2008FoundationStatistical,Linden2009QuantumMechanical,Short2011EquilibrationQuantum,Riddell2023NoresonanceConditions,Mark2024MaximumEntropy} that there are no high-order degeneracies in the energy spectrum (see Appendix~\ref{app:erlang_distribution}). The no-resonance conditions are believed to apply to any non-integrable quantum system. Their only known exceptions are systems of non-interacting spins and systems with exact solution by non-interacting fermions, such as the TFIM. These resonances are unstable to non-zero interactions, and it has been proven that almost-every local quantum system satisfies the no-resonance condition~\cite{Huang2021ExtensiveEntropy}.

Under this condition, we can define the higher order moments of the temporal ensemble in terms of the well-known moments of the Haar ensemble~\cite{Mark2024MaximumEntropy,Harrow2013ChurchSymmetric}. In particular, the $k$-th moment of the temporal ensemble is, up to corrections which are exponentially small in the system size, equal to
\begin{align}
\mathbb{E}_t[|\Psi(t)\rangle \langle \Psi(t)|^{\otimes k}] \approx \hat{\rho}_d^{\otimes k}\sum_{\sigma \in S_k} \text{Perm}(\sigma),
    \label{eq:temp_ens_moments_mt}
\end{align}
which, for $\hat{\rho}_d=\hat{I}/D$, are the moments of the Haar ensemble~\cite{Mark2024MaximumEntropy,Harrow2013ChurchSymmetric} (up to exponentially small corrections). Here $\text{Perm}(\sigma)$ is the permutation operator acting on $k$ copies of the Hilbert space with a permutation $\sigma$ in the symmetric group $S_k$ which takes any configuration $|i_1,i_2,\dots,i_k\rangle \in \mathcal{H}^{\otimes k}$ and maps it to $|i_{\sigma(1)},i_{\sigma(2)},\dots,i_{\sigma(k)}\rangle$. As an example, for $k=2$
\begin{align}
\mathbb{E}_t [ |\Psi(t) \rangle \langle \Psi(t)|^{\otimes 2}] \approx \hat{\rho}_d^{\otimes 2} (\hat{I} + \hat{S}),
\end{align}
where $\hat{I}$ is the identity and $\hat{S}$ is the swap operator between two copies of Hilbert space.

Crucially, this means that if we pick any random (late) time, then in a statistical sense the properties of the corresponding pure state $|\Psi(t)\rangle$ will be those of a state drawn from the uniform (i.e. Haar) random ensemble, reweighted as
\begin{align}
\label{eq:Haar_state_equivalency}
\ket{\Psi(t)} \approx\sqrt{D\hat{\rho}_d}\ket{\phi}~,~\ket{\phi}\sim \text{Haar}\,,
\end{align}
This is the general statement of HSE, that global quantum states uniformly randomly cover their \textit{available} Hilbert space, as constrained by conservation laws, symmetries, or a finite effective temperature, effects which are all encoded by the diagonal ensemble state $\hat{\rho}_d$. This factorization between the universal Haar behavior and the evolution-specific $\hat{\rho}_d$ then allows for straightforward evaluation of high-order, otherwise non-trivial quantities.

\begin{figure*}[t!]
	\centering
	\includegraphics[width=180mm]{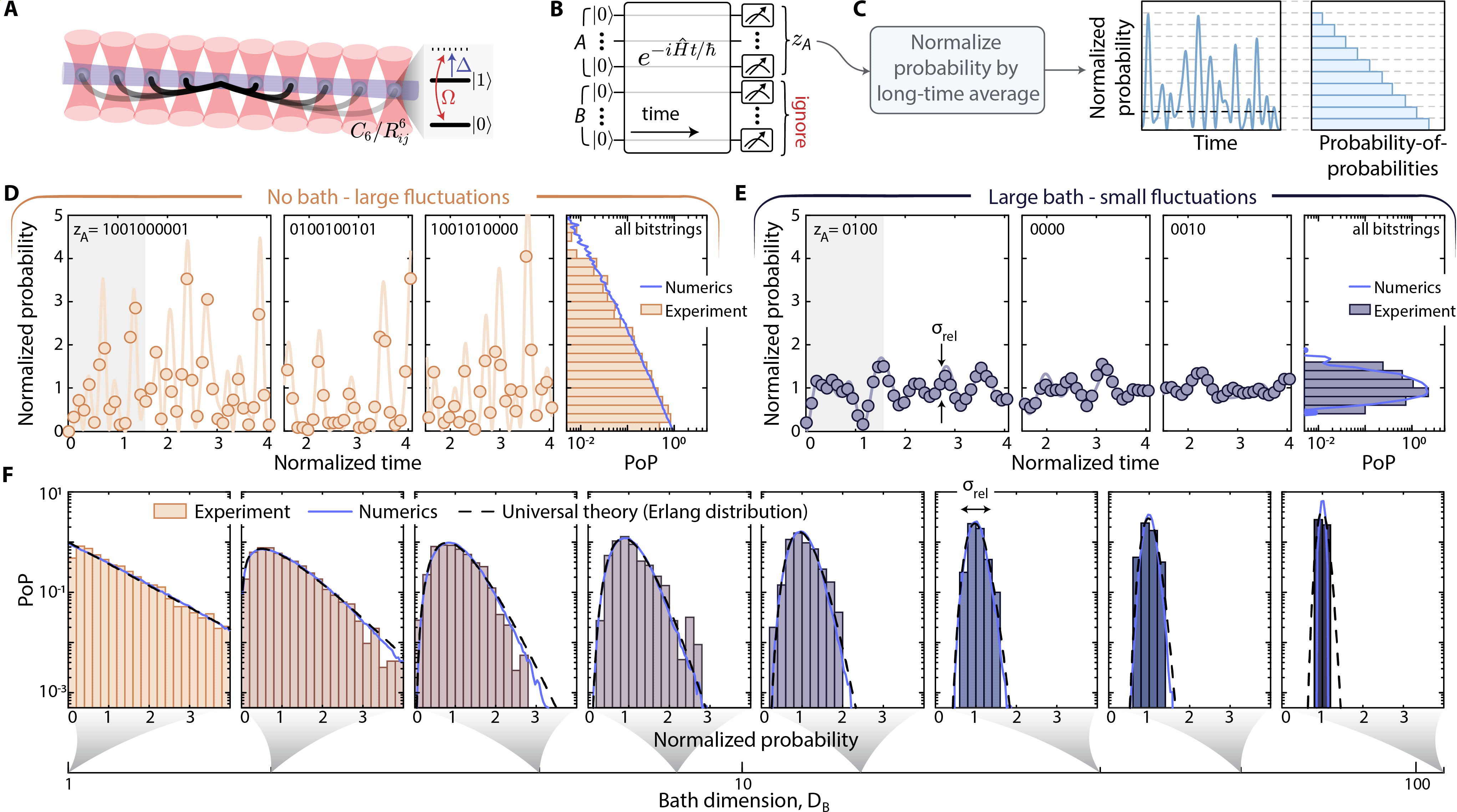}
	\caption{\textbf{HSE for system-bath interactions: a smooth quantum-to-classical transition for closed systems.} \textbf{A.} We employ a Rydberg quantum simulator with $N=10$ atoms to simulate an Ising-like Hamiltonian parameterized by detuning $\Delta$, Rabi frequency $\Omega$, interaction constant $C_6$, and interatomic distance $R$ (model details are in Appendix~\ref{app:experiment_details}). \textbf{B.} The all-zero state is time-evolved and measured producing a bitstring; in post-processing we sub-select the reduced bitstring $z_A$ in subsystem $A$, ignoring the measurement in subsystem $B$. \textbf{C.} To characterize relative fluctuations of the bitstring probabilities, probabilities are normalized by their long-time average and then binned into a histogram (indicated by dashed lines). This is the probability-of-probabilities (PoP) distribution. \textbf{D.} When $A$ is the entire system (i.e. there is no intrinsic bath, $N_A=10$ qubits, $N_B=0$ qubits), normalized bitstring probabilities are highly fluctuating past a short onset timescale (grey fill before this time, time is rescaled by the Rabi frequency). \textbf{E.} In contrast, when the bath is large ($N_A=4$ qubits, $N_B=6$ qubits), fluctuations are substantially reduced (e.g. typified by the temporal variance, $\sigma^2_\text{rel}$), and the PoP distribution is narrow. \textbf{F.} PoP distributions for many bath Hilbert-space dimensions, $D_B$, are in excellent agreement with our prediction that the PoP follows a universal form, the \textit{Erlang distribution} (see text). The Erlang distribution starts as an exponential distribution for $D_B=1$ (an existing prediction of HSE~\cite{Mark2024MaximumEntropy}) before becoming progressively more narrow and Gaussian as $D_B$ increases. $D_B$ values (indicated by grey zoom-out from lower axis) are constrained by the Rydberg blockade mechanism, see Appendix~\ref{app:experiment_details}.}
	\vspace{-0.5cm}
	\label{Fig2}
\end{figure*}

Specifically, we consider the probability $p(z,t)$ to measure a given bitstring $z$ over time, where the time-average probability $p_\text{avg}(z)$ is given by $p_\text{avg}(z)=\langle z|\hat{\rho}_d|z\rangle$. We then are interested not just in the time-average, but in temporal \textit{fluctuations} about this mean, i.e. higher order moments like the variance, etc. If we contract Eq.~\eqref{eq:temp_ens_moments_mt} with $\langle z|^{\otimes k}$ and $|z\rangle^{\otimes k}$ we find $\mathbb{E}_t[p(z,t)^k] = k!~ p_\text{avg}(z)^k$. where each of the $k!$ permutations have equal contribution. If we then divide out the $p_\text{avg}(z)$, and instead study a normalized bitstring probability $\tilde{p}(z,t)=p(z,t)/p_\text{avg}(z)$, we find that HSE predicts \textit{relative} fluctuations follow a universal form of simply
\begin{align}
\label{eq:kth_moment_PT}
\mathbb{E}_t[\tilde{p}(z,t)^k] = k!\,.
\end{align}
Intuitively, normalization by the mean eliminates the effects of symmetries and conservation laws which constrain HSE and make the $p_\text{avg}(z)$ non-equal.

\section{Testing HSE: The Probability-\\of-Probabilities distribution}
Having reviewed the basics of HSE, we now move to test whether these theoretical predictions are consistent with the behavior of an experimental quantum system. In particular we use an experimental Rydberg quantum simulator~\cite{Bernien2017ProbingManybody,Browaeys2020ManybodyPhysics} to implement a time-independent, global quench under the one-dimensional Ising-like Rydberg Hamiltonian, with parameters chosen such that the initial all-zero state, $|\Psi_0\rangle=|0\rangle^{\otimes N}$, thermalizes at infinite effective temperature within the Rydberg-blockaded subspace (Fig.~\ref{Fig2}A) - see Appendix~\ref{app:experiment_details} for experimental details. We consider a system size of $N{=}10$ qubits, which are measured after evolving for a time $t$, producing a global bitstring $z{\in}\{0,1\}^{N}$, and over many measurements we build up the global probability distribution $p(z,t)$. We then normalize these by $p_\text{avg}(z)$ (obtained from numerics) to obtain relative fluctuations $\tilde{p}(z,t)$ (Fig.~\ref{Fig2}B). We note that for numerical simulations, we consider the full Rydberg Hamiltonian, described in Appendix~\ref{app:experiment_details}, taking into account dynamics outside the blockaded Hilbert space. 

A first approach to testing HSE might be to directly calculate the moments $\mathbb{E}_t[\tilde{p}(z,t))^k]$, and compare against the prediction of Eq.~\eqref{eq:kth_moment_PT} order-by-order. However, we can directly compare at all orders simultaneously if instead of considering individual moments, we study the distribution which they encode. This so-called \textit{probability-of-probabilities} (PoP) is the central object of interest for the remainder of our work (Fig.~\ref{Fig2}C). Written as a function of the bitstring probability, $\tilde{p}_0$, we write the PoP as (where we drop the $z$ for notational simplicity)
\begin{align}
    P(\tilde{p}') = \mathbb{E}_{t}[\delta(\tilde{p}(t)-\tilde{p}')]&\equiv
    \lim_{T\rightarrow 
    \infty}\frac{1}{T}\int_0^T \delta(\tilde{p}(t)-\tilde{p}') dt\,,
\end{align}
where $\delta$ is the $\delta$-function. We construct the experimental PoP distribution by forming a histogram of $\tilde{p}(z,t)$ aggregated over all times, using a finite bin width $\Delta\tilde{p}$. 


The key property of the PoP is that the moments of the PoP are equal to the temporal fluctuations of $\tilde{p}(z,t)$ at all orders, 
\begin{align}
\int_0^\infty \tilde{p}^{'k} P(\tilde{p}')d\tilde{p}'
&=\lim_{T\rightarrow \infty}\frac{1}{T}\int_0^T \tilde{p}(t)^k dt\equiv\mathbb{E}_t[\tilde{p}(t)^k]\,.
\end{align}
In other words, the PoP encodes the full distribution of the bitstring probability values; this is in contrast to typical quantities like the cross-entropy~\cite{Boixo2018CharacterizingQuantum} which only encode second-order fluctuations  (and which require extensive classical computation). From Eq.~\eqref{eq:kth_moment_PT}, we identify the PoP for normalized bitstring probabilities under HSE as the Porter-Thomas~\cite{Porter1956FluctuationsNuclear,Boixo2018CharacterizingQuantum}  (exponential) distribution
\begin{equation}
    \label{eq:pt_emergence_mt}
    P(\tilde{p}) = \exp(-\tilde{p})\,,
\end{equation}
the same as for the case of random quantum dynamics~\cite{Boixo2018CharacterizingQuantum,Mullane2020SamplingRandom}.

For several representative choices of $z$, we observe the experimental $\tilde{p}$ are highly fluctuating (Fig.~\ref{Fig2}D, left). We can then aggregate probabilities from all bitstrings from intermediate evolution times, from which we find the PoP is a nearly exponential distribution (Fig.~\ref{Fig2}D, right). For the experiment here we focus on intermediate time dynamics such that the global fidelity remains high ($\gtrsim0.8$~\cite{Shaw2024BenchmarkingHighly}), i.e. such that the influence of the external environment is small. This is critical, as external decoherence can deform the PoP, as is visible in the slightly lowered experimental probability of $P(0)\approx0.5$ compared to the theoretical prediction of $P(0)=1$, visible in (Fig.~\ref{Fig2}D, right). We will elucidate the importance of this choice in the latter half of this work, where we shall study the deformation of the PoP at late times under the influence of the noisy environment. This deformation effect also forces our aggregation over many bitstrings as we can only experimentally sample a few tens of distinct early times, which is not sufficient to fully resolve the PoP on a single bitstring level. Aggregating over bitstrings does not modify the PoP because we can treat the probabilities $p(z,t)$ as independent random variables, both over time as well as between different bitstrings~\cite{Mark2023BenchmarkingQuantum}. These random variables are predicted to follow the same Porter-Thomas distribution, up to an overall factor of $p_\text{avg}(z)$ which is negligible here as the evolution is at infinite effective temperature. Further, in Fig.~\ref{EFig:singles} of the Appendix we numerically show the agreement between the PoP and the exponential distribution persists even on the level of a single bitstring sampled over time for clean, noise-free dynamics.

Further, while Eq.~\eqref{eq:pt_emergence_mt} is rigorously true only in the infinite time limit, remarkably we find the key signatures of our theory are visible experimentally already at these early times (see also Fig.~\ref{EFig:time_resolved} in the Appendix). Thus, the results of Fig.~\ref{Fig2}D stand as a concrete experimental observation consistent with Eq.~\eqref{eq:pt_emergence_mt} and HSE in ergodic Hamiltonian dynamics more generally.

\section{Generalizing HSE to subsystems}
While we have tested HSE in the established setting of global observables, the experimental data we have on hand allows us to investigate more than just the PoP of global bitstrings. By bipartitioning the system into two subsystems $A$ and $B$ (of sizes $N_A$ and $N_B$, and Hilbert-space dimensions $D_A$ and $D_B$ respectively), we can discard measurements in $B$ and examine the PoP formed only from \textit{local} bitstrings measured in $A$ (Fig.~\ref{Fig2}B). This allows us to directly study the fluctuations of a system coupled to a bath, where the dimension of the bath is readily controllable in post-processing. Explicitly, we consider the marginal probability, $p(z_A,t)$, to measure a given bitstring $z_A$ in $A$ at a particular time $t$. As in the global case, we will normalize probabilities by their long-time average (obtained from numerics), and then study their \textit{relative} fluctuations, e.g. $\tilde{p}(z_A,t)\equiv p(z_A,t)/p_\text{avg}(z_A)$. We then seek to study how the temporal fluctuations of $\tilde{p}(z_A,t)$ differ from the global case.

As an example, we take $N_A=4$ and $N_B=6$, and plot $\tilde{p}(z_A,t)$ for a few choices of $z_A$ (Fig.~\ref{Fig2}E). We see the bitstring fluctuations are greatly reduced (compared to those from the global system), though are still not completely damped out. Accordingly, when we form the PoP by aggregating over multiple experimental measurement times, we see the resultant PoP is much narrower than in the global case. Indeed, the width of the PoP decreases monotonically with increasing bath dimension $D_B$ (Fig.~\ref{Fig2}F), progressing from a wide exponential distribution for $D_B=1$ to a narrower, Gaussian-like distribution for $D_B\gg1$.

To understand this behavior, we apply HSE to derive the behavior of local observables. In particular, under HSE we predict the PoP to be
\begin{align}
P_\textrm{Erlang}(\tilde{p}; D_B) = \frac{D_B^{D_B}}{(D_B-1)!} \exp(-D_B\tilde{p})\ \tilde{p}^{D_B-1}\, ,
\label{eq:erlang}
\end{align}
which is the \textit{Erlang distribution}, parameterized solely by the intrinsic bath Hilbert-space dimension, $D_B$ (see Appendix~\ref{app:erlang_distribution} for the full derivation). 

Here we have assumed a large global Hilbert-space dimension, $D{\rightarrow}\infty$, and thermalization at infinite effective temperature. Thus, in the thermodynamic limit Eq.~\eqref{eq:erlang} dictates that temporal fluctuations of observables in subsystem $A$ are entirely controlled by the dimension of the bath, and \textit{not} by the dimension of the subsystem in which they are measured. The Erlang distribution is the convolution of $D_B$ independent exponential distributions (a viewpoint which we will return to and generalize in the following section concerning open system dynamics); for $D_B=1$, the Erlang is simply the exponential distribution, while for \mbox{$D_B{\rightarrow}\infty$} it approaches a narrow Gaussian distribution~\cite{Leemis2008UnivariateDistribution}. In other words, when the bath is small, large quantum fluctuations dominate (and recover the expected prediction of the exponential distribution), but when the bath is large, quantum fluctuations are damped out, leaving only small, Gaussian fluctuations. We identify this smooth quantum-to-classical crossover as integral to the transition from quantum mechanical to statistical mechanical descriptions of generic, closed, quantum many-body systems as the number of bath configurations increases.

\begin{figure}[t!]
	\centering
	\includegraphics[width=89mm]{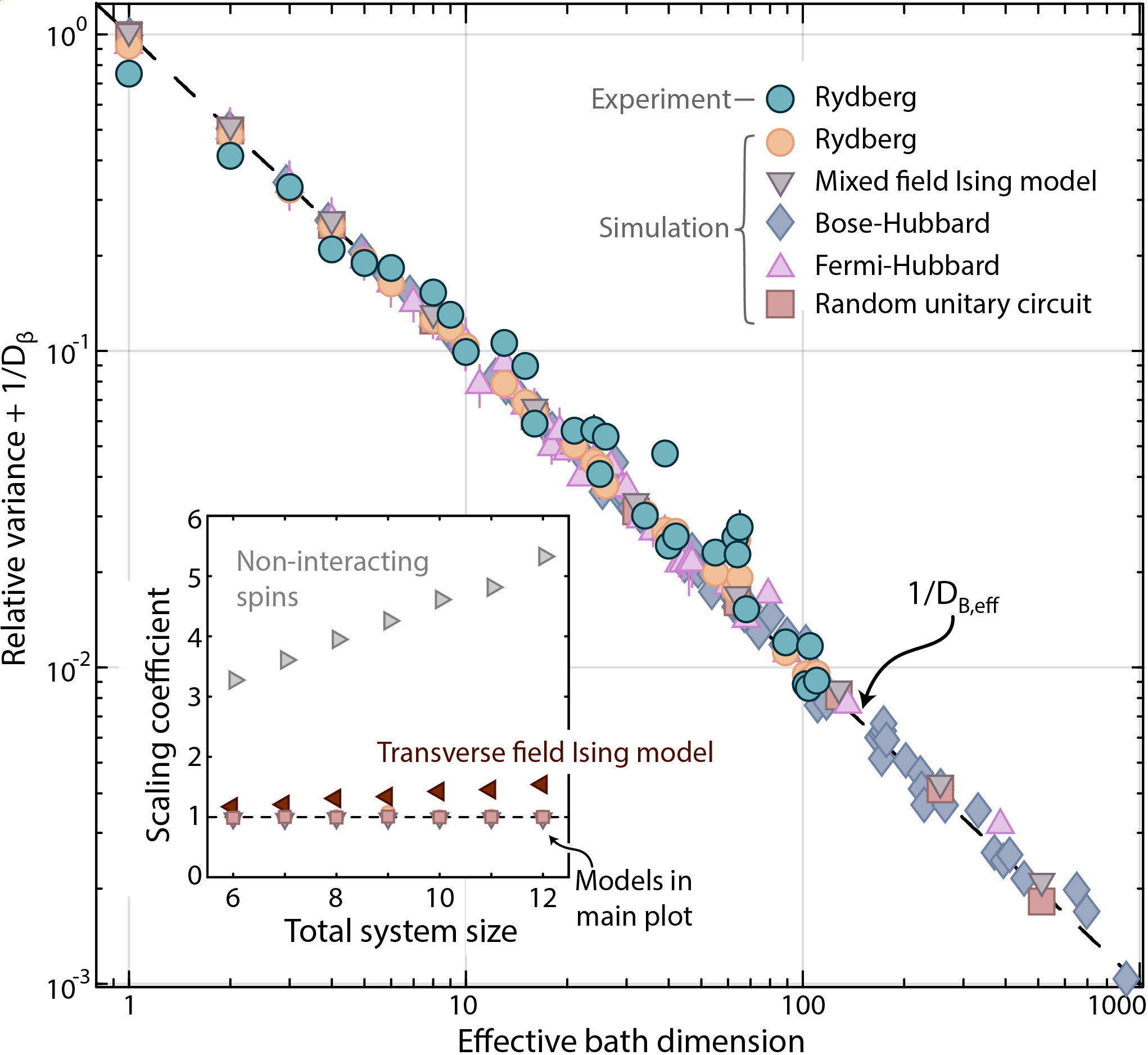}
	\caption{\textbf{Universality of the PoP variance.} The PoP variance (i.e. the variance of temporal fluctuations) scales inversely with the bath dimension for both the experiment and various one-dimensional numerical models exhibiting Hilbert-space ergodicity, in agreement with Eq.~\eqref{eq:effective_bath_dimension_mt}.  Markers are averaged over all bitstring realizations with an effective bath dimension in an interval from $[i-0.5,i+0.5)$, for $i\in\mathbb{Z}$, and error bars are the standard deviations over these realizations. Inset: The scaling coefficient $\langle D_{B,\text{eff}}(\sigma^2_\text{rel}+1/D_\beta)\rangle$ ($\langle\cdot\rangle$ denotes an average over choices of $D_{B,\text{eff}}$) is uniformly 1 for the models studied in the main plot, but deviates as a function of total system size for models for which Hilbert-space ergodicity does not hold. 
 }
	\vspace{-0.5cm}
	\label{Fig3}
\end{figure}

Across various choices of $D_B$, we see the Erlang distribution is in excellent agreement with experimental data across several orders-of-magnitude (Fig.~\ref{Fig2}F). We do observe slight deviations from the Erlang distribution in Fig.~\ref{Fig2}F for $D_B{\approx}D$ as our derivation assumed $D{\rightarrow}\infty$. Still, using HSE we can rigorously calculate arbitrary moments of the PoP accounting for finite $D$ and finite temperatures, as we demonstrate explicitly for low order moments in Appendix~\ref{app:erlang_distribution}. In particular, we predict the variance (second moment) of the PoP is
\begin{align}
\sigma^2_\text{rel}(z_A) \equiv \frac{1}{D_{B,\text{eff}}(z_A)}- \frac{1}{D_\beta(z_A)}\,,\label{eq:effective_bath_dimension_mt}
\end{align}
where $D_{B,\text{eff}}(z_A)$ and $D_\beta(z_A)$ are effective dimensions that appropriately generalize a $z_A$-dependent notion of Hilbert space dimension at finite effective temperature for the subsystem $B$ and the global system, respectively (see Appendix~\ref{app:erlang_distribution} for their detailed definitions). Importantly, at infinite temperature \mbox{$D_\beta\rightarrow D$} and \mbox{$D_{B,\text{eff}}\rightarrow D_B$}; then taking \mbox{$D{\rightarrow}\infty$} returns the expected variance of the Erlang distribution, $1/D_B$.

Significantly, the predictions of Eq.~\eqref{eq:erlang} for infinite temperature systems and Eq.~\eqref{eq:effective_bath_dimension_mt} for more general systems are not unique to the Rydberg Hamiltonian, but are instead universal for \textit{any} quantum many-body system which exhibits HSE (Fig.~\ref{Fig0}B). To observe this, we plot $\sigma^2_\text{rel}+1/D_\beta$ for various systems expected to exhibit HSE including our experimental Rydberg system, and also numerical simulations of one-dimensional systems including RUCs, a mixed field Ising model, and itinerant Hubbard models (Fig.~\ref{Fig3}) - see Appendix~\ref{app:numerical_details} for details of the various numerics. In all cases, we find good agreement with the analytic prediction of Eq.~\eqref{eq:effective_bath_dimension_mt}. 

\begin{figure*}[t!]
	\centering
	\includegraphics[width=149mm]{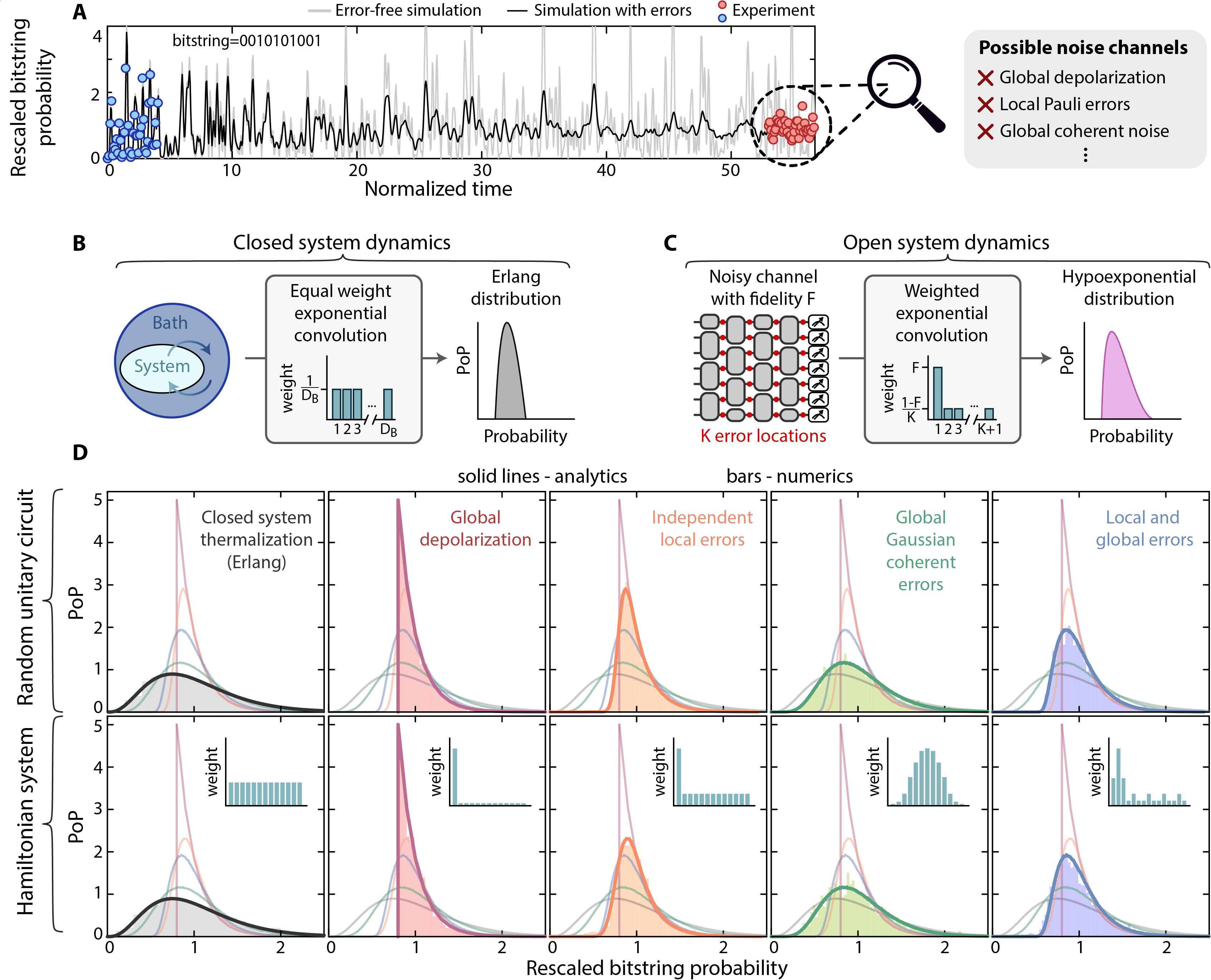}
	\caption{\textbf{Distinguishing noise channels using PoP distributions.} \textbf{A.} Interaction with the external environment through open system dynamics also leads to concentration of bitstring probabilities; as an example, we show bitstring probabilities following quench dynamics from experiment (markers), error-free simulation (grey line), and error-model simulation~\cite{Shaw2024BenchmarkingHighly} (black line). However, the microscopic error model is highly relevant to the manner of concentration, which allows us to discriminate different possible noise models directly from bitstring measurements. \textbf{B.} The Erlang distribution observed in the case of closed, infinite temperature thermalization can be alternatively derived as a convolution of independent exponential distributions with equal weights, one for each bath configuration. \textbf{C.} We extend this heuristic derivation to the case of open system dynamics, where the weights in the convolution are determined by the specifics of the noise channel, for example here a local error channel (see text), resulting in the PoP being a \textit{hypoexponential distribution}. \textbf{D.} Hypoexponential distributions for four different noise channels each with distinct hypoexponential weights (insets, see Appendix~\ref{app:numerical_details} for model details), with noise strengths set to result in the same fidelity $F\approx0.2$. Also shown is the Erlang distribution from closed system thermalization with an intrinsic bath of two qubits. PoP obtained from numerical simulations of RUC (top) and Hamiltonian evolution under a mixed field Ising model (bottom) show excellent agreement with  their corresponding analytical predictions (thick, dark lines), but can easily be distinguished from predictions of other bath types (thin, faint lines), both intrinsic and extrinsic.
        } 
	\vspace{-0.5cm}
	\label{Fig4}
\end{figure*}


On the contrary, for failure cases which do not exhibit HSE (because they do not fulfill the no-resonance conditions~\cite{Mark2023BenchmarkingQuantum}, see Appendix~\ref{app:erlang_distribution}), such as the transverse field Ising model (TFIM) and systems of non-interacting spins, the prediction of Eq.~\eqref{eq:effective_bath_dimension_mt} is violated (Fig.~\ref{Fig3} inset). Specifically, the estimated scaling coefficient with bath dimension -- $\langle D_{B,\text{eff}}(\sigma^2_\text{rel}+1/D_\beta)\rangle$ where $\langle\cdot\rangle$ is an average over choices of $D_{B,\text{eff}}$ -- is uniformly 1 for systems exhibiting HSE, but shows non-uniform and progressively greater deviations from 1 as a function of system size for non-ergodic cases. Both the TFIM and the 1D Fermi-Hubbard model are integrable (the former via Jordan-Wigner transformation~\cite{Pfeuty1970OnedimensionalIsing}, the latter by Bethe ansatz~\cite{Essler2005OneDimensionalHubbard}), but HSE holds for the latter and not the former because of the no-resonance conditions~\cite{Mark2023BenchmarkingQuantum}, and accordingly the 1D Fermi-Hubbard model exhibits the expected universal behavior in Fig.~\ref{Fig3} while the TFIM does not. These results support the universality of HSE, and demonstrate its testability in low-order observables, even incorporating the effects of finite system size.

\section{HSE principles for open systems\\and applications in noise learning}

So far we have studied HSE for closed system dynamics through the lens of the PoP distribution; for both local and global systems, we found the PoP has a universal, quantitative form. Now we ask how this behavior changes under the influence of an external, uncontrollable environment, as is the setting for all modern quantum experiments. 

Under such noisy open system dynamics the PoP distribution narrows~\cite{Boixo2018CharacterizingQuantum}, i.e. fluctuations of observables damp out (Fig.~\ref{Fig4}A), analogous to how fluctuations damped out for local subsystems as the bath dimension increased. We introduce a quantitative description of this open system behavior through an appropriate generalization of the Erlang distribution, leveraging our understanding of baths from closed systems into open ones. This provides a tool to distinguish between the different types of noise in the system even when the system and environment are far from equilibrium. --- applying it to experimental and numerical data yields predictions in agreement with \textit{ab initio} noise models of the system.

We first revisit our understanding of the Erlang distribution for closed system thermalization. Under HSE, the global bitstring probability, $\tilde{p}(z)$, is randomly sampled from an exponential distribution. The marginal probability, $\tilde{p}(z_A)$, is then a sum of $D_B$ uncorrelated exponential random variables, each with equal weight, $\omega{=}1/D_B$, as the infinite temperature bath has no preferred configuration. This process leads precisely to the Erlang distribution, i.e. the Erlang distribution is the convolution of $D_B$ exponential distributions~\cite{Leemis2008UnivariateDistribution} (Fig.~\ref{Fig4}B).

We can make a similar assumption for open system dynamics -- that for each bath configuration the bitstring distribution is sampled from an independent exponential distribution -- but now allow for non-equal weights that are determined by the microscopic details of the extrinsic coupling mechanisms. In other words, for each bath state, we take the system to be in an independently random state. As an example, we consider the case of RUC evolution where a single error can occur on a random qubit sometime during the evolution (Fig.~\ref{Fig4}C). For a circuit with $N$ qubits run for a depth $D$, there are \mbox{$K=N\times D$} potential error locations, and the circuit executes with a finite global fidelity of $F$. Thus, with a probability of $F$ the circuit will execute without error and will sample bitstring probabilities from an exponential distribution. However, if an error does occur at one of the $K$ locations, assuming the circuit is sufficiently scrambling~\cite{Boixo2018CharacterizingQuantum} it will sample a separate \textit{independent} exponential distribution. Without the ability to explicitly keep track of all errors, the result is an incoherent sum of independent exponential variables with weights $\omega_1=F$ and \mbox{$\{\omega_{2},\cdots,\omega_{K+1}\}=(1{-}F)/K$} (Fig.~\ref{Fig4}C).

For a given set of weights ${\omega_i}$ (which are positive and sum to 1), and associated probability distributions $\{p_i(z)\}$, the overall bitstring probability distribution is given by the weighted sum:
\begin{equation}
    p(z) = \sum_i \omega_i p_i(z)\,.
\end{equation}
When all weights are equal, the associated PoP is the Erlang distribution, but when weights are allowed to be non-equal, we identify the associated PoP as a \textit{hypoexponential distribution}~\cite{Leemis2008UnivariateDistribution,Nadarajah2008ReviewResults}, or generalized Erlang distribution, with PDF given by 
\begin{equation}
    P_\text{Hypo}(x) =  \Conv_i P_\text{Exp}(x/\omega_i)\label{eq:hypoexp_mt}
\end{equation}
where $\circledast$ denotes repeated convolution. While Eq.~\ref{eq:hypoexp_mt} does not have a closed form for general weights ${\omega_i}$, it can be analytically simplified for many relevant special cases and can always be numerically solved. Thus, we can quantitatively characterize largely arbitrary noise channels which couple the quantum system to the external environment simply by adjusting the weight vector. (Fig.~\ref{Fig4}C). See Appendix~\ref{app:hypoexponential_distribution} for explicit analytical constructions of the hypoexponential distribution for various error models, and Algorithm 1 therein for an example of how to numerically apply this approach for more bespoke error models. Note that the hypoexponential weights can be related to the Kraus operators of a given noise channel, see Appendix~\ref{app:kraus}.

To demonstrate the power of this formalism, we perform numerical simulations of RUC evolution and mixed field Ising model Hamiltonian evolution; explicitly we study the global PoP distribution measured at a fixed time, for which the results of HSE also apply~\cite{Mark2024MaximumEntropy} (see Fig.~\ref{EFig:singles} of the Appendix). In both cases we apply either: 1) global depolarizing error, 2) local amplitude damping errors, 3) global Gaussian coherent errors, or 4) a combination of errors (2) and (3). For each, the error strength is tuned so the final fidelity is $F{\approx}0.2$. By \textit{coherent error}, we mean a static, shot-to-shot under/over-rotation in either the qubit gates or the amplitude of the Hamiltonian terms, which here we take to be Gaussian-distributed and applied to all qubits globally. Crucially, determining the weight vectors and associated hypoexponential distributions for the different noise models does not require explicit simulation, and instead follows directly from the model definitions, see Appendix~\ref{app:hypoexponential_distribution}. We study these representative noise models as they are highly relevant to various aspects of quantum science. For instance, global depolarizing error is often assumed in studies of error-correcting codes, local amplitude damping is prevalent in T1-limited systems like many superconducting circuits~\cite{Burnett2019DecoherenceBenchmarking}, global Gaussian coherent errors are limiting for certain analog simulations including for Rydberg systems~\cite{Shaw2024BenchmarkingHighly}, and the combination of local and global errors is likely universal for all quantum platforms.

We note that the weights for the global depolarizing channel are the same as those of the local error channel with $K{\rightarrow}\infty$, consistent with Ref.~\cite{Dalzell2021RandomQuantum}. By contrast, however, the PoP for global coherent errors does not converge to the PoP of global depolarizing noise for increasing system sizes at fixed fidelity. Coherent errors could arise from non-Markovian Hamiltonian parameter variations~\cite{Shaw2024BenchmarkingHighly,Paladino2014NoiseImplications}, or systematic gate miscalibrations~\cite{Arute2019QuantumSupremacy,Burnett2019DecoherenceBenchmarking}; discriminating such noise sources from Markovian errors is crucial as they lead to different functional forms of fidelity decay which may be otherwise difficult to distinguish~\cite{Shaw2024BenchmarkingHighly,Cai2023StochasticError}, and their mitigation necessitates different models of error correction~\cite{Terhal2015QuantumError,Cafaro2010QuantumStabilizer,Clader2021ImpactCorrelations,Carroll2024SubsystemSurface}.

In all cases, we see excellent agreement between the numerical PoP distributions and their corresponding analytical hypoexponential predictions, while being visually distinct from those of other noise channels, and from thermalization with an infinite temperature bath with $D_B=4$ (Fig.~\ref{Fig4}D). Thus, on a qualitative level we can discriminate between a wide range of noise models through many-body bitstring probability measurements, even when those models lead to an equal final fidelity.

\begin{figure}[t!]
	\centering
	\includegraphics[width=89mm]{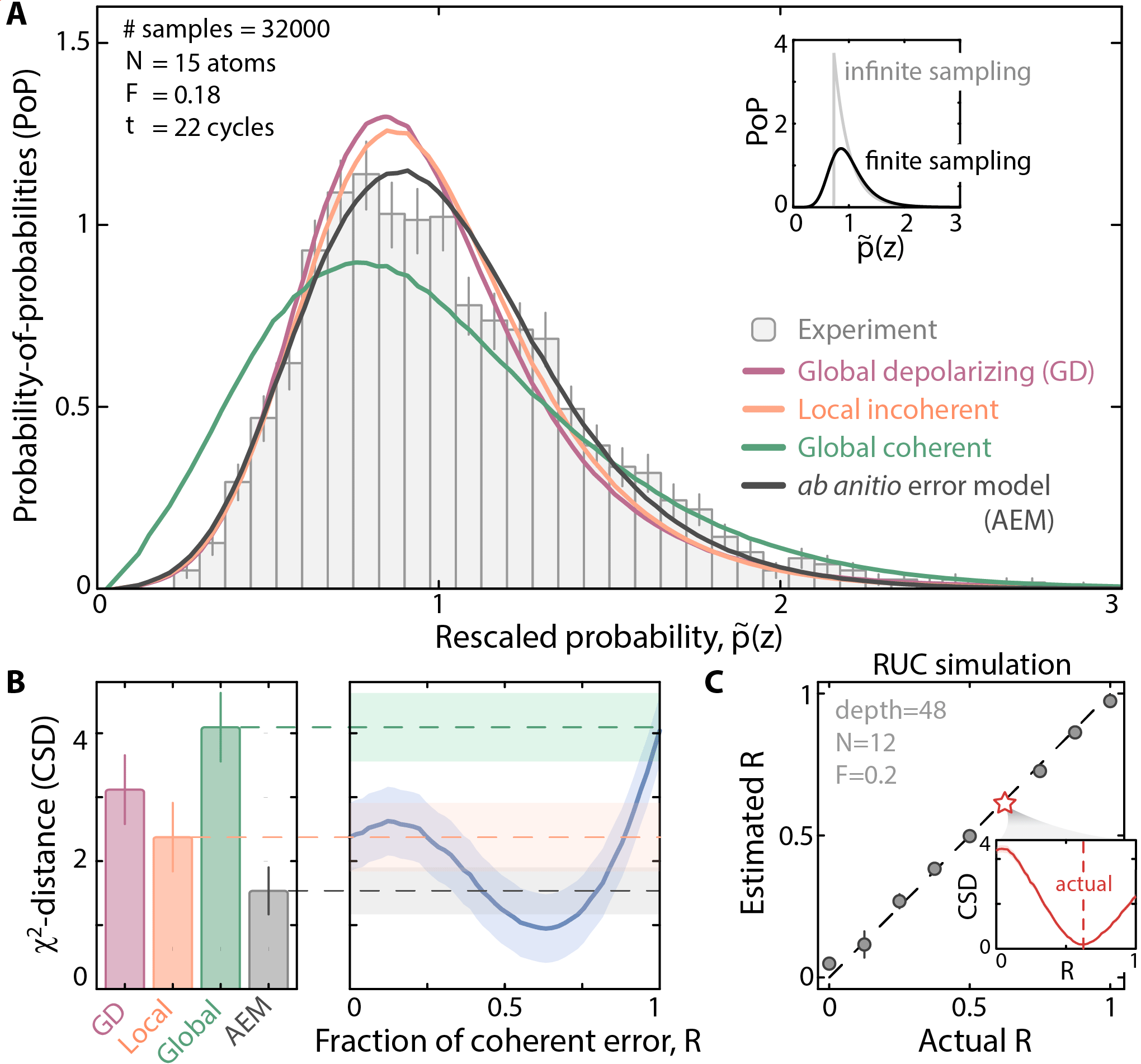}
	\caption{\textbf{Characterizing experimental noise. A.} We compare the PoP distribution obtained from the experiment following long-time evolution against various noise models, including our own \textit{ab initio} error model. Model PoPs are nontrivially modified by finite sampling effects (inset, see SM). \textbf{B.} We compute the symmetric $\chi^2$-distance (CSD) between the experimental distribution and the model PoPs (left), including to a mixture error model composed of global coherent error and local incoherent error with varying global coherent error fraction, $R$. \textbf{C.} We perform noisy RUC simulations with varying $R$, and find that estimating $R$ from the CSD minimum is well-correlated with the actual $R$ (SM).
        } 
	\vspace{-0.5cm}
	\label{Fig5}
\end{figure}

We then turn to apply this method quantitatively to our experimental quantum simulator. We program the same time-independent dynamics as we studied in the case of intrinsic thermalization, but now evolve out to late times with $N{=}15$ atoms such that the global fidelity is $F{\approx}0.18$, as estimated via the $F_d$ cross-entropy variant~\cite{Mark2023BenchmarkingQuantum,Shaw2024BenchmarkingHighly}. We then compare the experimental PoP distribution against the predictions for various noise models including our in-house \textit{ab initio} noise model. The use of a finite number of samples modifies the predicted PoP distributions (inset of Fig.~\ref{Fig5}A, see Appendix~\ref{app:finite_sampling_PoP}), which can potentially be circumvented by applying post-processing to rectify the distributions (Fig.~\ref{fig:ridge_regression}). Further, even with limited sampling the different noise models can be robustly discriminated by studying more sample-efficient low-order moments of the PoP (Appendix~\ref{app:PoP_moments}). 

Importantly, in Appendix~\ref{app:PoP_moments} we also derive the fidelity dependence of the low order moments of the global PoP under various noise channels. For instance, we find that the second moment of the PoP scales as $F^2$ for global depolarizing noise, but $F$ for global Gaussian coherent noise. These sharp dependencies underscore the necessity of operating the experimental system in a high fidelity regime when studying the effects of thermalization on the PoP, as we do in the early part of this work.

A qualitative evaluation (Fig.~\ref{Fig5}A) already suggests our \textit{ab initio} noise model is the best match to the experiment. To quantify this statement, we compute the normalized symmetric $\chi^2$-distance (CSD) between the experiment and the model PoPs, defined as 
\begin{align}
\label{eq:chi2distance}
\chi^2(\vec{x},\vec{y})=\frac{1}{B}\sum\limits_i^B \frac{(x_i-y_i)^2}{(x_i+y_i)/2}\,,
\end{align}
where $B$ is the number of bins in the PoP, $x_i$ is a bin drawn from experiment, and $y_i$ is a bin drawn from the model distribution. We find the distance is indeed minimized for the \textit{ab initio} noise model (Fig.~\ref{Fig5}B).

Further, we apply this technique to error models with a mixture of global Gaussian coherent error and local incoherent error. We vary the fractional contribution of global coherent error, $R$, and analytically construct the model PoP to compare against. We find the lowest CSD is found for $R\approx2/3$, suggesting room for future improvement of our \textit{ab initio} error model. This technique is applicable to both analog quantum simulators and digital quantum processors. For instance, we simulate noisy RUC dynamics for different choices of $R$, and for each compare the $R$ estimated from the CSD minimum versus the actual $R$ (Fig.~\ref{Fig5}C). We find the estimated and actual values are well-correlated (which improves with deeper circuits, see Fig.~\ref{fig:estimating_r}) across the entire range of $R$. This result implies that one can quantitatively estimate the relative contributions of coherent versus incoherent errors in noisy circuit dynamics using our methods.

Importantly, our analysis requires no circuit or Hamiltonian simulation, and relies only on the emergence of the Porter-Thomas distribution, which applies to both time-dependent and time-independent Hamiltonian evolution, as well as random quantum circuits. Further, it makes no assumption about the underlying noise model besides its weight vector. However, this also means it cannot discriminate between noise models with the same weights, such as distinguishing local bit flips from local phase flips. Thus, it should be considered a tool for efficiently identifying  experimentally-consistent models with bespoke noise distributions, which can then be further refined via additional measurements to distinguish different models with the same weights. More generally, beyond its potential practical applications this approach opens a new suite of theoretical analysis for understanding open system dynamics, as for instance the weights of the hypoexponential distribution are related to the system's Kraus operators, see Appendix~\ref{app:kraus}.

\section{Discussion}

The evolution of a quantum system interacting with its environment is of both fundamental and practical interest. Here, we have studied this process through the lens of Hilbert-space ergodicity~\cite{Mark2024MaximumEntropy} both for the case of interactions between complementary subsystems, and for noisy interactions with an external environment. 

For closed system thermalization at infinite temperature we have analytically, experimentally, and numerically found the PoP for bitstring measurements is universally the Erlang distribution. As a corollary, we showed experimental evidence consistent with existing predictions of HSE, namely the emergence of the Porter-Thomas distribution for global systems. Accordingly, there is a smooth quantum-to-classical transition with bath size from large, Porter-Thomas fluctuations to small, Gaussian fluctuations, highlighting the transition between quantum mechanical and statistical mechanical descriptions of physical systems as the number of bath configurations grow. This transition was observed universally for systems obeying HSE, even for systems which were integrable, which evolved at finite temperature, or which were constrained by conservation laws. While questions remain for calculating the PoP distribution at all orders for finite effective temperatures, we have made significant strides, showing how low-order moments of the PoP may be calculated as a function of an effective bath dimension and showing how the PoP may be obtained in some limits (Appendix~\ref{app:hypoexponential_distribution}). It remains open how our results extend to systems which do not exhibit Hilbert-space ergodicity, such as some non-interacting integrable systems~\cite{Pfeuty1970OnedimensionalIsing} or potentially non-thermalizing models like many-body localized systems~\cite{Nandkishore2015ManyBodyLocalization,Abanin2019ColloquiumManybody}.

For the case of open system dynamics, the hypoexponential model we introduce allows us to categorize different noise models via the weight vector and produce an analytical prediction for the PoP. This enables experimentally-accessible discrimination between candidate noise models purely through their effect on the many-body bitstring PoP, without expensive numerical simulations of quantum dynamics. We have demonstrated how bespoke noise models can be found through a quantitative comparison between model predictions and experimental statistics, as has been explored in the context of random circuit sampling~\cite{Liu2022BenchmarkingNearterm,Kim2021HardwareefficientRandom,Harper2020EfficientLearning}. Our technique is computationally inexpensive and applies equally to analog quantum simulators and digital quantum processors. Additional quantitative tests, such as comparing the low-order moments of the PoP distributions, may enhance discrimination power while being more sample-efficient. Fruitful questions remain, such as directly learning the hypoexponential weights via entanglement spectroscopy methods~\cite{Chung2014EntanglementSpectroscopya,Song2012BipartiteFluctuations}. Further, learning the weights of heavy-tailed and correlated noise distributions may guide improvements of quantum error correction schemes~\cite{Cafaro2010QuantumStabilizer,Clader2021ImpactCorrelations,Carroll2024SubsystemSurface}. As our method takes the fidelity as input, we note it is an explicit application of fidelity-estimation techniques~\cite{Choi2023PreparingRandom,Mark2023BenchmarkingQuantum,Boixo2018CharacterizingQuantum,Arute2019QuantumSupremacy,Shaw2024BenchmarkingHighly}.

Ultimately, our results probe new types of universal behavior visible to high precision quantum systems. In turn, the universality we uncover has applications for both improving quantum experiments and studying fundamental quantum dynamics. In total, our work demonstrates the power of applying a unified approach to studying many-body quantum systems, both for open and closed dynamics, through HSE principles. 

\vspace{3mm}
\begin{acknowledgements}
We acknowledge insightful discussions with Bill Fefferman, Andreas Elben, Gil Refael, and Federica M. Surace, and feedback on this manuscript from Elie Bataille, Richard Tsai, Xiangkai Sun, and Gyohei Nomura. We acknowledge support from the NSF QLCI program (OMA-2016245), the Institute for Quantum Information and Matter, an NSF Physics Frontiers Center (NSF Grant PHY-1733907), the Center for Ultracold Atoms, an NSF Physics Frontiers Center (NSF Grant PHY-1734011), the DOE (DE-SC0021951), the DARPA ONISQ program (W911NF2010021), the NSF CAREER awards (1753386) and (DMR-2237244), the AFOSR YIP (FA9550-19-1-0044), and the AFOSR (FA9550-23-1-0625). Support is also acknowledged from the U.S. Department of Energy, Office of Science, National Quantum Information Science Research Centers, Quantum Systems Accelerator. JC acknowledges support from the Terman Faculty Fellowship at Stanford. RF acknowledges support from the Troesh postdoctoral fellowship. PS acknowledges support from the IQIM postdoctoral fellowship. 
\end{acknowledgements}

\newpage
\setcounter{section}{0}
\onecolumngrid

\renewcommand\appendixname{APPENDIX}
\appendix
\renewcommand\thesection{\Alph{section}}
\renewcommand\thesubsection{\arabic{subsection}}

\section{Details of the experiment}
\label{app:experiment_details}
Our experiment is a Rydberg atom array quantum simulator~\cite{Browaeys2020ManybodyPhysics} trapping individual strontium-88 atoms in optical tweezers~\cite{Cooper2018AlkalineEarthAtoms,Norcia2018MicroscopicControl}. Up-to-date details of our apparatus may be found in previous works~\cite{Scholl2023ErasureConversion,Shaw2024BenchmarkingHighly}. In brief, one-dimensional defect-free~\cite{Endres2016AtombyatomAssembly,Barredo2016AtombyatomAssembler} arrays of atoms are initialized in the $5s^2$ $^{1}S_{0}$ state, and are cooled on the narrow-line $5s^2$ $^{1}S_{0}$ $\leftrightarrow$ $5s5p$ $^{3}P_{1}$ transition close to their motional ground state. Atoms are prepared into the long-lived $5s5p$ $^{3}P_{0}$ \textit{clock state} with a preparation fidelity of $\approx0.996$ per atom, which we then treat as a metastable ground state, $|0\rangle$. Atoms are then driven globally to the $5s61s$ $^{3}S_{1},m_{J}{=}0$ \textit{Rydberg state}, $|1\rangle$ while the traps are briefly blinked off.

Following Hamiltonian evolution, state projection is performed by autoionizing~\cite{Madjarov2020HighfidelityEntanglement} the Rydberg atoms, leaving them dark to our fluorescence imaging, with a fidelity of ${\sim}0.9996$ per atom. Atoms in the clock state are pumped into the imaging cycle, from which we map atomic fluorescence to qubit state~\cite{Covey20192000TimesRepeated,Madjarov2020HighfidelityEntanglement} with a detection fidelity ${\gtrsim}0.9995$ per atom~\cite{Covey20192000TimesRepeated,Scholl2023ErasureConversion}. For an initially defect-free array, this results in a series of bitstrings associated with the measured qubit states in the array; note that we discard any experimental bitstrings for which initial rearrangement failed.

\subsection{System Hamiltonian}

The Hamiltonian of our system is given by:
\begin{align}
\hat{H}/\hbar=\Omega\sum_i \hat{S}_i^x -\Delta\sum_i \hat{n}_i + \frac{C_6}{a^6} \sum_{i>j} \frac{\hat{n}_i \hat{n}_j}{|i-j|^6}, \label{eq:RydbergHam}
\end{align}
which describes a set of interacting two-level systems, labeled by site indices $i$ and $j$, driven by a laser with Rabi frequency $\Omega$ and detuning $\Delta$. The interaction strength is determined by the $C_6$ coefficient and the lattice spacing $a$. Operators are $\hat{S}^x_i =(\ket{1}_i\bra{0}_i + \ket{0}_i \bra{1}_i)/2$ and $\hat{n}_i = \ket{1}_i\bra{1}_i$, where $\ket{0}_i$ and $\ket{1}_i$ denote the electronic ground and Rydberg states at site $i$, respectively. Both $\Omega$ and $\Delta$ are tunable. 

We calibrate Hamiltonian parameters first via independent measurements, and then optimize further through maximum-likelihood-estimation (MLE) based on fidelity estimation. This technique is described in detail in our previous work~\cite{Choi2023PreparingRandom}; in brief we use the accumulated experimental bitstrings to calculate the many-body fidelity estimator~\cite{Mark2023BenchmarkingQuantum,Choi2023PreparingRandom,Shaw2024BenchmarkingHighly}, $F_d$, through cross-correlation with a numerical simulation. We then vary the simulation parameters to find the optimal fidelity overlap with the experiment.

For the $N=10$ data at approximately infinite temperature shown in Figs.~\ref{Fig2} and~\ref{Fig3} of the main text, the Hamiltonian parameters are chosen to be: $\Omega/2\pi=5.3$ MHz, $\Delta/2\pi=0.6$ MHz, $C_6/2\pi=254$ GHz$\times\mu\text{m}^6$, and $a=3.77\ \mu$m (the lattice spacing is calibrated with a laser-based ruler~\cite{Shaw2024MultiensembleMetrology}). For the $N=15$ data at approximately infinite temperature shown in Fig.~\ref{Fig5} of the main text, the Hamiltonian parameters are the same, but with $\Omega/2\pi=5.4$ MHz. For the $N=10$ data at finite temperature in Fig.~\ref{EFig:finite}, the parameters are the same, but with $\Omega/2\pi=5.3$ MHz, and $\Delta/2\pi=4.6$ MHz.

We typically normalize Hamiltonian evolution times by the Rabi frequency, with a corresponding unit of \textit{cycles}.

\subsection{Data analysis}
Through measurement, we accrue a set of time-labeled bitstrings, $\{z\}$. When normalizing bitstring probabilities, we obtain the temporal long time average directly from the \textit{diagonal ensemble}~\cite{Deutsch1991QuantumStatistical,Rigol2008ThermalizationIts,Mark2023BenchmarkingQuantum},
\begin{align}
\hat{\rho}_d = \sum_n |c_n|^2|E_n\rangle\langle E_n|\,,
\end{align}
where $|E_n\rangle$ are energy eigenstates, and \mbox{$c_n = |\langle E_n | \Psi_0\rangle|^2$} are overlaps with the initial state. Any observable, including bitstring probabilities, of $\hat{\rho}_d$ is equal to the same observable obtained from an infinite time average:
\begin{align}
p_\text{avg}(z)=\langle z|\hat{\rho}_d|z\rangle=\int_0^\infty dt\  |\langle z|\Psi(t)\rangle|^2\,.
\end{align}

A major feature of our system Hamiltonian is the Rydberg blockade mechanism~\cite{Browaeys2020ManybodyPhysics}, whereby the energy eigenspectrum is naturally stratified into a number of widely spaced sectors. In particular, the nearest-neighbor interaction is ${\approx}20\times$ greater than the next largest energy scale, so cases where neighboring pairs of atoms are both excited to the Rydberg state are greatly suppressed. The resulting effective Hilbert-space dimension, $D$, of a system with $N$ qubits is given by $\text{Fib}(N+2)\approx1.6^N$, where $\text{Fib}$ is the Fibonacci function. To be clear, statements in this work that the Rydberg system thermalizes at infinite effective temperature refer only to thermalization within the blockade-satisfying subspace, not across the entire Hilbert space.

For $N=10$, we find ${\approx}99\%$ of all experimental bitstrings are in the blockade-satisfying energy sector at short times; for data shown in the main text, we refine $\{z\}$ by discarding all realizations not in this sector. For the late-time $N=15$ data the number of blockade-satisfying bitstrings drops to ${\approx}75\%$, which we attribute mainly to spontaneous decay~\cite{Scholl2023ErasureConversion}, though leave careful investigation to future work. We stress that the number of blockade-satisfying bitstrings decreases only mildly as a function of system size~\cite{Shaw2024BenchmarkingHighly}, and is not a major impediment to the studies of either closed-system thermalization or open system decoherence that we carry out in this work.

The effective dimension of the intrinsic bath, $D_B$, is altered by the blockade constraint. Specifically, the blockade must be accounted for at the boundary between subsystems $A$ and $B$, such that $D_B$ then depends on the bitstring measured in $A$. As an example, consider the system $A$ as the leftmost atoms in a one dimensional array with ten total atoms, and $B$ as the remaining $N_B=9$ atoms, such that the global bitstring is written as $z=z_Az_B$. For the subsystem bitstring $z_A=0$, there are no possible blockade violation across the boundary between $A$ and $B$, and so the bath dimension is given by $D_B(z_A=0)=\text{Fib}(N_B+2)=89$. However, for the subsystem bitstring $z_A=1$, the blockade constraint forbids the leftmost qubit in $B$ from also being a 1. Thus the bath dimension is $D_B(z_A=1)=\text{Fib}((N_B-1)+2)=55$. In general, we allow $A$ to be any contiguous block of atoms in the array (meaning $B$ may be discontiguous), and for instance in the analysis of Fig.~\ref{Fig2}C of the main text we aggregate over bitstrings in $A$ with equal effective bath dimension.

The variance for the random unitary circuit in Fig.~\ref{Fig3} is taken as the median over 1000 circuit realizations. In Fig.~\ref{Fig5}, The error model prediction is calculated via a Monte Carlo wavefunction approach~\cite{Molmer1993MonteCarlo,Choi2023PreparingRandom,Shaw2024BenchmarkingHighly} from which bitstrings are sampled at an equal number as the experiment. Experimental error bars in Fig.~\ref{Fig5} are analytic based on the Poisson sampling errors: for a histogram bin with $m$ counts, its standard error is $\sqrt{m}$ (see Section~\ref{subapp:finite_sampling_PoP} of this SI).

\section{Numerical simulation parameters}
\label{app:numerical_details}
Here we summarize the parameters used in numerical simulations throughout the main text. In general, PoP distributions from numerics are aggregated over several hundred cycles of evolution, except where otherwise noted.

\subsection{Error free simulations}

\textbf{Rydberg:} Rydberg simulation parameters are the same as for experiment, as given in a previous section.

\textbf{MFIM:} For the mixed field Ising model simulations in Fig.~\ref{Fig3} and Fig.~\ref{EFig:finite}, we utilize the following Hamiltonian~\cite{Kim2014TestingWhether}:
\begin{align}
\label{eq:ham_mfim}
\hat{H}/\hbar  = \sum_{i=1}^N (h_x\hat{S}_i^x + h_y \hat{S}_i^y - \zeta \hat{S}_i^z)  + \sum_{i=1}^N J \hat{S}_i^x\hat{S}_{i+1}^x,
\end{align}
where $h_x = 0.41, h_y = 0.45$, and $J=1$. The field $\zeta=0$ corresponds to infinite temperature evolution and $\zeta=0.86$ to finite temperature evolution. In Fig.~\ref{fig:scrooge_hypoexp}, we instead set $\zeta=0$ but tune the initial state as $\ket{\Psi_0} = \left[\exp(i \theta \hat{S}^x)|0\rangle\right]^{\otimes N}$ for various angles $\theta$.

\textbf{TFIM:} For the transverse field Ising model simulations in the inset of Fig.~\ref{Fig3} we consider a one-dimensional open chain under the following Hamiltonian:
\begin{align}
\label{eq:ham_tfim}
\hat{H}/\hbar  = -\sum_i \hat{S}_i^x  - 3\sum_{i\neq j} \hat{S}_i^z\hat{S}_j^z.
\end{align}
We use as an initial state of $|\Psi(t=0)\rangle=|+\rangle^{\otimes N}$.

\textbf{Non-interacting spins:} Non-interacting spins studied in the inset of Fig.~\ref{Fig3} are initialized in their ground state, and then evolved with equal Rabi frequency and no detuning, thus undergoing disconnected, correlated Rabi oscillations.

\textbf{Bose-Hubbard:} 
We simulate the Bose-Hubbard model with $N$ particles on a one-dimensional open chain with length $M$:
\begin{align}
    \label{eq:ham_BH}
    \hat{H}/\hbar = -\Omega \sum_{j=1}^{M-1} \left(\hat{b}^\dagger_j \hat{b}_{j+1} + \text{h.c.}\right) + \frac{U}{2} \sum_{j=1}^M \hat{n}_j\left(\hat{n}_j-1\right),
\end{align}
where $\hat{n}_j = \hat{b}^\dagger_j \hat{b}_j$. We choose parameter values $(\Omega,U) = (1,2.87)$. In our simulations, we have $M=N$, and we take the initial state to be the
one with one particle at each site, i.e.~$\ket{1,1\dots,1}$.

\textbf{Fermi-Hubbard:}
The 1D Fermi-Hubbard model Hamiltonian is:
\begin{equation}
    H_\text{FH} =  \sum_{j=1}^N \left[-\Omega\sum_{\sigma = \uparrow,\downarrow}  \left(c^\dagger_{j,\sigma} c^{\mathstrut}_{j+1,\sigma} + \text{h.c.}\right) + U n_{j,\uparrow} n_{j,\downarrow}\right]~,
\end{equation}
where $c^\dagger_{j,\sigma}$ is the creation operator of a fermion at site $j$, with spin $\sigma \in \{\uparrow, \downarrow\}$, and $n_{j,\sigma} = c^\dagger_{j,\sigma}c^{\mathstrut}_{j,\sigma} \in \{0,1\}$ is the fermion occupation number. We use the parameter value $(\Omega,U) = (1,1)$. We take as initial state the anti-ferromagnetic state at half filling, $\ket{\uparrow,\downarrow, \dots}$.

\textbf{Random Unitary Circuit:} 
For RUC simulations we consider a brickwork pattern of SU(4) gates.

\subsection{Open system simulations}

\textbf{Rydberg Hamiltonian dynamics:}
Our \textit{ab initio} error model for Rydberg Hamiltonian dynamics has been described previously~\cite{Choi2023PreparingRandom}, and shown to be accurate to experiment in several regimes~\cite{Choi2023PreparingRandom,Scholl2023ErasureConversion,Shaw2024BenchmarkingHighly}. We utilize stochastic wavefunction evolution~\cite{Molmer1993MonteCarlo} with around a thousand trajectories, incorportating various pre-calibrated noise sources such as laser noise, atomic motion, and spontaneous decay, amongst others. We note that many of these noise sources generically are highly non-Markovian with important impact on the many-body fidelity~\cite{Shaw2024BenchmarkingHighly}. Further, spontaneous decay channels are potentially non-unital, and predominantly dominated by leakage errors out of the qubit subspace~\cite{Scholl2023ErasureConversion}.

\textbf{Spin chain Hamiltonian dynamics:} \textit{Amplitude damping --} To obtain the bottom row of Fig.~\ref{Fig4} of the main text, we simulate the open system dynamics of the mixed Field Ising model [Eq.~\eqref{eq:ham_mfim}] in the presence of amplitude damping, or spontaneous decay $|0\rangle\!\langle 1|$ errors, with the quantum trajectory method~\cite{Molmer1993MonteCarlo,Daley2014QuantumTrajectories}. In all trajectory methods, we perform a sufficient number of trajectories, typically 1000 for a system size of $N=12$, to see convergence of the PoP.

\textit{Global coherent errors --} To simulate global coherent errors, we again perform a trajectory method, where in each trajectory, we sample the parameter $J$ in Eq.~\eqref{eq:ham_mfim} from a normal distribution $N(J_0,\sigma_J^2)$, with $J_0 = 1$ and $\sigma_J$ tuned to achieve the desired fidelity at a given time.

\textit{Mixed error sources --} To simulate both amplitude damping and global coherent errors, we simply use a quantum trajectory method with both noise sources enabled.

\textbf{Random unitary circuits:} \textit{Local errors --}
We simulate the open system dynamics of a random unitary circuit (RUC) in the presence of amplitude damping errors. We evolve the density matrix by applying a layer of random unitary gates, before applying Kraus operators of the amplitude damping channel (for the top row of Fig.~\ref{Fig4})  independently at every qubit. 

\textit{Global coherent errors --} In order to model global coherent errors in a RUC, we parameterize each ideal random two-qubit gate as an $SU(4)$ rotation about some axis and with some angle. In each run we model the angles of each gate to be systematically over- or under-rotated by a factor which is constant over all gates within each run, but sampled from a normal distribution over many runs. For the mixed error sources in Fig.~\ref{Fig4}, we perform this stochastic over- and under-rotation with the density matrix evolution method outlined above.

\textit{Mixed error sources --} For the mixed error sources in Fig.~\ref{Fig5}, we use a quantum trajectory method with both bitflip errors and stochastic over- and under-rotation errors (outlined above).

\section{Deriving the Erlang distribution from Hilbert-space ergodicity}
\label{app:erlang_distribution}
In this work, we studied closed system dynamics and predicted the distribution functions of bitstring probabilities (or, probability-of-probabilities, PoP) in a subsystem which is strongly coupled to the rest of the system that serves as a bath. Our main quantitative predictions were Eq.~\eqref{eq:erlang}, the Erlang distribution, for such systems at infinite effective temperature and large total system sizes, as well as Eq.~\eqref{eq:effective_bath_dimension_mt} for the variance of the PoP at arbitrary temperatures. 

Here, we provide a rigorous derivation of the Erlang distribution, repeating some of the statements from Section~\ref{sec:background} for completeness. First we formally define the probability-of-probabilities (PoP) distribution and the no-resonance conditions upon which Hilbert-space ergodicity depends. Then, we derive the Erlang distribution which describes the PoP of subsystems thermalizing at infinite effective temperature, during closed system dynamics and in the large system size limit. In more technical sections, we then relax some of our assumptions and show how arbitrary order moments of the PoP may be calculated even at finite effective temperatures and finite system sizes. We also note that we provide a different, more heuristic, derivation of the Erlang distribution in Section~\ref{subsubsec:erlang}.

\subsection{The probability-of-probabilities distribution}
The PoP distribution for a single bitstring, $z$, is a histogram of the bitstring probabilities, $p(z,t)$, formally over infinite time. For a given probability, $p_0$, we write it as (where we drop the $z$ for notational simplicity)
\begin{align}
    P(p_0) &= \mathbb{E}_{t}[\delta(p(t)-p_0)]\equiv
    \lim_{T\rightarrow 
    \infty}\frac{1}{T}\int_0^T \delta(p(t)-p_0) dt\,.
\end{align}
The PoP encodes all temporal fluctuations of the time-series $p(t)$ as its moments,
\begin{align}
\int_0^1 p_0^k P(p_0)dp_0
&=\lim_{T\rightarrow \infty}\frac{1}{T}\int_0^T \int_0^1 p_0^k\delta(p(t)-p_0) dp_0 dt\\
&=\lim_{T\rightarrow \infty}\frac{1}{T}\int_0^T p(t)^k dt\equiv\mathbb{E}_t[p(t)^k]\,.
\end{align}
The PoP is defined here for a single bitstring over all evolution times, but it can alternatively be defined for a single time over all bitstrings, a distinction which we will revisit below.

In this study, we are chiefly interested in studying \textit{relative} temporal fluctuations, i.e. normalizing the time-series $p(t)$ to have mean 1. For random-circuit dynamics, this is simply the rescaling $\tilde{p} \equiv Dp$, while in Hamiltonian dynamics, this is $\tilde{p}(z) \equiv p(z)/p_\text{avg}(z)$, where $p_\text{avg}(z) \equiv \mathbb{E}_t[p(z,t)]$ is the time-averaged probability of $z$. This accordingly rescales the PoPs, which we will write as $P(\tilde{p}) d\tilde{p}$ (occasionally omitting $d\tilde{p}$). The $p_\text{avg}(z)$ may be found from the diagonal ensemble:
\begin{align}
\hat{\rho}_d = \sum_n |c_n|^2|E_n\rangle\langle E_n|\,,
\label{eq:diagonal_ensemble}
\end{align}
where $|E_n\rangle$ are energy eigenstates, and \mbox{$c_n = |\langle E_n | \Psi_0\rangle|^2$} are overlaps with the initial state. Any observable, including bitstring probabilities, of $\hat{\rho}_d$ is equal to the same observable obtained from an infinite time average:
\begin{align}
\label{eq:pdiag}
p_\text{avg}(z)=\langle z|\hat{\rho}_d|z\rangle=\lim_{T\rightarrow \infty}\frac{1}{T}\int_0^T dt\  |\langle z|\Psi(t)\rangle|^2\,.
\end{align}

\subsection{The no-resonance conditions}
\label{subsubsec:no_resonance}

As described in Section~\ref{sec:background}, the chief object of our interest is the \textit{temporal ensemble}, the set of pure states, $|\Psi(t)\rangle$, evaluated at different times $t$ for evolution under a time-independent Hamiltonian, $\hat{H}$. The temporal ensemble has been previously studied~\cite{Nakata2012PhaserandomStates,Mark2023BenchmarkingQuantum,Mark2024MaximumEntropy}, where at late times it was connected to the \textit{random phase ensemble}, the ensemble of quantum states with well-defined amplitudes in the energy eigenbasis, but random phases. This connection depends on so-called \textit{no-resonance conditions}~\cite{Reimann2008FoundationStatistical}, which are colloquially the absence of high-order degeneracies in the energy eigenspectrum. More concretely, a Hamiltonian $\hat{H}$ satisfies the $k$-th no-resonance condition if for any two sets of $k$ eigenvalues $\{E_{\alpha_j}\}_{j=1}^k$ and $\{E_{\beta_j}\}_{j=1}^k$ the equation
\begin{equation}
    E_{\alpha_1} + E_{\alpha_2} + \cdots + E_{\alpha_k} = E_{\beta_1} + E_{\beta_2} + \cdots + E_{\beta_k}~ \label{eq:k_no_resonance}
\end{equation}
is true if and only if the sets of indices $(\alpha_1, \alpha_2, \dots, \alpha_k)$ and  $(\beta_1, \beta_2, \dots, \beta_k)$ are equal up to reordering. For example, the 2nd-no resonance condition states that the relation $E_a+E_b=E_c+E_d$ is true if and only if $a = c, b=d$ or $a=d,b=c$. That is, there are no ``accidental resonances". Existence of $k$-th no-resonance conditions for any order $k$ guarantee the temporal ensemble is equivalent to the random phase ensemble in the infinite time limit~\cite{Mark2024MaximumEntropy}; as shown in Ref.~\cite{Mark2024MaximumEntropy} this then is equivalent to the statement that Hilbert-space ergodicity holds. We note that the while the statements of HSE require the no-resonance conditions to hold at all orders, this is not necessary for trivial 1st order degeneracies, which can be eliminated by working with representative states from projections onto degenerate subspaces~\cite{Mark2023BenchmarkingQuantum}.

\subsection{The Erlang distribution from Hilbert-space ergodicity}
In Section~\ref{sec:background} we described how the Porter-Thomas distribution could be derived for global systems using Eq.~\eqref{eq:temp_ens_moments_mt}, the moments of the temporal ensemble. Similarly, we can also derive that the rescaled marginal distributions $p(z_A)/p_\text{avg}(z_A)$ follow the Erlang distribution, as stated in the main text. To do so, we consider the $k$-th cumulants of $p(z_A)$, which is related to the $k$-th moment, but subtracts off trivial contributions from lower order moments. The $k$-th cumulant of the Erlang distribution can be simply calculated as $\kappa_k=(k-1)!/D_B^{k-1}$, since the Erlang distribution is the convolution of multiple exponential distributions~\cite{Leemis2008UnivariateDistribution}, and cumulants of convolutions are additive~\cite{Novak2012ThreeLectures}.

To calculate the $k$-th cumulant we consider the permutations in Eq.~\eqref{eq:temp_ens_moments_mt}. The permutation operator swaps between different copies of the Hilbert space, and can be represented by a directed graph indicating the action of the permutation on $k$ elements. Importantly, this graph can be disconnected, e.g. for three copies of the Hilbert space the permutation operator can leave one unchanged while swapping the other two. However, in this case the permutation will only contain terms which are related to lower-order moments. The cumulant explicitly subtracts off such lower order moments, and thus it is related only to \textit{cyclic} permutations, i.e. permutations with only a single cycle~\cite{Novak2012ThreeLectures}.

Applying this to our observable of interest, we write the general formula for the $k$-th cumulant of the time series of $p(z_A)$, $\kappa_k$, as:
\begin{align}
  \kappa_k &= \sum_{\sigma \in C_k}\sum_{z_B^{(1)}}\dots\sum_{z_B^{(k)}} \Big(\bigotimes_{i=1}^k \langle z_A, z_B^{(i)}| \Big)\hat{\rho}_d^{\otimes k}\text{Perm}(\sigma) \Big(\bigotimes_{i=1}^k | z_A, z_B^{(i)}\rangle \Big)
\end{align}
where $\bigotimes$ represents a repeated Kronecker product, and $C_k$ is the subset of $S_k$ containing only cyclic permutations, which importantly has $(k-1)!$ elements. This is a general statement for cumulants of a local bitstring probability, $p(z_A)$, in the large total system size limit; it is much like Eq.~\eqref{eq:kth_moment_PT}, but is subselected only to the cyclic permutations and with the inclusion of sums over different choices of $z_B$ in the contraction.

To simplify this equation, it is helpful to first consider its lowest non-trivial form for $k=2$, for which the only contributing permutation is the swap operator, $\hat{S}$, with explicit steps
\begin{align}
    \kappa_2&=\sum_{z_B,z_B'}(\langle z_A,z_B|\otimes\langle z_A,z_B'|)\hat{\rho}_d^{\otimes 2}\hat{S} (|z_A,z_B\rangle\otimes|z_A,z_B'\rangle)\\
    &=\sum_{z_B,z_B'}\langle z_A,z_B|\hat{\rho}_d|z_A,z_B'\rangle\langle z_A,z_B'|\hat{\rho}_d|z_A,z_B\rangle\\
    &=\sum_{z_B}\langle z_A,z_B|\hat{\rho}_d |z_A,z_B\rangle^2+\sum_{z_B\neq z_B'}|\langle z_A,z_B|\hat{\rho}_d |z_A,z_B'\rangle|^2\\
    &\approx\sum_{z_B}\langle z_A,z_B|\hat{\rho}_d |z_A,z_B\rangle^2\,.
\end{align}
In the third line, we have split the sum over different choices of $z_B$ and $z_B'$ into cases where $z_B=z_B'$ and cases where $z_B\neq z_B'$. In the final line, we have made the infinite temperature approximation that $\hat{\rho}_d\approx\hat{I}/D$ -- up to small variations along the diagonal due to the varied $p_\text{avg}(z)$ -- which nullifies the second term, where $z_B\neq z_B'$. In other words, we assume off-diagonal terms are negligible.

Carrying out a similar calculation -- and again only keeping terms where all $z_B$ are equal -- for the $k$-th cumulant yields
\begin{align}
    \kappa_k&\approx (k-1)!\sum_{z_B}\langle z_A, z_B|\hat{\rho}_d| z_A, z_B\rangle^k\,,\label{eq:generalmoment}
\end{align}
where the coefficient $(k-1)!$ comes precisely from the number of elements in the set of cyclic permutations. To be explicit, by selecting cases where all $z_B$ are equal (because of the infinite temperature condition) the cyclicity of the permutations does not actually matter for our proof, beyond the fact that there are $(k-1)!$ of them.

Then to further simplify Eq.~\eqref{eq:generalmoment}, we assume that any $z_B$ dependence can be ignored and we can approximate 
\begin{align}
\sum_{zB} \langle z_A, z_B| \hat{\rho}_d |z_A, z_B \rangle^k &\approx \langle z_A|\text{tr}_B(\hat{\rho}_d)|z_A\rangle^k/D_B^{k-1}\\
&=p_\text{avg}(z_A)^k/D_B^{k-1}. 
\end{align}
This is an assumption that the basis $\{z_B\}$ is \textit{energy non-revealing}, see Section~\ref{subsec:finite_temp_hypoexp}. 

Thus, when considering cumulants of relative quantities, i.e dividing the $k$-th cumulant by $p_\text{avg}(z_A)^k$, we find the $k$-th cumulant is $(k-1)!/D_B^{k-1}$. This is precisely equal to the $k$-th cumulant of the Erlang distribution. Since all cumulants match, all moments must match too, and we conclude that the marginal probabilities $p(z_A)$ follow an Erlang distribution for systems thermalizing at infinite temperature, up to corrections exponentially small in the system size.


In this derivation, we have made two approximations. First, we have assumed that the basis $z_B$ is energy non-revealing, and second, we have assumed that the off-diagonal terms of the form $\langle z_A z_B|\hat{\rho}_d|z_A z_B'\rangle$ are negligible. Remarkably, we may perform an exact calculation of the $k$-th moments $\mathbb{E}_t[p(z_A,t)^k]$ to estimate the effect of these approximations. We outline these more technical results below.

\begin{figure}
     \centering
     \includegraphics[width=120mm]{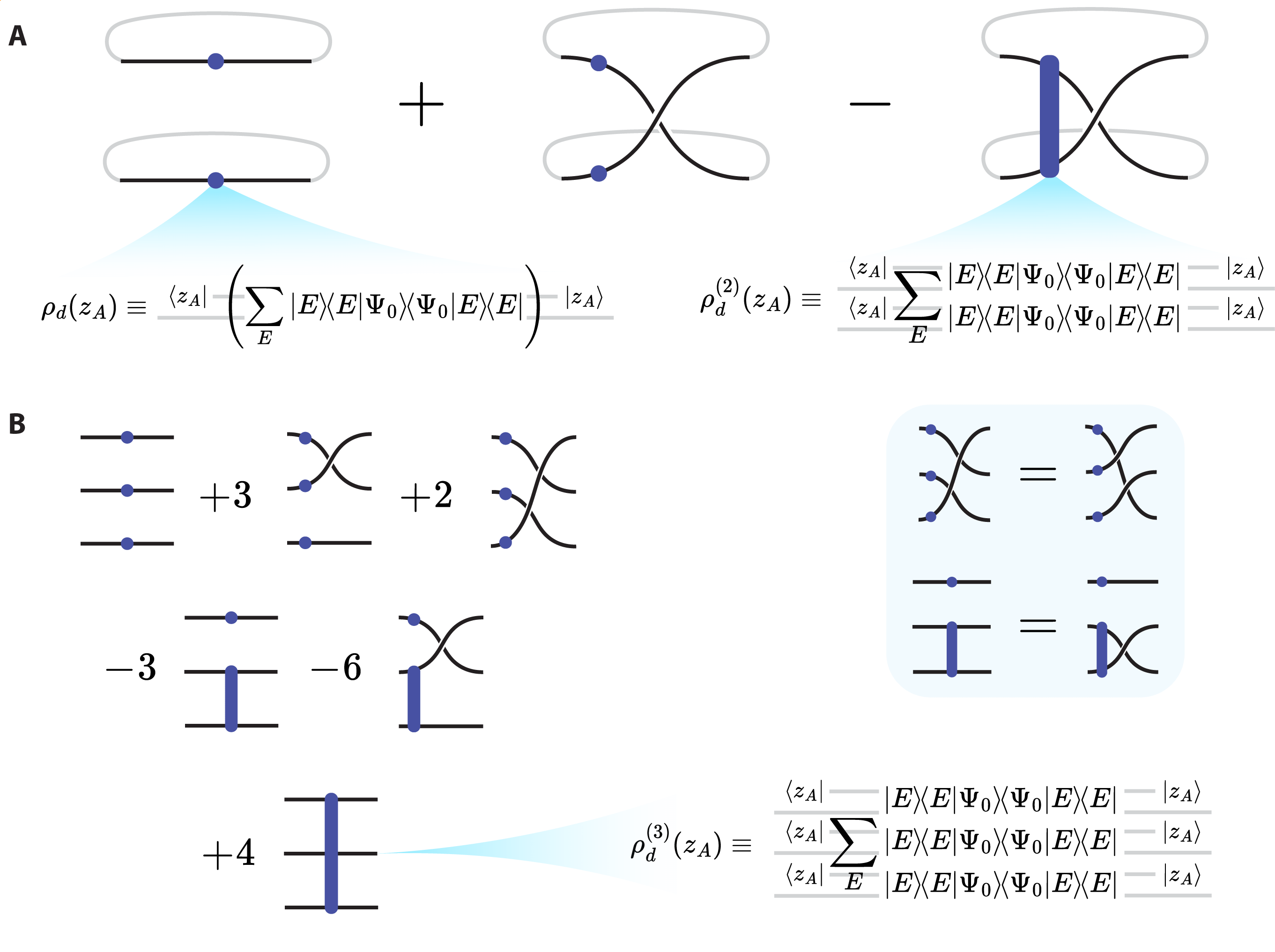}
    \caption{\textbf{Diagrams for subsystem bitstring probability fluctuations}. We diagrammatically organize our computations for the second and third moments $\mathbb{E}_t[p(z_A,t)^k]$, which we show are equal to the low moments of the Erlang distribution at infinite temperature. \textbf{A.} We organize the terms in the second moment of the temporal ensemble diagrammatically, which translates to the equation~ $\mathbb{E}_t[p(z_A,t)^2] = \text{tr}[\hat{\rho}_d(z_A)]^2+\text{tr}[\hat{\rho}_d(z_A)^2]-\text{tr}[\hat{\rho}_d^{(2)}(z_A)]$ [Eq.~\eqref{eq:subsystem_bitstring_second_moment}]. The blue dot represents $\hat{\rho}_d(z_A)$, which is an operator on $\mathcal{H}_B$, and can be interpreted as the initial state, dephased in the energy eigenbasis, and projected onto a subsystem bitstring $\ket{z_A}$. The blue bar represents $\hat{\rho}_d^{(2)}(z_A)$, which acts on the double Hilbert space $\mathcal{H}^{\otimes 2}_B$, and is a subleading correction term to the first two terms. Each diagram is ultimately traced out (gray lines), which we subsequently omit for simplicity. \textbf{B}. The diagrammatic framework greatly simplifies the computation of the third moment [Eq.~\eqref{eq:temporal_bitstring_third_moment}]. The six permutations of $S_3$ give rise to three terms, since permutations with the same cycle type produce the same terms (blue box). Correction terms may be expressed in terms of diagrams with blue bars. Furthermore, diagrams where the lines with a blue bar are swapped are equivalent. Therefore, there are only two inequivalent diagrams involving $\hat{\rho}_d^{(2)}(z_A)$. The number of equivalent diagrams, along with double-counting coefficients from the twirling identity gives rise to the coefficients $-3$ and $-6$. Finally, there is a higher-order diagram that accounts for triple-counting, involving $\hat{\rho}^{(3)}(z_A)$.}
     \label{fig:diagrams_for_calculations}
 \end{figure}

\subsection{Universal variance of subsystem bitstring probabilities, including $1/D$ corrections} 

\begin{figure*}[t!]
	\centering
	\includegraphics[width=110mm]{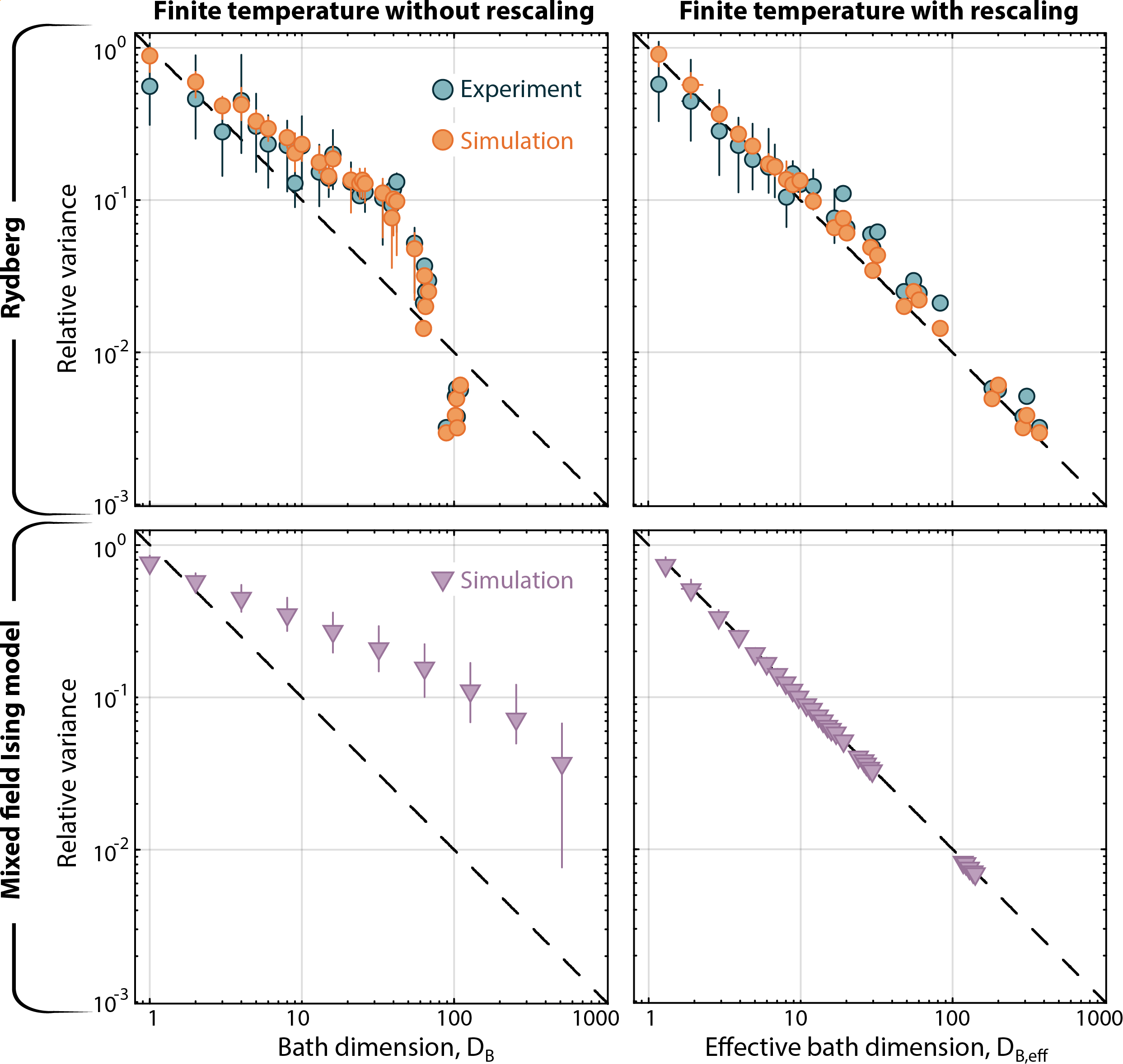}
	\caption{\textbf{Rescaling by effective bath dimension for finite temperature systems.} We show relative variance plots for Rydberg (both experiment and simulation) as well as for mixed field Ising model (simulation). In each case the quench evolution is at an finite effective temperature. In the left plots we show the relative variance (without the addition of a $1/D$ term) versus the bath dimensional, showing non-universal behavior. In the right plots we show the relative variance (plus the $1/D_\beta$ term, Eq.~\ref{eq:dbeta}) versus the effective bath dimension, $D_{B,\text{eff}}$ (Eq.~\eqref{eq:dbeff}), which recovers the scaling expected from Eq.~\eqref{eq:effective_bath_dimension}. In all plots, the markers are averaged over all bitstring realizations with an effective bath dimension in an interval from $[i-0.5,i+0.5)$, for $i\in\mathbb{Z}$, and error bars are the standard deviations over these realizations.
 }
	\label{EFig:finite}
\end{figure*}

We study the variance of bitstring probabilities by using explicit expressions for the second moment of the temporal ensemble when Hilbert-space ergodicity applies~\cite{Mark2023BenchmarkingQuantum,Mark2024MaximumEntropy}. 
Specifically, we study the relative temporal fluctuation $\sigma_\text{rel}^2(z_A)$, and we find an expression that has a natural interpretation as an effective dimension of the finite-temperature bath.

We define the relative temporal fluctuation $\sigma_\text{rel}^2(z_A)$ of the probability of a particular bitstring $z_A$ in subsystem $A$ as follows:
\begin{align}
\sigma_\text{rel}^2(z_A) &\equiv \frac{\mathbb{E}_t[ p(z_A, t)^2]  - \mathbb{E}_t[p(z_A, t)]^2}{\mathbb{E}_t[p(z_A, t)]^2} = \frac{\mathbb{E}_t[ p(z_A, t)^2]}{\mathbb{E}_t[p(z_A, t)]^2} - 1. \label{eq:rel_fluc_def}
\end{align}
Here $p(z_A, t) = \text{tr}\big[(|z_A \rangle \langle z_A | \otimes \hat{I}_B) | \Psi(t) \rangle \langle \Psi(t) | \big]$ is the marginal probability of measuring $z_A$ at time $t$, where $\hat{I}_B$ is the identity operator acting on subsystem $B$. The denominator of Eq.~\eqref{eq:rel_fluc_def} can be simplified to  
\begin{align}
\mathbb{E}_t[p(z_A, t)] &= \text{tr} \bigg[ |z_A \rangle \langle z_A| \otimes \hat{I}_B \left( \lim_{T \rightarrow \infty} \frac{1}{T} \int_0^T dt\; | \Psi(t) \rangle \langle \Psi(t)| \right) \bigg]= \langle z_A | \text{tr}_B(\hat{\rho}_d) | z_A \rangle \equiv \text{tr}_B \big[ \hat{\rho}_d(z_A) \big],
\end{align}
where in the second-last equality $\text{tr}_B(\hat{\rho}_d)$ is a mixed state on subsystem $A$, and in the last equality we instead consider the quantity $\hat{\rho}_d(z_A) \equiv (\langle z_A|\otimes \hat{I}_B) \hat{\rho}_d (|z_A\rangle \otimes \hat{I}_B)$, which is a mixed state on the subsystem $B$. 

Meanwhile, the numerator of Eq.~\eqref{eq:rel_fluc_def} can be evaluated from the second moment of the temporal ensemble~\cite{Mark2023BenchmarkingQuantum,Mark2024MaximumEntropy}. This yields three terms: the permutation terms from Eq.~\eqref{eq:temp_ens_moments_mt}, as well as a correction term illustrated in Fig.~\ref{fig:diagrams_for_calculations}.
\begin{align}
\mathbb{E}_t [p(z_A, t)^2 ] &= \mathbb{E}_t \bigg[ \big(\langle z_A | \text{tr}_B \left(| \Psi(t) \rangle \langle \Psi(t) |\right) | z_A \rangle \big)^2  \bigg] \\
&= \text{tr} \bigg[ \big( |z_A \rangle \langle z_A| \otimes \hat{I}_B \big)^{\otimes 2} \hat{\rho}_d^{\otimes 2} \bigg] + \text{tr} \bigg[ \big( |z_A \rangle \langle z_A| \otimes \hat{I}_B \big)^{\otimes 2} \hat{\rho}_d^{\otimes 2} \hat{S} \bigg] - \text{tr} \bigg[ \big( |z_A \rangle \langle z_A| \otimes \hat{I}_B \big)^{\otimes 2} \hat{\rho}_d^{(2)} \bigg].\\
&=\text{tr}_B \big [ \hat{\rho}_d (z_A) \big]^2 + \text{tr}_B \big[ \hat{\rho}_d (z_A)^2 \big] - \sum_E q_{z_A}^2(E)
\label{eq:subsystem_bitstring_second_moment}
\end{align}
where \mbox{$\hat{\rho}_d^{(2)} = \sum_E |E, E \rangle \langle E, E | \hat{\rho}_0^{\otimes 2} | E, E \rangle \langle E, E |$} is a two-copy generalization of the diagonal ensemble, and in the last line we have defined $q_{z_A}(E) \equiv  \sum_{z_B} | \langle z_A z_B | E \rangle|^2 |\langle E | \Psi_0 \rangle |^2$ (here we have dropped the subscript on energy eigenstates for notational simplicity). The inclusion of $\hat{\rho}_d^{(2)}$ prevents double counting of certain terms in the sum of permutations (see Fig.~\ref{fig:diagrams_for_calculations}), and is explicitly included here while previously (e.g. Eq.~\eqref{eq:temp_ens_moments_mt}) we disregarded it as negligible in the large system size limit.

Combining all of these simplified results, we obtain the relative fluctuation as follows:

\begin{align}
\sigma^2_\text{rel}(z_A) &= \frac{\text{tr}_B \big [ \hat{\rho}_d (z_A) \big]^2 + \text{tr}_B \big[ \hat{\rho}_d (z_A)^2 \big] - \sum_E q_{z_A}^2(E) }{ \text{tr}_B \big [ \hat{\rho}_d (z_A) \big]^2} -1\\
&\equiv \frac{1}{D_{B,\text{eff}}}- \frac{1}{D_\beta}\,,\label{eq:effective_bath_dimension}
\end{align}
where
\begin{align}
    D_{B,\text{eff}}^{-1}&\equiv\frac{\text{tr}_B \big[ \hat{\rho}_d (z_A)^2 \big]}{\text{tr}_B \big [ \hat{\rho}_d (z_A) \big]^2}\label{eq:dbeff}\\
    D_\beta^{-1}&\equiv \frac{\sum_E q_{z_A}^2(E)}{\text{tr}_B \big [ \hat{\rho}_d (z_A) \big]^2}\label{eq:dbeta}\,.
\end{align}
Here $D_{B,\text{eff}}^{-1}$ is the purity of the conditional reduced density matrix $\hat{\rho}_d (z_A)/\text{tr}_B \big [ \hat{\rho}_d (z_A) \big]$. It measures the effective size of the bath, hence we dub it the \textit{effective bath dimension}. Meanwhile, $D_\beta^{-1}$, is the purity of the classical distribution~\cite{Mark2023BenchmarkingQuantum} (of size $D$) over the energy eigenstates, which has typical value $O(1/D)$. At infinite temperature, we can approximate $\hat{\rho}_d(z_A) \approx \hat{I}_B/D$ and $q_{z_A}(E) \approx D_B/D^2$, where $D$ and $D_B$ are the Hilbert-space dimension of the total system and subsystem $B$ respectively, leading to $D_\beta\rightarrow D$ and $D_{B,\text{eff}}\rightarrow D_B$.

Applying these infinite temperature approximations, we immediately arrive at
\begin{equation}
 \sigma^2_\text{rel}(z_A) = \frac{1}{D_B} - \frac{1}{D}.
\label{eq:subleading_correction_to_variance}   
\end{equation}
Explicitly, in Fig.~\ref{Fig3} of the main text, in the main plot for the Rydberg, mixed field Ising model, and random unitary circuits the x-axis is $D_B$ as those systems are evolved at infinite effective temperature, while for the Fermi-Hubbard and Bose-Hubbard systems the x-axis is $D_{B,\text{eff}}$ to account for particle number conservation. Further, when calculating the scaling coefficients for all models in the inset of Fig.~\ref{Fig3}, we use $D_{B,\text{eff}}$ instead of $D_B$ - this is to be maximally fair to all models in case there are slight non-infinite temperature conditions.

The prediction of Eq.~\eqref{eq:effective_bath_dimension} is also matched by both numerical and experimental data at finite effective temperature. In Fig.~\ref{EFig:finite}, we plot the relative variance, $\sigma^2_\text{rel}(z_A)$ versus either the bath dimension, $D_B$, or the effective bath dimension, $D_{B,\text{eff}}$ (Eq.~\eqref{eq:dbeff}), for finite temperature evolution with both Rydberg (experiment and simulation) and mixed field Ising model (simulation) systems. We observe the relative variance scales inversely with the effective bath dimension, as expected. 

To be explicit, for each bitstring $z_A$ we calculate the corresponding $D_{B,\text{eff}}$ and $D_{\beta}$ numerically, which will be different generically for every $z_A$. For the experimental data, we select only blockade-satisfying bitstrings as we do not have sufficiently many experimental measurements to resolve probabilities with high confidence outside this space. However, we note in general that rescaling the bath dimension naturally takes into account the Rydberg blockade, as it is essentially a finite temperature effect. In other words, while we can operate in an infinite effective temperature regime within the blockade-satisfying subspace, within the global Hilbert space the system is at effectively low temperature; this temperature is naturally accommodated via the bath dimension rescaling for each $z_A$.


\subsection{Higher moments of subsystem bitstring probabilities}
\label{subsec:higher_moments_intrinsic}
Eq.~\eqref{eq:effective_bath_dimension} shows that the relative temporal variance, i.e. the second moment of the PoP, approximately satisfies $\sigma^2_\text{rel} = 1/{D_B} - 1/D$. This agrees with the second moment of the Erlang distribution, up to the $1/D$ term. Here, we explicitly compute the third moment, demonstrate that it is similarly consistent with the Erlang distribution and discuss leading order corrections.

\begin{figure*}[t!]
	\centering
	\includegraphics[width=140mm]{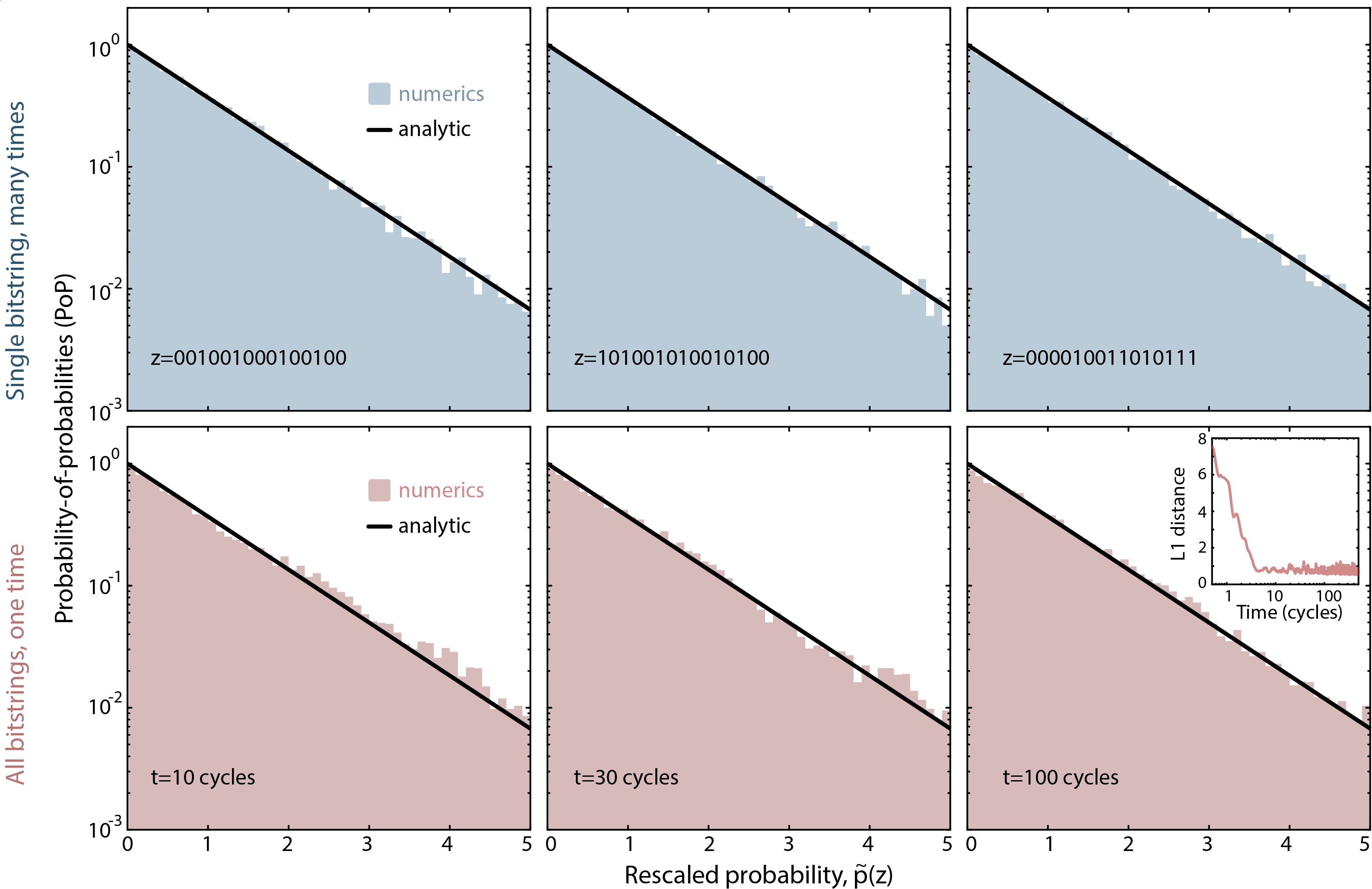}
	\caption{\textbf{Observation of Porter-Thomas statistics on the level of a single bitstring or time.} Probability-of-probabilities (PoP) distributions obtained from numerical simulations for an $N=15$ Rydberg atom array with the same parameters as used in the main text. Top: PoP distributions for the probability to observe a single bitstring, aggregated over evolution times ranging from 0 to 500 cycles, and normalized by the average temporal probability. Bottom: PoP distributions aggregated over all $2^{15}$ bitstrings at a single fixed time, with probabilities normalized by their average temporal probability. In all cases, the observed PoP distributions agree with the exponential Porter-Thomas distribution, as is expected from theoretical arguments related to Hilbert-space ergodicity~\cite{Mark2024MaximumEntropy}. In the case of averaging over all bitstrings at a fixed time, this agreement is observed to set in past $\approx4$ cycles, as quantified by the L1-norm between the numerical and analytic PoP as a function of time (inset). Note that in all plots we consider not just the lowest energy non-blockaded sector, but the entire Hilbert space. 
 }
	\label{EFig:singles}
\end{figure*}

Higher moments of the temporal ensemble may be evaluated~\cite{Mark2023BenchmarkingQuantum,Mark2024MaximumEntropy}. We find:
\begin{align}
\mathbb{E}_t[p(z_A,t)^3] &= \text{tr}[\hat{\rho}_d(z_A)]^3 + 3 \text{tr}[\hat{\rho}_d(z_A)] \text{tr}[\hat{\rho}_d(z_A)^2] + 2 \text{tr}[\hat{\rho}_d(z_A)^3]  - 3 \text{tr}[\hat{\rho}_d(z_A)] \text{tr}[\hat{\rho}_d^{(2)}(z_A)] \label{eq:temporal_bitstring_third_moment}\\
&~~~- 6 \text{tr}[\hat{\rho}_d^{(2)}(z_A)(\hat{I}\otimes \hat{\rho}_d(z_A))] + 4 \text{tr}[\hat{\rho}_d^{(3)}(z_A)]\nonumber
\end{align}
where $\hat{\rho}_d^{(k)}(z_A) \equiv (\langle z_A| \otimes \hat{I}_B)^{\otimes k}\sum_{E} |c_E|^{2k} |E\rangle \langle E|^{\otimes k} ( |z_A\rangle  \otimes \hat{I}_B)^{\otimes k}$ are higher-order correction terms. We diagrammatically illustrate this expression in Fig.~\ref{fig:diagrams_for_calculations}B. We emphasize that this expression is exact, and further approximations are often required to make headway.

The relevant quantity we wish to compare is the third cumulant $\kappa_3$
\begin{align}
\kappa_3 &\equiv \frac{\mathbb{E}_t[p(z_A,t)^3]}{\mathbb{E}_t[p(z_A,t)]^3} - 3 \frac{\mathbb{E}_t[p(z_A,t)^2]}{\mathbb{E}_t[p(z_A,t)]^2} + 2= 2 \frac{\text{tr}[\hat{\rho}_d(z_A)^3]}{\text{tr}[\hat{\rho}_d(z_A)]^3} - 6 \frac{\text{tr}[\hat{\rho}_d^{(2)}(z_A)(\hat{I}\otimes \hat{\rho}_d(z_A))]}{\text{tr}[\hat{\rho}_d(z_A)]^3} + 4 \frac{\text{tr}[\hat{\rho}_d^{(3)}(z_A)]}{\text{tr}[\hat{\rho}_d(z_A)]^3}\label{eq:third_cumulant_finite_temp}\\
&\approx \frac{2}{D_B^2} - \frac{6}{D_B D} + \frac{4}{D^2}
\label{eq:subleading_correction_to_third_cumulant}
\end{align}
where in the last step we have used the same approximations above at infinite temperature, that $\hat{\rho}_d(z_A) \approx \hat{I}_B/D$ and $q_{z_A}(E) \approx D_B/D^2$ (and note that $\text{tr}(\hat{\rho}_d^{(3)}(z_A)) = \sum_E q_{z_A}(E)^3$). As with the variance, this agrees with the third cumulant of the Erlang distribution (which is simply $2/D_B^2$) up to corrections $O(1/D)$. It is also close, but not exactly equal to $2(1/D_B - 1/D)^2$, i.e.~replacing $1/D_B$ with $1/D_B-1/D$ as the parameter in the Erlang distribution.

We note that both Eqs.~\eqref{eq:subleading_correction_to_variance} and~\eqref{eq:subleading_correction_to_third_cumulant} agree with the (average) second and third cumulants of Haar random states~\cite{Bauer2020UniversalFluctuations,Harrow2013ChurchSymmetric}, i.e.~for marginal distributions $p_\psi(z_A) \equiv \langle z_A| \text{tr}_B(|\psi\rangle\langle \psi|)|z_A\rangle$ of Haar random states $|\psi\rangle$,
\begin{align}
    \mathbb{E}_{\ket{\psi}\sim \text{Haar}}[D_A^2 p^2_\psi(z_A)] - 1&= \frac{D+D_A}{D+1} - 1 \approx \frac{1}{D_B} - \frac{1}{D}\\
    \mathbb{E}_{\ket{\psi}\sim \text{Haar}}\left[D_A^3 p^3_\psi(z_A) - 3D_A^2 p^2_\psi(z_A) + 2\right]&=\frac{D^3+ 3D^2 D_A+2D D_A^2}{D(D+1)(D+2)} - 3\frac{D+D_A}{D+1} + 2\approx \frac{2}{D_B^2} - \frac{6}{D_B D} + \frac{4}{D^2}
\end{align}

We next note that the substitutions $\hat{\rho}_d(z_A) \approx \hat{I}_B/D$ and $q_{z_A}(E) \approx D_B/D^2$ leading up to Eq.~\eqref{eq:subleading_correction_to_third_cumulant} are heuristic in nature and they may not appropriately capture the relevant higher-order fluctuations to the desired level of precision (here, up to $1/D^2$). Nevertheless, they appear to give the correct answer; we have verified that Eqs.~\eqref{eq:subleading_correction_to_variance} and~\eqref{eq:subleading_correction_to_third_cumulant} agree with a more careful calculation which treats the eigenbasis $\{\ket{E}\}$ as a random basis, and averages over such bases with an integral over Haar-random unitaries.


\subsection{Equivalence of PoP distributions over bitstrings or over time, and time-dependence of the PoP}
So far, we have shown the global PoP forms the Porter-Thomas distribution when aggregated over many measurement times. However, under Hilbert-space ergodicity, Refs.~\cite{Mark2023BenchmarkingQuantum,Mark2024MaximumEntropy} state the Porter-Thomas distribution should also be apparent when forming the PoP over many choices of bitstrings at a fixed measurement time.

\begin{figure*}[t!]
	\centering
	\includegraphics[width=47mm]{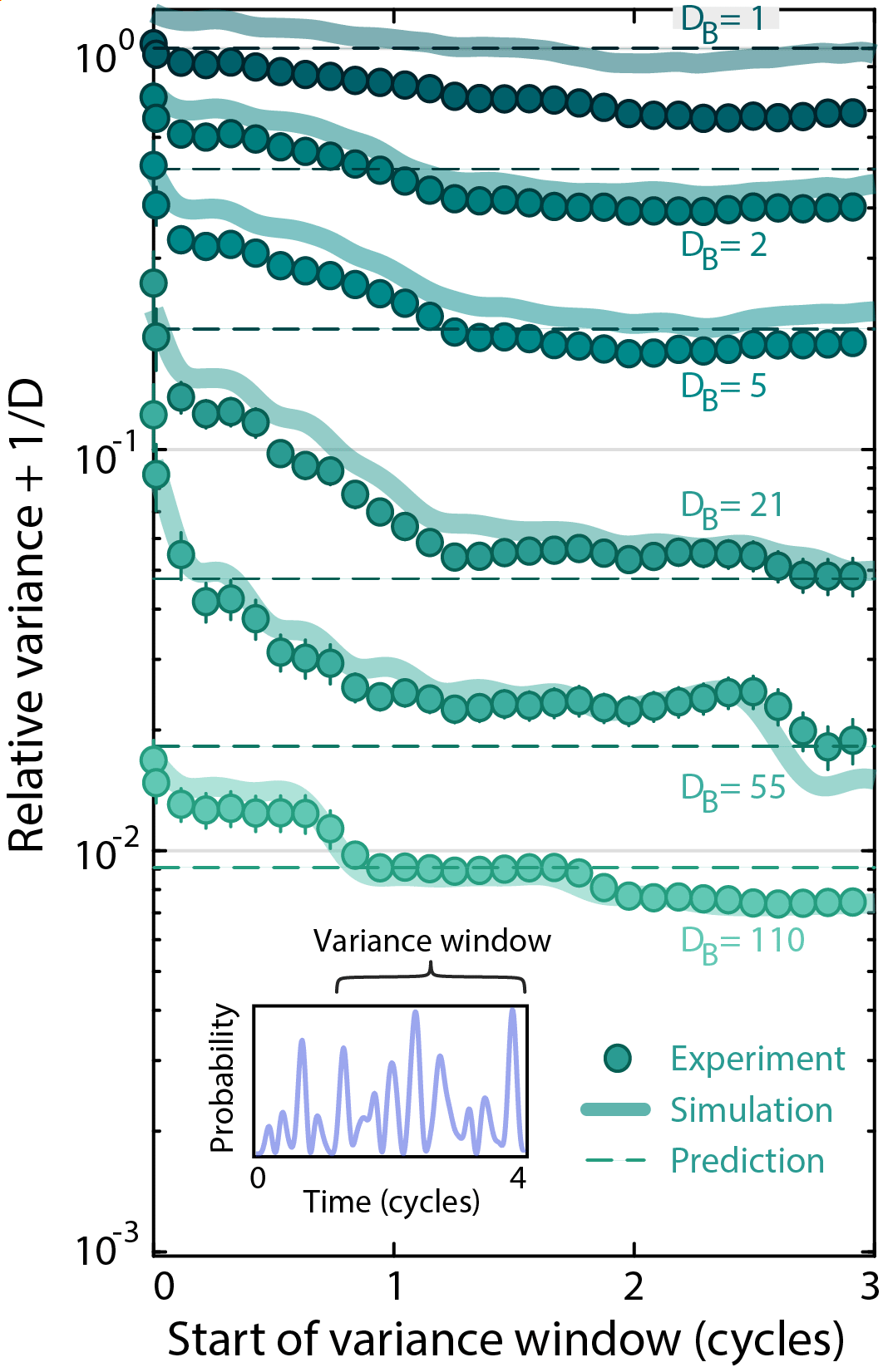}
	\caption{\textbf{Time-resolved saturation of the temporal variance.} We plot $\sigma^2_\text{rel}+1/D$ for both experiment and numerical simulations evolved at infinite temperature, for various choices of bath dimension, $D_B$. The variance is calculated in a temporal window of varying length, where the endpoint is fixed at $4$ Rabi cycles of evolution, the latest time measured experimentally, while the start of the window is varied (inset). We see for both the numerics and the experiment the variance settles to the prediction of $1/D_B$ for start times of $\approx1.5$ cycles. We note that when simulating out to many hundreds of cycles -- to emulate the infinite time limit -- the convergence is observed without excluding any early time transient behavior (not shown), but we do not have the experimental resolution to evolve to such late times without incurring fidelity loss.
 }
	\label{EFig:time_resolved}
\end{figure*}

In Fig.~\ref{EFig:singles} we present numerical evidence for this by simulating the Rydberg Hamiltonian with no decoherence out to several hundred cycles of evolution for $N=15$ atoms. We then form the global PoP either by a) aggregating normalized probabilities across many times for a single bitstring, or b) aggregating normalized probabilities over many bitstrings at a fixed time. Explicitly, in both cases the normalization is done by dividing probabilities by their respective long-time averages. Note that in the main text for the Rydberg systems we only consider the non-blockaded Hilbert space as it is difficult to gather sufficient experimental statistics to reconstruct bitstring probabilities in the higher-energy sectors, but in Fig.~\ref{EFig:singles} we consider all bitstrings, not just those drawn from the non-blockaded subspace.

An important aspect visible in the inset of Fig.~\ref{EFig:singles} is that the fixed-time PoP converges close to the Porter-Thomas distribution over a few cycles of evolution. In Fig.~\ref{Fig2} of the main text -- where we observe the experimental global PoP is roughly consistent with the Porter-Thomas distribution -- we aggregated over both bitstring index and measurement times from 1.5 to 4 cycles of evolution to gather sufficient statistics. Evolving out to longer times experimentally modifies the PoP distribution due to interaction with the external environment, as discussed at length in the main text. To verify we study sufficiently late dynamics in the experiment, we consider $\sigma^2_\text{rel}$ calculated in a temporal window of varying length, but fixed endpoint at $4$ Rabi cycles of evolution (Fig.~\ref{Fig3}), the latest time measured experimentally. We find $\sigma^2_\text{rel}$ appears roughly converged for a start time of $\approx1.5$ cycles, for which we observe both the experimental and numerical $\sigma^2_\text{rel}$ closely matches the $1/D_B-1/D$ prediction. We note some disagreement for the experimental data at $D_B=1$, which again we attribute to early signatures of decoherence.

\subsection{Failure cases}
\begin{figure*}[t!]
	\centering
	\includegraphics[width=74mm]{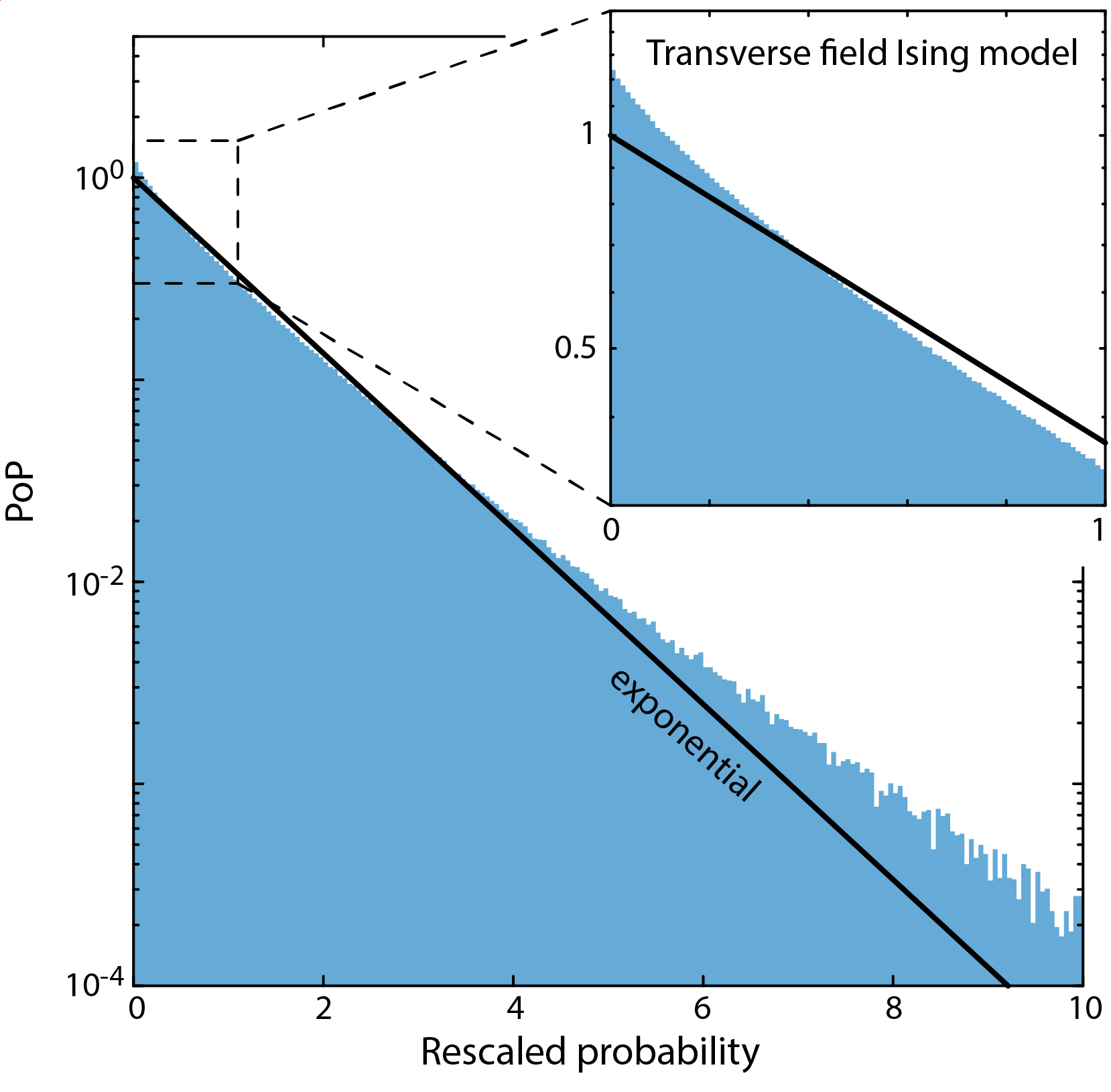}
	\caption{\textbf{Non-Porter-Thomas behavior of the transverse field Ising model.} We evolve an $N=12$ atom system under the transverse field Ising model out to several hundred cycles, as in the inset of Fig.~\ref{Fig3}, and aggregate all global bitstring probabilities over all bitstrings and at different times (to obtain better statistics), normalized by their respective long time averages. The resulting PoP shows non-exponential and non-universal signatures. Accounting for such effects may allow us us to better describe models not exhibiting Hilbert-space ergodicity.
 }
	\vspace{0.5cm}
	\label{EFig:tfim}
\end{figure*}

As shown in the inset of Fig.~\ref{Fig3} of the main text, not all models follow our analytic predictions, in particular those that do not fulfill the \textit{no-resonance conditions} such as the transverse field Ising model and systems of non-interacting spins. As a further example of this failure we plot the PoP aggregated over global bitstring probabilities from many times for the transverse field Ising model (Fig.~\ref{EFig:tfim}). Probabilities are normalized by their long time average, and are measured out to several hundred cycles. We observe clear non-exponential signatures in the PoP.

\section{Hypoexponential distributions for probability-of-probabilities}
\label{app:hypoexponential_distribution}

In Section \ref{app:erlang_distribution}, we formally derived the emergence of the Erlang distribution from results in Hilbert-space ergodicity for closed system dynamics. Using the intuition gained from that derivation, in this section we build models of how the probability-of-probabilities (PoP) should behave under a variety of conditions, including open system dynamics, explicitly comparing them to numerical simulations. We find that the PoP qualitatively and quantitatively is changed in the presence of different types of noise, which we used in the main text both to discriminate between noise channels as well as learn their relative strengths.

We first give an overview of how the exponential, or Porter-Thomas distribution emerges from ergodic unitary dynamics. We then introduce hypoexponential distributions, which parameterize all the PoP distributions studied in this work, including the Erlang distribution which describes closed system dynamics at infinite temperature. For closed system dynamics at finite temperature, we describe the hypoexponential weights and discuss the conditions on the bath size under which these weights are the eigenvalues of the system (or bath) reduced density matrix. 

We then move on to the case where our system is coupled to an external environment, describing our models for local errors, global depolarizing noise, Gaussian coherent errors, and a mixture of local and Gaussian coherent errors. 

In Section~\ref{app:finite_sampling_PoP}, we describe the effects of finite sampling on empirically estimating the PoP. This naturally leads us to more sample-efficient methods of characterizing the PoP via its cumulants (Section~\ref{app:PoP_moments}), which, by a chain of approximate equalities summarized in Eq.~\eqref{eq:approximate_chain}, quantify the decoherence strength (Kraus operators) of the quantum channel.

In this section, all probability distribution functions (PDFs) will be supported on the non-negative real line $x\geq 0$.

\subsection{Exponential (Porter-Thomas) distribution from random probability distributions} Our basic building block will be the exponential distribution $\text{Exp}(\lambda)$, with PDF
\begin{equation}
    P_\text{Exp}(x,\lambda) dx = \lambda \exp(- \lambda x) dx
\end{equation}
We will refer to variables which follow the exponential distribution with $\lambda = 1$ (or, \textit{Porter-Thomas distribution}) 
as ``standard exponential variables." 

The Porter-Thomas distribution is a universal feature of random quantum states~\cite{Porter1956FluctuationsNuclear,Boixo2018CharacterizingQuantum,Mullane2020SamplingRandom}. To see this, note that the probability distributions $p(z) = |\langle z|\Psi \rangle|^2$ obtained from Haar-random states are random probability distributions $p(z)$. To be precise, such distributions are drawn from the \textit{flat Dirichlet distribution}, which is a distribution over the space of probability distributions (the probability simplex). Such random distributions can be generated by i.i.d. sampling each probability value $p(z)\sim \text{Exp}(1/D)$, and then normalizing all entries such that $\sum_z p(z) = 1$. This final step introduces weak correlations $\text{cov}[p(z),p(z')] = \delta_{z,z'}/2D - 1/2D^2$, which vanish in the limit of large Hilbert-space dimension $D$. 

We are primarily interested in the marginal distributions $P[p]$ of individual probability values, namely, the distribution of the values of $p(z)$. This is known to be the \textit{Beta distribution} $\text{Beta}(1,D-1)$ \cite{Balakrishnan2003PrimerStatistical,Leemis2008UnivariateDistribution}, which converges to $\text{Exp}(1/D)$ in the large system limit $D\rightarrow \infty$. Crucially, the marginal distribution $P[p(z)]$ can be observed from a single \textit{typical} distribution $p(z)$, and one does not have to sample many Dirichlet random distributions. Since the entries $p(z)$ of a typical distribution $\{p(z)\}$ are very weakly correlated, the histogram of their values, or probability-of-probabilities (PoP), is also approximately exponentially distributed. In this context, the exponential distribution is referred to as the Porter-Thomas distribution.

In the above discussion, we have motivated the Porter-Thomas distribution through Haar-random states. However, the Porter-Thomas distribution is present in states which can be practically produced. Specifically, it is present in the states output from random unitary circuits (RUCs)~\cite{Boixo2018CharacterizingQuantum}. It is convenient to rescale the probability values as $Dp(z)$, which follow the standard exponential distribution (i.e. with mean 1). 

Furthermore, as shown in Section~\ref{sec:background}, the Porter-Thomas distribution appears in states obtained from unitary Hamiltonian evolution~\cite{Mark2023BenchmarkingQuantum}, i.e.~$\ket{\Psi(t)} = \exp(-i\hat{H}t)\ket{\Psi_0}$ for an arbitrary initial state $\ket{\Psi_0}$. We prove~\cite{Mark2024MaximumEntropy} this in the limit of exponentially long times, and observe that in practice the required time is small, only growing weakly with system size~\cite{Mark2023BenchmarkingQuantum}. Specifically, we consider the probabilities $p(z,t) \equiv |\langle z | \Psi(t)\rangle |^2$ and their time-average values $p_\text{avg}(z) \equiv \mathbb{E}_t[p(z,t)]$. Hamiltonian dynamics differs from RUC dynamics by energy conservation, which introduces non-trivial $p_\text{avg}(z)$. The probabilities $p(z,t)$ fluctuate about their average value, but the rescaled probabilities $\tilde{p}(z) \equiv p(z)/p_\text{avg}(z)$ follow the Porter-Thomas distribution.

\subsection{Hypoexponential distributions}
As described in the main text, the hypoexponential distribution is the probability distribution of a weighted sum of independent standard exponential variables. For a given set of weights $\omega_i$ (which are positive and sum to 1), and associated i.i.d. Dirichlet probability distributions $\{p_i(z)\}$, the overall distribution is given by the weighted sum:
\begin{equation}
    p(z) = \sum_i \omega_i p_i(z)\,.
\end{equation}

When the number of weights is small (i.e. for a small bath dimension), the weights can be related to the eigenvalues of the density matrix $\hat{\rho}$. Their properties may be inferred from each other (Section~\ref{subsubsec:weight_vector_and_mixed_state_moments}). For example, the state purity is related to the second moment of the weights $\text{tr}(\hat{\rho}^2) = \sum_i \omega_i^2$, which we utilize heavily in Section~\ref{app:PoP_moments}.

The corresponding PoP distribution is given by $\text{Hypo}(\{\omega_i\})$ (Note that the hypoexponential distribution is conventionally parameterized in terms of the inverse weights $\omega_i^{-1}$, which are known as \textit{rates}). When the $\omega_i$ are distinct, the hypoexponential distribution has probability distribution function (PDF)~\cite{Nadarajah2008ReviewResults}
\begin{equation}
    P_\text{Hypo}(x) =  \Conv_i P_\text{Exp}(x/\omega_i)= \sum_i \omega_i^{-1} \exp(-x/\omega_i) \prod_{j \neq i} \frac{\omega_i}{\omega_i-\omega_j},
    \label{eq:hypoexp}
\end{equation}
where $\circledast$ denotes convolution, an operation of functions defined by $(f\circledast g)(x) = \int dy f(y) g(x-y)$. Eq.~\eqref{eq:hypoexp} does not hold when there are degenerate $\omega_i$, which is often the case for our purposes. Hence, we will rarely directly use Eq.~\eqref{eq:hypoexp} (see Ref.~\cite{Nadarajah2008ReviewResults} for an expression valid for degenerate $\omega_i$). 

Different types of noise give different weights $\omega_i$. Below, we state types of noise we consider, their weights $\omega_i$ and their resultant PoP distributions.

\subsection{Intrinsic bath at infinite temperature} 
\label{subsubsec:erlang}
We first consider the PoP distribution in the context of closed system quantum thermalization: a subsystem effectively thermalizes due to its entanglement with the rest of the system, which serves as a bath. In this section, we let the bath be at infinite effective temperature. The quantities of interest are the subsystem bitstring probabilities $p(z_A)$, which are marginal probabilities $p(z_A) = \sum_{z_B} p(z_A\cdot z_B)$. As discussed in the main text, we find that the PoP of the marginal probabilities are described by Erlang distributions. In Section~\ref{app:erlang_distribution}, we gave a rigorous derivation of this, including small deviations from the perfect Erlang distribution. Here, we give an intuitive, heuristic derivation.

Consider an infinite temperature state in the absence of symmetries for simplicity. We take the average probabilities $p_\text{avg}(z)$ to be trivially $1/D$. Therefore, the joint probabilities $p(z_A\cdot z_B)$ are essentially i.i.d. exponential variables, and the (rescaled) sum $D_A p(z_A)$ is the average of $D_B$ i.i.d. standard exponential variables, i.e.~$\omega_i = 1/D_B$ for $i=\{1,2,\dots,D_B\}$. This special case of the hypoexponential distribution is precisely the \textit{Erlang distribution}, parameterized by a positive integer $k$ (here $k=D_B$). Because the weights are degenerate, its PDF cannot be directly obtained from Eq.~\eqref{eq:hypoexp}, but can be independently derived as
\begin{equation}
    P_\text{Erlang}(x;k) = \Conv_{i=1}^k P_\text{Exp}(kx) = \frac{k^k x^{k-1}}{(k-1)!}\exp(-kx),
\end{equation}
where in our case of interest the variable $x = D_A p(z_A)$, and $k=D_B$.

\subsection{Intrinsic bath at finite temperature}
\label{subsec:finite_temp_hypoexp}
Having established the PoP distribution for an intrinsic bath at infinite effective temperature, a natural next step is to investigate the PoP distribution at finite effective temperatures. Specifically, we consider the following setting: a system is partitioned into a larger subsystem $A$ and smaller one $B$, which we treat as the ``bath."

We first restrict ourselves to a special type of measurement basis $z_A$, whose conditions are explained later. Here, we claim that when the bath $B$ is small, the PoP of the bitstrings on $A$ is the hypoexponential distribution with coefficients $\lambda_i^{-1}$, where $\{\lambda_i\}_{i=1}^{D_B}$ are the eigenvalues of the reduced density matrix $\hat{\rho}_A$ (or $\hat{\rho}_B$). That is,
\begin{equation}
    \tilde{p}(z_A) \equiv p(z_A)/p_\text{avg}(z_A) \sim \text{Hypo}(\{\lambda_i\}),
    \label{eq:finite_temp_hypo}
\end{equation}
where $p(z_A) = \sum_{z_B}p(z_A\cdot z_B)$ and $p_\text{avg}(z_A) = \sum_{z_B}p_\text{avg}(z_A\cdot z_B)$ are the marginal distributions of $z_A$. 
\begin{figure}
    \centering
    \includegraphics[width=131mm]{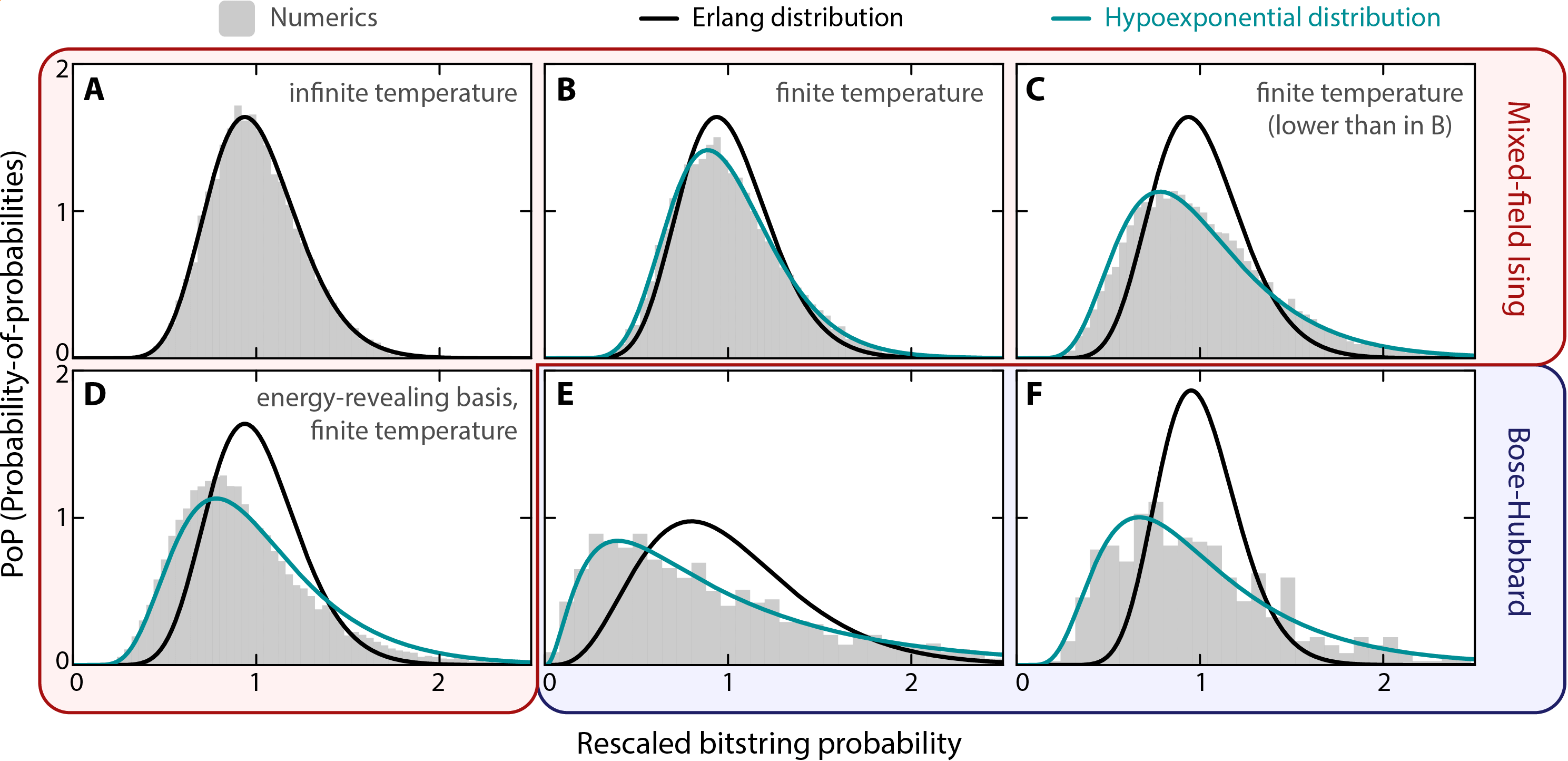}
    \caption{\textbf{Hypoexponential distributions for finite temperature closed system dynamics.} In (\textbf{A-D}), we plot the PoP of the conditional probabilities of $\tilde{p}(z_A) = p(z_A)/p_\text{avg}(z_A)$ formed by states time-evolved by the mixed field Ising model (MFIM). Here, the system is a one-dimensional $N=18$ open chain and the subsystem $B$ is four sites at the center of the chain. (\textbf{ABC}) The temperature is tuned by rotating the initial state $\ket{\Psi_0} = \left[\exp(i \theta \hat{S}_x)|0\rangle\right]^{\otimes N}$ at various angles $\theta = 0, -0.6, -1.2$. In \textbf{ABC}, $B$ is measured in the $\{z_B\}$ basis. Only at infinite temperature does the PoP agree with the Erlang distribution (green), while at finite temperatures, the PoP agrees with the hypoexponential distribution [Eq.~\eqref{eq:finite_temp_hypo}] instead. In \textbf{D}, $B$ is measured in the energy-revealing $\{x_B\}$ basis, showing small deviations from the hypoexponential distribution. 
    In \textbf{EF}, we form the PoP from the state $\ket{1,1\dots 1}$ evolved by the Bose-Hubbard model [Eq.~\eqref{eq:ham_BH}] with $M=10$ sites and $N=10$ particles, measured in the particle number basis, and with a subsystem of $3$ sites. Due to the number-conservation of the evolution, we restrict to the measurement outcomes where $z_B$ has \textbf{(E)} $4$ and \textbf{(F)} $5$ particles, and form the symmetry-resolved density matrix accordingly. The PoPs agree better with the hypoexponential prediction instead of the Erlang distribution with the appropriate symmetry-resolved subsystem dimension.
    }
    \label{fig:scrooge_hypoexp}
\end{figure}

This result may be understood through the lens of \textit{projected ensembles}~\cite{Choi2023PreparingRandom,Cotler2023EmergentQuantum}. These are the ensembles of states defined on a subsystem Hilbert space (here, $\mathcal{H}_B$) by measuring a global state $|\Psi\rangle$ in its complement ($A$). For concreteness, we let $\ket{\Psi} = \ket{\Psi(t)}$ be generated by many-body dynamics for time $t$. Each of the exponentially many measurement outcomes $z_A$ projects $|\Psi\rangle$ into a distinct pure state $|\Psi(z_A)\rangle \in \mathcal{H}_B$. The statistical properties of projected ensembles translate to properties of the PoP.

In companion work~\cite{Mark2024MaximumEntropy}, we study projected ensembles at finite temperature, relevant to our setting here. When two conditions are satisfied, $\hat{\rho}_d(z_A)$ is approximately the reduced density matrix $\hat{\rho}_B(t) \equiv \text{tr}_A\left[|\Psi(t)\rangle \!\langle \Psi(t)|\right]$, which gives the stated claim Eq.~\eqref{eq:finite_temp_hypo}. These two conditions are (i) that the measurement basis should be ``energy non-revealing", and (ii) the bath size must be sufficiently small.



\textbf{Effect of the measurement basis --- } The result Eq.~\eqref{eq:finite_temp_hypo} applies to specific measurement bases $|z_A\rangle$ that are ``energy non-revealing," i.e.~do not have significant correlations with the Hamiltonian that generated the state $\ket{\Psi(t)}$ (and which defines the notion of temperature). One example of an ``energy non-revealing basis" is the $Z_A$ basis for the mixed field Ising model with fields and couplings in the $X$ and $Y$ directions, where the measurement outcomes $z_A$ are uncorrelated with the energy density. For other bases, such as the $X_A$ basis for the mixed field Ising model discussed above, such correlations may exist. Specifically, due to energy conservation of the overall state, measurement outcomes with higher values of energy (on subsystem $A$) on average lead to projected states with lower energy (on subsystem $B$), and vice-versa. Therefore the projected ensemble $\{|\Psi(z_A)\rangle\}$ may not be taken to be samples from \textit{the same distribution}, independent of $z_A$. Among other consequences~\cite{Mark2024MaximumEntropy}, the simple prediction Eq.~\eqref{eq:finite_temp_hypo} does not hold in this setting.

\textbf{Effect of bath size --- } When the measurement basis is energy non-revealing, we may conclude that the \textit{time-averaged} reduced density matrix $\mathbb{E}_t[\hat{\rho}_B(t)]$ is proportional to any of the $\hat{\rho}_d(z_A)$. However, $\hat{\rho}_B(t)$ has fluctuations over time, and therefore the eigenvalues of $\mathbb{E}_t[\hat{\rho}_B(t)]$ (the weights of the hypoexponential PoP) may not agree with the eigenvalues of $\hat{\rho}_B(t)$. To see this, it is instructive to consider the case of infinite temperature, where we model the global state as a Haar-random state. Indeed, for Haar-random states, the eigenvalues of the reduced state follow the so-called Marchenko-Pastur distribution~\cite{Shapourian2021EntanglementNegativity,Znidaric2006EntanglementRandom}, which is supported on the interval $[D_B^{-1}(1-\sqrt{D_B/D_A})^2,D_B^{-1}(1+\sqrt{D_B/D_A})^2]$ and is controlled by the parameter $D_B/D_A = 2^{2|B|-N}$. Meanwhile, the time-averaged reduced state is equal to $\hat{I}_B/D_B$, with all eigenvalues equal to $1/D_B$. When $|B| = N/2$, the second moment of the Marchenko-Pastur distribution is $\sum_j\lambda_j^2 \approx 2/D_B$, while the flat spectral distribution has second moment $1/D_B$, a clear disagreement. This feature is not particular to our results, and instead reflects the conditions under which the time-dependent reduced density matrix may be approximated by a time-independent thermal state. Thus, the size of the bath should be sufficiently small; specifically, $|B|$ should be a constant amount less from half the system size $N/2$.\\

In Fig.~\ref{fig:scrooge_hypoexp}, we confirm some of our predictions. In Fig.~\ref{fig:scrooge_hypoexp}ABC we plot the PoP of a state time-evolved under the MFIM [Eq.~\eqref{eq:ham_mfim}], with a total system size of $N=18$ qubits, and a bath size of $N_B=4$ qubits. We tune the effective temperature by rotating the initial state $\ket{\Psi_0} = \left[\exp(i \theta 
\hat{S}_x)|0\rangle\right]^{\otimes N}$. At infinite temperature (A, $\theta=0$), the PoP is well described by the Erlang distribution. Tuning through finite temperature (BC, $\theta = -0.6,-1.2$), the PoP is instead well described by the hypoexponential distribution~[Eq.~\eqref{eq:finite_temp_hypo}]. Measuring in the energy-revealing $\{x_B\}$ basis (D), the PoP shows small deviations from the Eq.~\eqref{eq:finite_temp_hypo}. To illustrate the ubiquity of our prediction, we illustrate this with a finite temperature state evolved under the Bose-Hubbard model. In the presence of number conservation $\langle\hat{N}\rangle =N_\text{tot.}$, we must form the projected ensemble by restricting to measurement outcomes with fixed number $\hat{N}_A$, and forming the according symmetry resolved density matrix $\hat{\rho}_B(\hat{N}_A) \propto \sum_{\{z_A|n(z_A) = \hat{N}_A\}} (\langle z_A|\otimes \hat{I}_B)|\Psi(t)\rangle \langle \Psi(t)|(|z_A\rangle \otimes \hat{I}_B)$, as detailed in Refs.~\cite{Cotler2023EmergentQuantum,Mark2024MaximumEntropy}.

\subsection{Extrinsic bath: Local errors}
\label{subsubsec:localerrors}
We now consider the appearance of the hypoexponential under coupling to an extrinsic bath, i.e. with weights controlled by a given noise model. 

\begin{figure*}[t!]
	\centering
	\includegraphics[width=165mm]{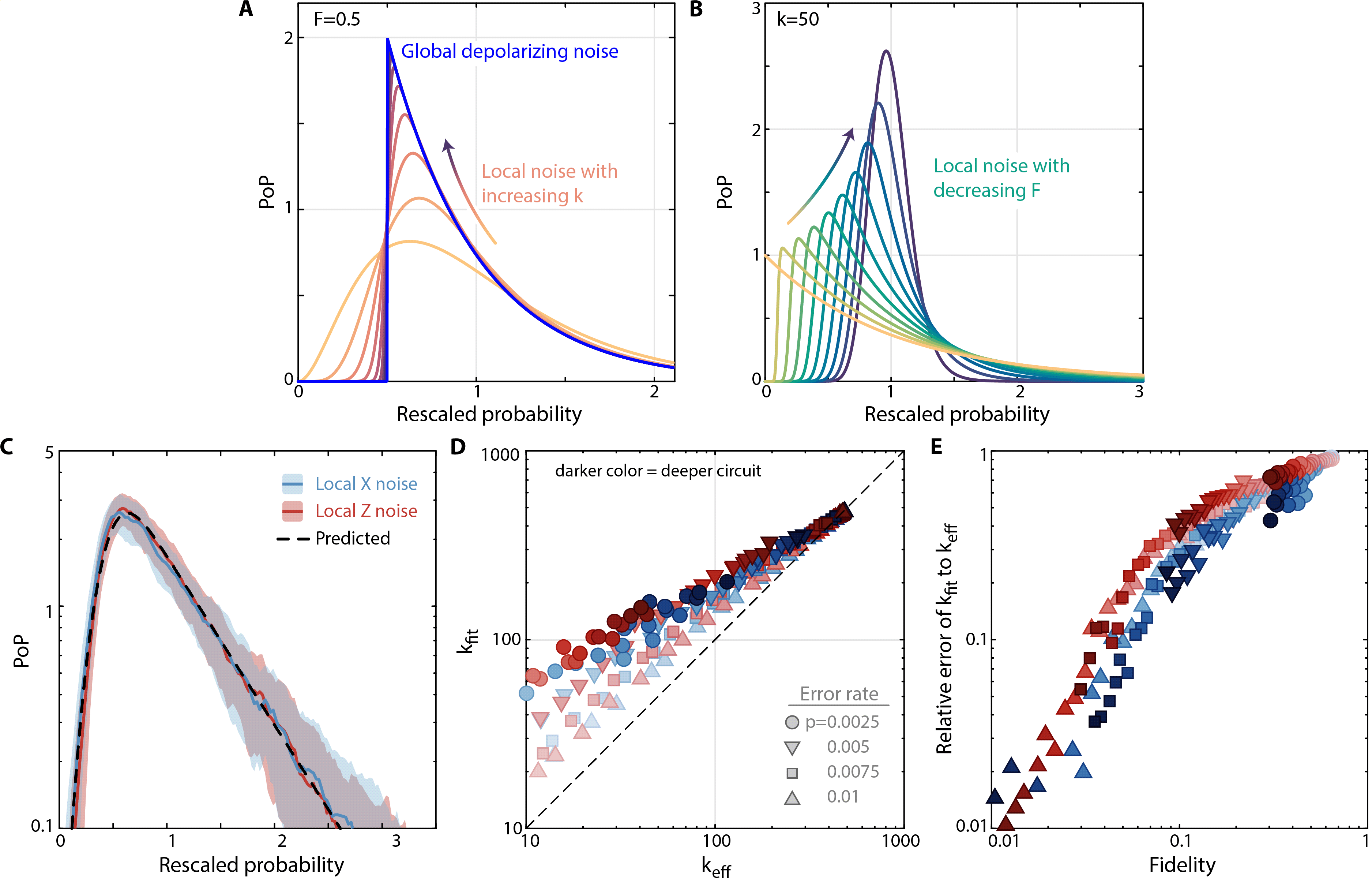}
	\caption{\textbf{Probability-of-probabilities (PoP) for local noise.} \textbf{A.} PoP distributions for local noise (see Eq.~\eqref{eq:pop_localnoise}) with $F=0.5$ and $k$ varying from $2$ to $10000$ logarithmically. As $k$ increases, the PoP distribution converges to the prediction for global depolarizing noise (see Eq.~\eqref{eq:globaldepol}) at the same fidelity, consistent with Ref.~\cite{Dalzell2021RandomQuantum}. \textbf{B.} PoP distribution for local noise with $k=50$ and $F$ varying from $1$ to $0.1$. \textbf{C.} PoP distribution for either local Pauli-X errors or Pauli-Z errors, both of which are roughly equally well described by the same effective $k$ value. \textbf{D.} Fitted value of $k$ versus value predicted from Eq.~\eqref{eq:local_noise_keff} for noisy simulation with either Pauli-X or Pauli-Z errors evaluated for varying circuit depths. \textbf{E.} The fitted value agrees best with the predicted value in the low fidelity regime, as quantified by the relative error between $k_\text{eff}$ and $k_\text{fit}$. In \textbf{C-E}, results are aggregated over 10 random circuit instances, blue points represent a local Pauli-$X$ error model, and red points represent a local Pauli-$Z$ error model.
 }
	\vspace{0.5cm}
	\label{EFig:localnoise}
\end{figure*}

First, we consider the case of local errors. As illustrated in Fig.~\ref{Fig4} of the main text, our model for the PoP obtained from local errors with $k$ locations is that it is the convolution of one exponential variable (speckle pattern) with large weight $F$ (the ``ideal trajectory"), and $k$ exponential variables with small weights $(1-F)/k$. In this case, the PDF is given by
\begin{align}
    \label{eq:pop_localnoise}
    P_\text{Local Err.}(x;F,k) = P_\text{Exp}(x/F) \circledast P_\text{Erlang}(x/(1-F);k) =  \left(\frac{k}{1-F^{-1}+k}\right)^k \gamma\left(k, \frac{F-1+kF}{F(1-F)} x\right) \frac{e^{-x/F}}{F (k-1)!}~,
\end{align}
where $\gamma(k,x) \equiv \int_0^x z^{k-1}e^{-z}dz$ is the lower incomplete Gamma function. In Fig.~\ref{EFig:localnoise}AB we plot these PDFs for various values of $k$ and $F$.

This model holds for any type of local error. In Fig.~\ref{Fig4}B of the main text, we illustrate the PoP with amplitude damping errors, i.e.~spontaneous decay of the $\ket{1}$ state to $\ket{0}$, while in Fig.~\ref{EFig:localnoise} we show it for local Pauli noise. In practice, however, the various assumptions made in the main text for the local noise channel may not hold, for instance because errors are not necessarily completely orthogonal and because multiple errors can occur in a single shot of the experiment. These largely result in a value of $k$ which is not simply the spacetime volume of the circuit. To address this, in practice we use an effective predicted $k$ value defined as
\begin{align}
\label{eq:local_noise_keff}
k_\text{eff} = \frac{(1-F)^2}{\kappa_2-F^2},
\end{align}
where $\kappa_2$ is the second cumulant of the experimentally measured PoP distribution (as discussed in greater detail in Section~\ref{app:PoP_moments}, see Table~\ref{ETab:moments}). Using Eq.~\eqref{eq:local_noise_keff} we can predict the $k$ value for a given measured PoP distribution at a given fidelity. In Fig.~\ref{EFig:localnoise}C we show the good agreement between Eq.~\eqref{eq:pop_localnoise} and numerical simulations using an effective $k$ found from Eq.~\eqref{eq:local_noise_keff}. Further, in Fig.~\ref{EFig:localnoise}D we compare $k_\text{eff}$ from Eq.~\eqref{eq:local_noise_keff} against $k$ value of best fit for various error rates and evolution depths, and generally find the two are in good agreement, particularly in the low fidelity regime (Fig.~\ref{EFig:localnoise}E).

We stress that this model is limited. As an example, it stipulates that the weights of the non-ideal trajectories be equal. This is not true if more than one error can occur during the dynamics: the probability of the trajectory decreases exponentially with the number of errors, while the number of such distinct trajectories increases exponentially. Furthermore, even the weights of single-error events may not be uniform. For example, the probability of amplitude damping errors depend on the magnetization of the state at each layer, which may change during the dynamics. These effects may lead to small deviations between our model and PoPs generated by realistic many-body dynamics, which may affect high-precision tasks such as determining the fraction of coherent-to-local error in Fig.~\ref{Fig5}BC of the main text. However, such limitations may be solved with straightforward model-level refinements (such as explicitly adjusting weights to account for multiple error occurrences).

\subsection{Extrinsic bath: Global depolarizing noise} Global depolarizing noise is a commonly used toy model for the effects of noise on quantum states. This channel maps states $\hat{\rho} \mapsto F \hat{\rho} + (1-F) \hat{I}/D$, where $\hat{I}$ is the identity and $F$ is a parameter which is approximately the fidelity of the channel. Under this channel, the bitstring distribution $p(z) \mapsto F p(z) + (1-F)/D$ and hence the PoP distribution of $x = D p(z)$ is shifted to the right and rescaled by $1/F$, with PDF
\begin{equation}
    P_\text{GD}(x;F) = \begin{cases}
        0 & \text{for } 0 \leq x < 1-F\\
        F^{-1} \exp\left(-\frac{x-(1-F)}{F}\right) & \text{for } x \geq 1-F\\
    \end{cases}
    \label{eq:globaldepol}
\end{equation}
As seen in Fig.~\ref{EFig:localnoise}, $P_\text{Local Err.}(x;F,k) \rightarrow P_\text{GD}(x;F)$ as $k\rightarrow \infty$, consistent with Ref.~\cite{Dalzell2021RandomQuantum}, which states that the output distribution of random circuit sampling under local noise converges to that of ``global white noise" when scaling the circuit spacetime volume.

\subsection{Extrinsic bath: Gaussian coherent errors} 
\label{subsubsec:gaussian}
Next, we discuss the case of Gaussian coherent errors, which describes shot-to-shot variation of a global parameter. 

For a circuit model, we model this error by systematically under- or over-rotating every two-qubit gate by a multiplicative, Gaussian-distributed factor. For Hamiltonian evolution, a global Hamiltonian parameter, such as the Rabi frequency $\Omega$, is taken to be a Gaussian random variable. Such a noise channel is linked to shot-to-shot parameter fluctuations observed in Hamiltonian evolution~\cite{Shaw2024BenchmarkingHighly}, and certain models of control errors. As found in Ref.~\cite{Shaw2024BenchmarkingHighly}, evolution for a time $t$ under these parameter values produces a continuous family of pure states $\ket{\Psi(t,\Omega)}$ which mutually overlap as $|\langle \Psi(t,\Omega)| \Psi(t,\Omega')\rangle|^2 \propto \exp(-N (\Omega-\Omega')^2 t^2/2 \sigma_\Omega^2)$. When $\Omega$ is normally distributed, i.e.~with $P(\Omega) = \exp(-(\Omega-\Omega_0)^2/(2\sigma'^2))/\sqrt{2\pi \sigma'^2}$ for some $\sigma'$, we obtain a normally-distributed mixture $\hat{\rho}(t) = \int d\Omega P(\Omega) |\Psi(t,\Omega)\rangle\langle \Psi(t,\Omega)|$.

Our model for the PoP converts this continuous mixture of states into a discrete sum of probability distributions, with discrete weights $\omega_i$. It is natural to let the weights follow a Gaussian profile when $\sigma_F$ is sufficiently large (see below for deviations) 
\begin{equation}
 \omega_i^\text{coh.} = \frac{1}{\mathcal{Z}(\sigma_F)}\exp\left[-\frac{i^2}{2\sigma^2_F}\right],   
 \label{eq:Gaussian_hypoexponential_model}
\end{equation}
where $i$ ranges from $-\infty$ to $\infty$. These weights are parameterized by the width $\sigma_F$, and $\mathcal{Z}(\sigma_F) \equiv \sum_{i=-\infty}^\infty \exp[-i^2/(2\sigma^2_F)]$ is a normalization constant. This sum is equal to the Jacobi theta function~\cite{NIST:DLMF} $\theta_3(\exp[-1/(2\sigma^2_F)])$, which is well approximated by $\sqrt{2\pi} \sigma_F$ when $\sigma_F \gg 1$, but approaches 1 as $\sigma_F\rightarrow 0$. Incidentally, since $\omega_0 = 1/\mathcal{Z}(\sigma_F)$ has largest height, it corresponds to the fidelity $F$, i.e.~the maximum overlap $\langle\Psi|\hat{\rho}|\Psi\rangle$ for any pure state $\ket{\Psi}$. Therefore, we may parameterize the PoP in terms of $F$ instead, by solving for $\sigma_F$ such that $\theta_3(\exp[-1/(2\sigma_F^2)]) = F^{-1}$. 

A large fidelity corresponds to a small $\sigma$, and our model of discrete Gaussian weights shows quantitative differences in this regime. As an illustration, the second moment of the PoP can be computed analytically, and differs from the Gaussian hypoexponential model at small $\sigma$ (Section~\ref{app:PoP_moments} and Fig.~\ref{fig:moments_of_noise_models}\textbf{C}). This may be addressed by a non-Gaussian profile of weights in the narrow limit.



As a technical point, the weights $\omega_{-i}^\text{coh.} = \omega_i^\text{coh.}$ are degenerate and therefore we must modify Eq.~\eqref{eq:hypoexp} to
\begin{align}
     P_\text{Gauss. Coh.}(x;\{\omega_i\}) &= P_\text{Exp}(x/\omega_0) \circledast \left(\Conv_{i=1}^\infty P_\text{Erlang}\left(x/\omega_i;2\right)\right)\\
     &= \omega_0^{-1} e^{-x/\omega_0} \prod_{j > 0} \frac{\omega_0^2}{(\omega_0-\omega_j)^2} + \sum_{i = 1}^\infty \left[ \omega_i^{-2} \Bigg(x-\sum_{\substack{j\in \mathbb{Z}\\j \neq i}} \frac{\omega_i \omega_j}{\omega_i-\omega_j} \Bigg) e^{-x/\omega_i}\frac{\omega_i}{\omega_i-\omega_0} \prod_{\substack{j\in \mathbb{Z}\\j \neq i}} \frac{\omega_0}{\omega_0-\omega_j}\right]
\end{align}

\subsection{Extrinsic bath: Mixture of local and global coherent errors}
Finally, we turn to modeling the effects of both local and global coherent errors simultaneously. We take the errors to be independent of each other, and the weights are simply products of the local and global weights $\omega_{ij} = \omega_i^\text{loc.} \times \omega_j^\text{coh.}$. We further notice that in numerical simulations of such a mixture of errors, the fidelity factorizes into the fidelities arising from incoherent and coherent errors (Fig.~\ref{fig:moments_of_noise_models}A), i.e.~$F = F_\text{coh.}\times F_\text{loc.}$, i.e. $\log(F_\text{coh.}) + \log(F_\text{loc.}) = \log(F)$. This allows us to define the ``fraction" of coherent errors as  $f = \log(F_\text{coh.})/\log(F)$. We note that such a factorization is also approximately apparent for our \textit{ab initio} Rydberg error model~\cite{Shaw2024BenchmarkingHighly}.

For completeness, we outline our algorithm used in Fig.~\ref{Fig5}BC to generate the PoP arising from a mixture of local and global errors based on our model, with a finite number of samples. The effects of a finite number of samples are detailed on a later section. We use this algorithm to generate the local and global coherent finite-sample PoPs in Fig.~\ref{Fig5}A, as well as to provide a reference distribution to calculate the $\chi^2-$distances in Fig.~\ref{Fig5}BC. Note that in the case of RUCs, the average probability is $p_\text{avg}(z) = 1/D$, and the rescaled empirical probabilities will only take finite values $j D/M$ for positive integer $j$.

\RestyleAlgo{ruled}
\SetKwComment{Comment}{/* }{ */}
\begin{algorithm}
    \label{alg:algorithm}
    \caption{Numerical algorithm to generate model PoP under both local and global coherent noise}
    \KwData{Fidelity $F$, fraction $f$, number of samples $M$, Hilbert-space dimension $D$, number of trajectories $k$, averaged probabilities $p_\text{avg}(z)$, and bin width $\Delta \tilde{p}$}
    \KwResult{Empirical PoP $\vec{n}$}
    $F_\text{loc.} \gets F^f$\;
    $F_\text{coh.} \gets F^{1-f}$\;
    $\sigma_F \gets$ \text{Solution of } $\theta_3(\exp[-1/(2\sigma_F^2)]) = F^{-1}_\text{coh.}$\;
    $n(\sigma_F) \gets F_\text{coh.}$\;
    $p(z) \gets 0$\;
    \For{$j~\mathrm{in}~\{-\lceil5 \sigma_F \rceil,-\lceil5 \sigma_F \rceil+1,\dots,\lceil5 \sigma_F \rceil\}$\Comment*[r]{Infinite sum truncated at $\pm \lceil5 \sigma_F \rceil$}
    }{
        $p_\text{Rand}(z) \gets~\mathrm{Dirichlet~random~distribution}$\;
        $p(z) \gets p(z) + F_\text{inc.}n(\sigma_F)\exp[-j^2/(2\sigma_F^2)] p_\text{Rand}(z)$\;
        \For{$n ~\mathrm{in}~\{1,2,\dots,k\}$}{
            $p_\text{Rand}(z) \gets~\mathrm{Dirichlet~random~distribution}$\;
        $p(z) \gets p(z) + \frac{(1-F_\text{inc.})}{k}n(\sigma_F)\exp[-j^2/(2\sigma_F^2)] p_\text{Rand}(z)$\;
        }
    }
    $p(z) \gets p(z)\times p_\text{avg}(z)/\Vert p(z)\times p_\text{avg}(z) \Vert_1$ \Comment*[r]{Multiply mixture of Dirichlet random distributions by $p_\text{avg}$ and normalize}
    $n_\text{trial} \gets 100$\;
    $\vec{n} \gets \vec{0}$ \Comment*[r]{Initialize mean for finite sample PoP}
    \For{$j~\mathrm{in}~\{1,2,\dots,n_\text{trial}\}$}{
    $\{z_1,\dots, z_M\} \gets M \text{ samples from }p(z)$\;
    $N_z \gets \#(z_i = z)$\; 
    $\vec{n} \gets \vec{n} + \text{Hist}(\{\frac{N_z}{M p_\text{avg}(z)}\};\Delta\tilde{p})/n_\text{trial}$ \Comment*[r]{Form histogram of rescaled empirical frequencies $N_z/(Mp_\text{avg}(z))$, with bin width $\Delta\tilde{p}$, add to mean vector $\vec{n}$. Each entry is the expected number of bitstrings with empirical frequency in bin.}
    }
    
    
\end{algorithm}

\subsection{Other noise sources}
We briefly remark that our analysis here is by no means complete. While we have analyzed commonly considered noise sources, other sources such as qubit leakage and readout error will yield different PoP distributions. For example, readout error may be treated in a similar fashion to local error. However, here there are only $N$ possible error locations, and as such the parameter $k$ may not be taken to be large. This is particularly notable for our moment analysis in Eq.~\eqref{eq:local_error_second_moment}. We leave a detailed study of this, and other noise models, to future work. However, we expect that in general accounting for novel noise models will simply amount to judicious corresponding choice of the hypoexponential weight vector.

\newpage
\clearpage
\section{Effects of finite sampling on the probability-of-probabilities}
\label{app:finite_sampling_PoP}
It is important to understand the effect of a finite number of experimental samples on the probability-of-probabilities discussed in this work. Finite sampling must be carefully treated since its effects on the PoP can be drastic: the simplest estimation of the PoP does not only introduce statistical noise, but also systematic bias. Here we first discuss how experimental and model PoPs are compared in the main text, then describe how the model PoP is affected by finite sampling by effectively smearing out the distribution. Finally, we  suggest a method to potentially invert this finite sampling effect. In the next section we show that low-order moments of the PoPs can be estimated in an unbiased way which is robust to a finite number of samples. 

\subsection{Comparing experiment and finite-sampled model PoP}
\label{sec:comparison}
In the main text, we compare the experimental empirical PoP against PoPs obtained from analytical models, accounting for the finite sampling of the experiment. Specifically, we accrue roughly $32000$ blockade-satisfying~\cite{Browaeys2020ManybodyPhysics} bitstrings from the experimental evolution, which is $m\approx20.6\times$ the Hilbert-space dimension of $D=1597$, and compare against analytical PoPs which are modified to account for this degree of finite sampling (described below). For our \textit{ab initio} error model, we perform the full quantum simulation, then sample an equivalent number of bitstrings from the resultant state, then repeat this sampling many times to find the mean finite-sampling prediction.

For the analytical noise models, no direct quantum simulation is necessary. We first take the analytical model PoP, and sample a given normalized probability vector from it. We then sample this probability vector as many times as the experiment was sampled and reconstruct an empirical model PoP. We repeat this procedure many times, and take the average over iterations to define the numerically sampled model PoP. We emphasize that in doing so, no actual quantum simulation was needed, simply repeated resampling. The results of this resampling are in excellent agreement with analytical predictions where available (discussed below), as in Fig.~\ref{EFig:finitesampling}.

\begin{figure}
    \centering
    \includegraphics[width=89mm]{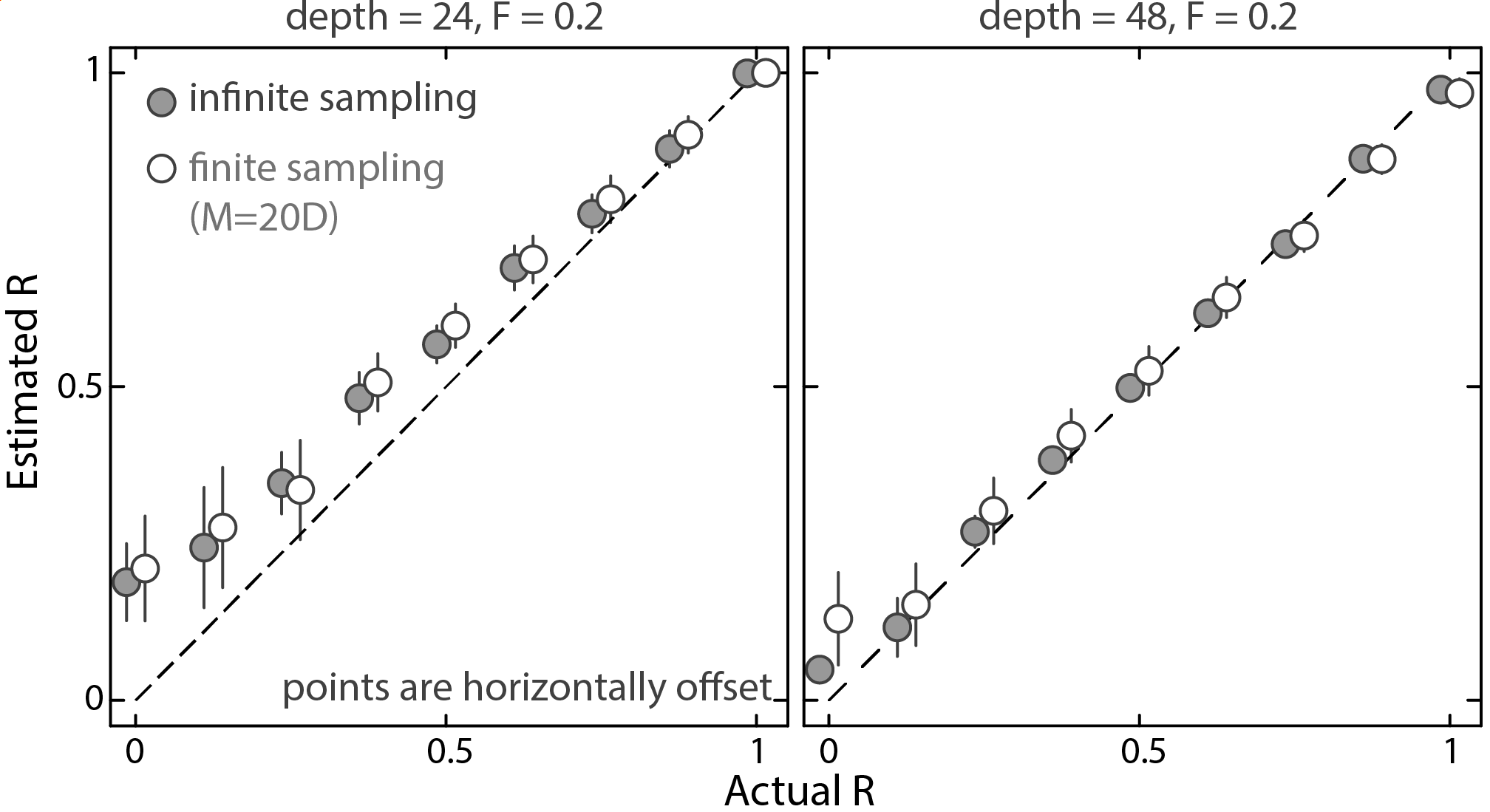}
    \caption{\textbf{Noise learning improves with greater depth.} We perform RUC numerics with varying fraction of global coherent error to local incoherent error, as in Fig.~\ref{Fig5}c of the main text, with parameters chosen such that the final fidelity is 0.2 after either depth 24 (left) or depth 48. For shorter depth circuits, the estimator correlation becomes worse, which we believe is due to a) limited scrambling, and b) not accounting for higher-order terms in the analytic noise channels (i.e. non-dilute effects of the local error channel). We also show the results for either infinite or finite sampling, which are horizontally offset from each other to better distinguish, which does not seem to have a dramatic effect on the estimation correlation.}
    \label{fig:estimating_r}
\end{figure}

For our comparison between the model PoPs and RUC numerics in Fig.~\ref{Fig5}C of the main text, we again employ the standardized $\chi^2$-distance (CSD) as our distance, and compare our noisy RUC simulations against model PoPs. For each point in Fig.~\ref{Fig5}, we average over 100 RUC circuit realizations. We show additional results in Fig.~\ref{fig:estimating_r}, where we see the estimation correlation improves with greater circuit depth at an equal fidelity, which we attribute to a) limited scrambling in shallower circuits and b) not-accounting for higher-order terms in the analytic noise channels; we note that the second of these effects is solvable with model-level refinements, discussed in Secs.~\ref{subsubsec:localerrors} and~\ref{subsubsec:gaussian}. We further find that finite sampling does not introduce significant bias in the estimator correlation, only increasing its statistical fluctuations.

In the following subsections, we describe an analytic description of finite sampling for the PoP distribution, and potential ways to undo these effects.

\subsection{Multinomial distribution and Poisson sampling} Taking $M$ i.i.d. samples from a probability distribution $p(z)$ yields a number $n_z$ of occurrences of each bitstring $z$. The vector $\vec{n} \equiv (n_{z_1},\dots, n_{z_D})$ of such numbers is a random variable following the \textit{multinomial distribution} $\vec{n} \sim \text{Multinomial}(M, p(z))$ with PDF
\begin{equation}
    P(\vec{n};M,p(z)) = \frac{M!}{n_{z_1}! \cdots n_{z_D}!}p(z_1)^{n_{z_1}}
 \cdots p(z_D)^{n_{z_D}},
 \label{eq:multinomial}\end{equation}
where we have indexed the bitstrings from 1 to $D$. This is known \textit{i.i.d. sampling}~\cite{Canonne2020SurveyDistribution}. 

While Eq.~\eqref{eq:multinomial} is an exact statistical description of the outcome of sampling, it is analytically easier to use the Poisson sampling model, which assumes that each number $n_z$ is an independent Poisson random variable with mean $M p(z)$, i.e.~$n_z \sim \text{Poi}(\mu=M p(z))$~\cite{Canonne2020SurveyDistribution}. 
Since each number variable is independent, the constraint
$\sum_z n_z = M$ is not enforced; in the multinomial distribution, there are weak anticorrelations between number variables $n_z$ and $n_{z'}$. The Poisson sampling approximation becomes increasingly accurate with increasing sample number $M$ and serves as a useful model for sampling~\cite{Canonne2020SurveyDistribution}. Nevertheless, the full multinomial distribution is necessary for quantitatively accurate formulae in some cases which we will indicate.

\subsection{Finite-sample estimate of PoP distribution}
\label{subapp:finite_sampling_PoP}
The Poisson sampling model will be essential to helping us understand the effects of finite sampling on reconstructing the PoP distribution. Here, we wish to estimate the distribution of probabilities $P(p)$ from empirical data. The challenge lies in the fact that the probabilities $p(z)$ are difficult to empirically estimate --- the \textit{plug-in estimate} $\hat{p}(z) \equiv n_z/M$ is a rescaled Poisson variable and is unbiased, but may exhibit large fluctuations. For simplicity of discussion, we first consider the case where $p_\text{avg}(z) = 1/D$ (as in RUCs) before tackling the general case. In this case, the possible values of $\hat{p}(z)$ are discrete; they are integer multiples of $1/M$, and the rescaled probability $\tilde{p}(z)$ is simply $Dp(z)$. Accordingly, we shall express our PoP as $P(Dp)$, which is system-size independent.

A standard practice to empirically estimate the PoP is to simply take the histogram of $\hat{p}(z)$. That is, one takes $M$ samples from $p(z)$, measuring each bitstring $n_z$ times. The empirical frequencies $\hat{p}(z) = n_z/M$ take discrete values in multiples of $1/M$ (when $p_\text{avg}(z)$ is nontrivial, the empirical rescaled frequencies are continuous) and estimates the PoP by the fraction of bitstrings with a given empirical frequency $\hat{P}(D\hat{p}) \equiv N_{\hat{p}}/D$.

For each bitstring $z$ with probability $p(z)$, the empirical frequency $n_z/M$ is a rescaled Poisson random variable. For a given empirical frequency $\hat{p}$, the number of bitstrings $ N_{\hat{p}}$ with that empirical frequency is also a Poisson random variable with expected value
\begin{equation}
    \hat{P}(D\hat{p}=nD/M) = \int_0^\infty d(Dp) P(Dp) p_\text{Poi}(n;pM),
    \label{eq:finite_sampling_PoP}
\end{equation}
where $p_\text{Poi}(n;pM) \equiv e^{-pM} (pM)^n/n!$ is the probability of a Poisson variable taking value $n$ when the expected number of counts is $\mu=pM$. This prediction determines the empirical PoP with a finite number of samples. We emphasize that Eq.~\eqref{eq:finite_sampling_PoP} is the expected value of the finite-sampling PoP, and a given experimental PoP will show statistical fluctuations about this mean. In practice, occurrences for large $Dp$ occur with sufficiently small probability that the number of potential values may be truncated at some level. Then, as long as $M/D$ is large enough, there will be enough outcomes that each PoP count is approximately independent (i.e. the normalization constraint is negligible), acting as a Poisson random variable with $\sqrt{N_{\hat{p}}}$ uncertainty.

When the factor $p_\text{avg}(z)$ is not constant, the empirical rescaled probability $\hat{p}/p_\text{avg}(z)$ takes on continuous values and we must perform histogram binning instead. Here, the expected value of the histogram is given by
\begin{equation}
    \hat{P}(\hat{p}) \approx \sum_{n=0}^{\infty}\frac{1}{D}\sum_z \int d \tilde{p} P(\tilde{p}) p_\text{Poi} (n;M p_\text{avg}(z) \tilde{p}) \delta\left(\hat{p}-\frac{n}{M p_\text{avg}(z)}\right).
\end{equation}

\begin{figure*}[t!]
	\centering
	\includegraphics[width=105mm]{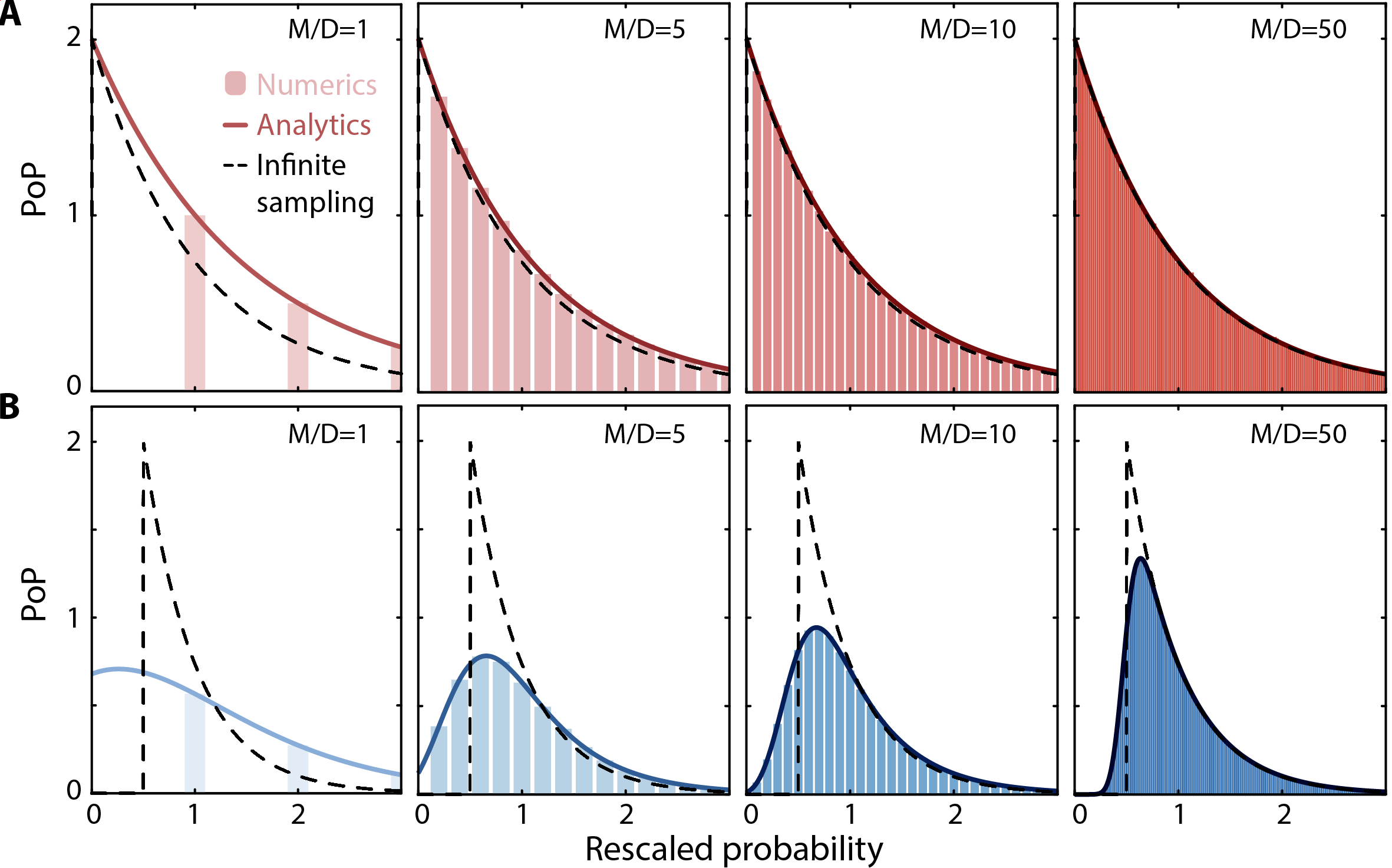}
	\caption{\textbf{Finite sampling effects on the probability-of-probabilities (PoP).} The PoP is nontrivially modified when reconstructed from a finite number of samples, as generally considered in Eq.~\eqref{eq:finite_sampling_PoP}. However, not all PoPs are as susceptible as others; for instance, \textbf{(A)} the Porter-Thomas distribution (Eq.~\eqref{eq:PT_finite_sampling}) shows relatively little sensitivity to $M/D$ (the ratio of number of measured bitstrings divided to the total Hilbert-space dimension), while \textbf{(B)} the PoP for global depolarization noise with $F=0.5$ (Eq.~\eqref{eq:GD_finite_sampling}) shows greater sensitivity due to the discontinuous PDF. In both cases, we see good agreement between numerical finite sampling and the analytical predictions Eqs.~\eqref{eq:PT_finite_sampling} and \eqref{eq:GD_finite_sampling}.
 }
	\vspace{0.5cm}
	\label{EFig:finitesampling}
\end{figure*}

The empirical PoP $\hat{P}(\hat{p})$ is akin to (but not exactly) a convolution of the underlying PoP with the Poisson PDF. It solely depends on the underlying PoP, $P(Dp)$, as well as the ratio of samples to Hilbert-space dimension, $m\equiv M/D$. In the limit $m \rightarrow \infty$, the estimate converges to the true value $\hat{P}(D\hat{p}) \rightarrow P(Dp)$. Finite values of $m$ show potentially substantial deviations between $\hat{P}(D\hat{p})$ and $P(Dp)$. In Fig.~\ref{EFig:finitesampling}, we illustrate the effects of finite sampling when the underlying PoP distribution is Porter-Thomas or arising from global depolarization errors [Eq.~\eqref{eq:globaldepol}]. The Porter-Thomas (PT) distribution is accurately reconstructed with modest ratios $m \gtrsim 1$. However, the global depolarization distribution shows a sharp discontinuity at $Dp = 1-F$, which is smoothed out under finite sampling. Even large sampling ratios such as $m = 30$ show considerable deviations. 

These two cases can be analytically solved. The PT distribution yields the following finite-sampling PoP
\begin{equation}
    \hat{P}(D\hat{p}) d(D\hat{p}) = \frac{1}{P_{> 0}} \left(\frac{m}{1+m}\right)^{1+ m D\hat{p}} d(D\hat{p}).
    \label{eq:PT_finite_sampling}
\end{equation}
where $P_{>0} = m/(1+m)$ is the probability of observing a non-zero value of $\hat{p}$ in Eq.~\eqref{eq:finite_sampling_PoP}, elaborated on below. Meanwhile, the global-depolarization PoP gives:
\begin{equation}
    \hat{P}(D\hat{p}) d(D\hat{p}) = \frac{1}{P_{> 0}} F^{-1} e^{F^{-1}-1} \left(\frac{Fm}{1+F m}\right)^{1+ m D \hat{p}} \frac{\Gamma(1+ m D\hat{p}, \frac{1-F}{F}(1+F m)}{\Gamma(1+ m D\hat{p})} d(D\hat{p}),
    \label{eq:GD_finite_sampling}
\end{equation}
where $P_{>0} = 1 - e^{m(F-1)}$, $\Gamma(k,x) \equiv \int_x^\infty z^{k-1} \exp(-z) dz$ is the upper incomplete Gamma function and $\Gamma(k)$ is the Gamma function.

In the literature, the probability $\hat{P}(\hat{p} =0)$ is typically not included in the empirical PoP, and we must renormalize the resulting histogram by the factor $P_{> 0}^{-1}$. Indeed, if one does include $\hat{P}(\hat{p} =0)$, the finite-sampling deviations from the PT distribution become more evident, the factor $P_{>0}^{-1}$ should not be included in Eqs.~\eqref{eq:PT_finite_sampling} and~\eqref{eq:GD_finite_sampling}, and Eq.~\eqref{eq:finite_sampling_PoP} is true without modification.

\subsection{Inverting finite sample effects}

\textbf{Proof of principle: inverting with Laguerre polynomials ---} Remarkably, for any value of $m$, it is theoretically possible to invert Eq.~\eqref{eq:finite_sampling_PoP} with Laguerre polynomials (making a technical assumption that the PoP $P(Dp)$ is square-integrable, which is true in any finite system). We expand $P(Dp)$ in terms of Laguerre polynomials: 
\begin{equation}
 P(Dp) = \sum_{k=0}^\infty c_k L_k(2 m Dp) \exp(-m Dp),  
 \end{equation}
where $L_k(x)$ is the $k$-th Laguerre polynomial, satisfying the orthogonality relation $\int_0^\infty dx e^{-x} L_n(x) L_m(x) = \delta_{nm}$ and $c_k$ are the unknown coefficients to be determined. Using the above orthogonality relation, $c_k$ can be extracted from the empirical PoP, $\hat{P}\left(D\hat{p}_n\right)$, with the formula
\begin{equation}
    c_k = 2 \sum_{n=0}^k \binom{k}{n} (-2)^n \hat{P}\left(D\hat{p}_n\right),
    \label{eq:solve_finite_sampling_PoP}
\end{equation}
where $\hat{p}_n= n/M$ are the discrete values of the empirical frequencies. However, this formula is not practical. It is not robust to any amount of statistical or measurement error since it weights the measured frequencies $\hat{P}\left(D\hat{p}= Dn/M\right)$ with exponentially increasing and oscillatory weights in $n$. Nevertheless, this serves as a theoretical basis that the ideal PoP can be reconstructed from any finite-sampling PoP. For a more practical procedure, we employ a regularized least squares approach.

\textbf{Inversion with regularized least squares --- }Our practical approach to undo the effects of finite sampling will be based on regularized least squares. We note that this method is not used for any data shown in the main text, and we instead present it for completeness and as a potential way to further improve our results.

We discretize the domain of the PoP into a discrete number of histogram bins centered at $\tilde{p}_j$ and with width $\Delta\tilde{p}$. For concreteness, it suffices to stop at $\tilde{p} = 5$. We also discretize the finite-sampling PoP into discrete and finite histogram bins. When $p_\text{avg}(z) = 1/D$, its domain is naturally discrete, but this is not so for nontrivial $p_\text{avg}(z)$. Working in the former case for simplicity, we can rewrite Eq.~\eqref{eq:finite_sampling_PoP} as a linear equation
\begin{align}
    \mathbf{\hat{P}} &= \mathcal{M} \mathbf{P}~, \label{eq:linear_equation_finite_PoP}\\
    \mathcal{M}_{n,j} &= 
\exp(-\tilde{p}_jM) \frac{(\tilde{p}_j M)^n}{n!}(\Delta \tilde{p}) 
\end{align}

When $p_\text{avg}(z)$ is nontrivial, one must perform histogram binning on both the range and the domain, and the matrix $\mathcal{M}$ is instead
\begin{equation}
    \mathcal{M}_{i,j} \approx \frac{1}{D}\sum_z\sum_{n=\lceil M (\hat{p}_i - \Delta \hat{p}/2) p_\text{avg}(z)\rceil}^{\lfloor M (\hat{p}_i + \Delta \hat{p}/2) p_\text{avg}(z)\rfloor} P(\tilde{p}_j) p_\text{Poi} (n;M p_\text{avg}(z) \tilde{p}_j)
\end{equation}

\begin{figure}
    \centering
    \includegraphics[width=120mm]{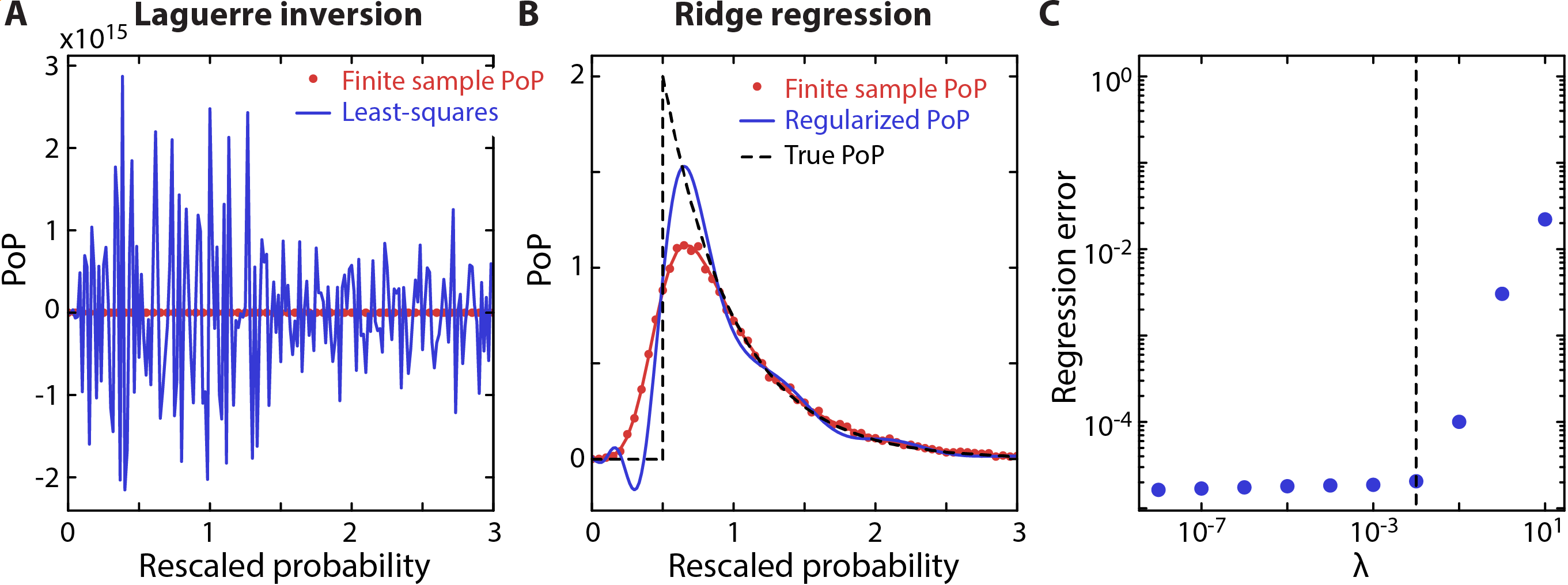}
    \caption{\textbf{Strategies to reverse finite-sampling effects on the PoP.} We simulate the finite-sampling effects on a global depolarization PoP, with a Hilbert-space dimension of $D=2^{15}$, and $M=20 D$ samples. \textbf{A.} Regular least squares regression [Eq.~\eqref{eq:linear_equation_finite_PoP}] is numerically unstable and leads to large, unphysical solutions (note scale of values). \textbf{B.} 
    Regularized regression penalizes this divergence and gives an approximate reconstruction (blue) of the infinite-sampling PoP (black dashed) from the finite-sampling PoP (red points). \textbf{C.} The free parameter $\lambda$ is chosen by plotting the regression error as a function of $\lambda$, displaying two regimes where the solution is respectively overfitted and overdamped. We choose $\lambda$ at the transition between these regimes. Nevertheless, the reconstructed PoP is imperfect; adding further constraints, that for instance force the PoP to be always positive, may improve agreement, but we leave such avenues to future work.}
\label{fig:ridge_regression}
\end{figure}

Our result Eq.~\eqref{eq:solve_finite_sampling_PoP} essentially gives an explicit result for $\mathcal{M}^{-1}$ in the Laguerre polynomial basis. Unfortunately, $\mathcal{M}^{-1}$ is extremely ill-conditioned and cannot accommodated the shot-noise in $\mathbf{\hat{P}}$. If one naively attempts to solve this linear equation, the solution is wildly oscillatory and large. In order to remedy this, we simply regularize our solution and solve the least-squares equation with a penalty term:
\begin{equation}
    \mathbf{P}_\text{est} = \text{argmin}_\mathbf{y}\left[ \left(\mathcal{M} \mathbf{y} - \mathbf{\hat{P}}\right)^T A \left(\mathcal{M}\mathbf{y} - \mathbf{\hat{P}}\right) + \lambda \mathbf{y}^T B \mathbf{y}\right]
\end{equation}
where we have presented this problem in its full generality and $A$ and $B$ are arbitrary symmetric positive-semidefinite matrices that we can suitably choose to regularize the regression. When $A$ and $B = \hat{I}$, the first term is the well-known squared error and the second is the $L_2$ norm. However, we may modify $A$ and $B$ to enforce other constraints, such as smoothness of the solution. This is known as ridge, or Tikhonov, regression~\cite{Hoerl2020RidgeRegression} and has explicit solution
\begin{equation}
    \mathbf{P}_\text{est} = \left( \mathcal{M}^T A \mathcal{M} + \lambda B \right)^{-1} \mathcal{M}^T A \mathbf{\hat{P}} 
\end{equation}
As an illustration, we demonstrate this in Fig.~\ref{fig:ridge_regression}.  The parameter $\lambda$ is tunable, and one must choose and appropriate value. While it can be deduced analytically, here we present a user-friendly and intuitive method. In Fig.~\ref{fig:ridge_regression}C we plot the squared error $\Vert\mathcal{M} \mathbf{P}_\text{est}(\lambda)-\mathbf{\hat{P}}\Vert_2^2$ as a function of the parameter $\lambda$. This shows two parameter regions. When $\lambda$ is small, the squared error is nearly independent of $\lambda$. In this regime, the solution is overfitted and enforcing the constraint effectively regularizes the solution without significant effect on the squared error. When $\lambda$ is large, the squared error grows quickly with $\lambda$. In this regime, the solution is overdamped and the constraint is too strong. We find that operating at the midpoint between both regimes is often optimal. Despite this, its reconstructed solutions are often imperfect, though potentially improvable through addition of additional constraints on the regression (like forcing the PoP to be always positive). Therefore, in the main text we rely on direct comparisons of the finite-sampling PoP, outlined below.

\newpage
\clearpage
\section{Cumulants of probability-of-probabilities distributions}
\label{app:PoP_moments}
As established above, due to finite-sampling effects, it can be challenging to empirically estimate the PoP. In the main text, we have compared directly to finite sampled version of model PoP predictions, and in addition we have discussed potential theoretical means of reversing the finite sampling effect. Here, we discuss how we can also estimate properties of the PoP to greater precision, without reconstructing the PoP distribution itself. Specifically, the moments of the PoP can be efficiently estimated, which we may use to learn about the type of noise present in the system. Further, the moments for different error models can have different functional dependencies on the fidelity, $F$, meaning that analysing how moments change over multiple values of fidelity may improve predictive power.

From Section~\ref{app:hypoexponential_distribution}, the probability distribution can be written as $\tilde{p}(z) = \sum_i \omega_i \tilde{p}_i(z)$, where $\tilde{p}(z) \equiv p(z)/p_\text{avg}(z)$ is the rescaled probability and $\tilde{p}_i(z)$ are independent Porter-Thomas (i.e.~Dirichlet random) distributions. Remarkably, some properties of the coefficients $\{\omega_i\}$ can be estimated from empirical samples from $p(z)$, even \textit{without knowledge} of the constituents $\tilde{p}_i(z)$. Specifically, cumulants of the PoP are related to moments of $\{\omega_i\}$.

As above, we first discuss the case when the factors $p_\text{avg}$ are constant, before generalizing. The $k$-th (raw, i.e. non-central) moment of the PoP $P(D p)$ is given by
\begin{equation}
    m_k \equiv D^{-1}\sum_z (Dp(z))^k \approx  \int (Dp)^k P(Dp) d(Dp),
\end{equation}
where the approximation follows by treating each value $Dp(z)$ as an independent sample from the PoP $P(Dp)$. We then consider the cumulants $\kappa_k$. For example, the first three cumulants are:
\begin{align}
    \kappa_1 &= m_1 \label{eq:first_cumulant}\\
    \kappa_2 &= m_2 - m_1^2\\
    \kappa_3 &= m_3 - 3 m_2 m_1 + 2 m_1^3 \label{eq:third_cumulant}
\end{align}
A key property of the cumulants are that they add for sums of independent random variables~\cite{Novak2012ThreeLectures}. That is, for $\tilde{p} = \sum_i \omega_i \tilde{p}_i$, $\kappa_k(\tilde{p}) = \sum_i \omega_i^k \kappa_k(\tilde{p}_i)$. Since each $\tilde{p}_i$ is an independent Porter-Thomas distribution, they have identical cumulants $\kappa_k = (k-1)!$. Therefore, these $k$-th cumulants of the PoP are equal to the raw $k$-th moment of $\{\omega_i\}$
\begin{equation}
    \frac{\kappa_k(p)}{(k-1)!} \approx \sum_i \omega_i^k~,
\end{equation}
As an illustration, we concretely demonstrate this with $k=2$. We have
\begin{equation}
    \kappa_2(p) = D\sum_z p(z)^2 - 1 \approx \sum_i \omega_i^2
\end{equation}
To see this, note that $\sum_z p(z)^2 = \sum_z \omega_i \omega_j \sum_{ij} p_i(z) p_j(z)$. Since $p_i(z), p_j(z)$ are independent Porter-Thomas distributions, they are approximately orthogonal, in the sense $D\sum_z p_i(z) p_j(z) = 1+ \delta_{ij} + O(D^{-1/2})$. Using this relation, we obtain:
\begin{equation}
    D\sum_z p(z)^2 = 1+ \sum_i \omega_i^2 + O(D^{-1/2}),
\end{equation}
where we have used the fact that $\sum_i \omega_i = 1$. Remarkably, due to this approximate orthogonality, the second moment of the PoP only depends on the coefficients $\omega_i$, up to exponentially small fluctuations.
When $p_\text{avg}(z)$ is nontrivial, we should replace $Dp(z)$ with $\tilde{p}(z) \equiv p(z)/p_\text{avg}(z)$ and the average over $z$ as a weighted mean $\mathbb{E}_z[\cdot] = \sum_z p_\text{avg}(z) [\cdot]$. This modifies the moments to $m_k = \sum_z p(z)^{k-1}/p_\text{avg}(z)^{k-1}$.

\subsection{Relation of hypoexponential moments to mixed state moments}
\label{subsubsec:weight_vector_and_mixed_state_moments}
The moments give key information about the mixed state. In the context of closed system dynamics, we found that the moments of the weight vector satisfy
\begin{equation}
    \sum_i \omega_i^k \approx \text{tr}(\hat{\rho}_A^k),
    \label{eq:moments_matching}
\end{equation}
as long as the bath $B$ is sufficiently small.

We conjecture that the above equation also holds in the context of open system dynamics, as long as the fidelity is sufficiently high. Indeed, Eq.~\eqref{eq:moments_matching} is true for Haar-random measurement bases $\{|z\rangle\}$, up to subleading $O(D^{-1/2})$ corrections. If the eigenstates of $\hat{\rho}$ look ``random" with respect to the computational basis $|z\rangle$, we get the desired equivalence.

\subsection{Moments under different noise sources}

The moments $\sum_i \omega_i^k$ behave differently under different noise channels. In Table~\ref{ETab:moments}, we summarize the low-order cumulants ($\kappa_2$ and $\kappa_3$) of the PoP predicted for various noise channels and their dependence on the fidelity $F$. By independently measuring the fidelity $F$ (using, for example, cross-entropy benchmarking~\cite{Arute2019QuantumSupremacy,Mark2023BenchmarkingQuantum,Shaw2024BenchmarkingHighly}), and the cumulants $\kappa_k(p)$, one would be able to discriminate between such noise models.
  
 Consider first a local error channel, we have $\{\omega_i^\text{loc.}\} \approx \{ F, (1-F)/k, (1-F)/k,\dots, (1-F)/k\}$. Therefore, its second moments
 \begin{equation}
    \sum_i \left(\omega_i^\text{loc.}\right)^2 = F^2 + \frac{(1-F)^2}{k} \overset{k\rightarrow \infty}{\approx} F^2. 
    \label{eq:local_error_second_moment}
 \end{equation}
This relationship generalizes to higher moments: the largest weight $F$ dominates, and we have $\sum_i \left(\omega_i^\text{loc.}\right)^k \approx F^k$.

\begin{figure}
    \centering
    \includegraphics[width=120mm]{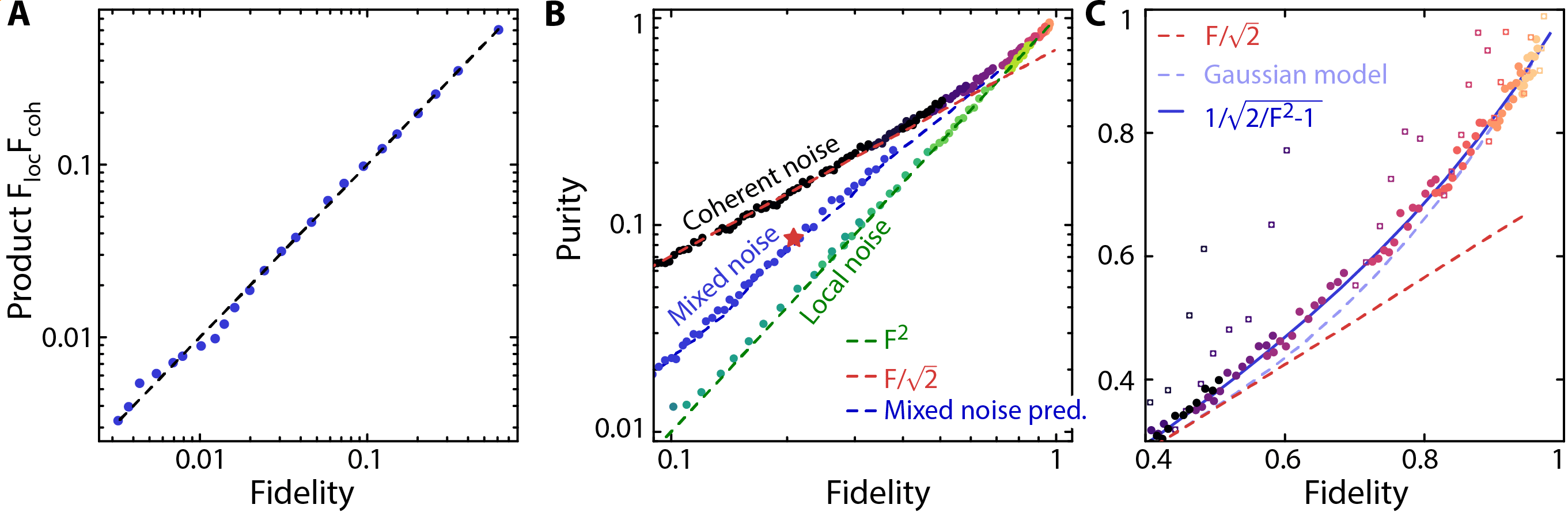}
    \caption{\textbf{RUC simulations of mixed local and coherent errors.} \textbf{A.} We first verify that errors factorize over a large range of fidelities. Here, a single instance of a random unitary circuit is presented, at a fixed proportion of local error and coherent error. As a function of circuit depth, its fidelity $F$ decreases. We simulate the same circuit in the presence of only local or coherent errors at the same rate, obtaining the ``local" and ``coherent" fidelities $F_\text{loc.}$ and $F_\text{coh.}$. The overall fidelity is well approximated by $F\approx F_\text{loc.}F_\text{coh.}$. This makes the concept of an error source ``fraction" meaningful. \textbf{B.} Plots of the purity, or second moment, $D\sum_z p(z)^2$ for noisy RUC dynamics in the presence of global coherent or local noise, or a mixture of the two noise sources (Table~\ref{ETab:moments}). The purity due to global coherent noise asymptotically behaves as $F_\text{coh.}/\sqrt{2}$, while the purity due to local noise behaves as $F^2_\text{loc.}$. In the presence of both types of noise, the purity behaves as $F^2_\text{loc.}F_\text{coh.}/\sqrt{2}$ (blue dashed line), enabling efficient estimation of the noise content from the second moment. Our mixed error simulation is a depth-96 RUC with 12 qubits, using the same error rates as in Fig.~\ref{Fig5}c of the main text (precise depth marked with a red star), while our local and global coherent errors are simulations of RUCs at several error rates. Data points displayed are at depths of 24 and above. \textbf{C.} Purity from global coherent noise at high fidelities. Here, we see deviations from the asymptotic behaviour $F/\sqrt{2}$ (red dashed). In particular, it takes a value of 1 at $F=1$. At intermediate values $0.5<F<0.9$, the purity deviates from the prediction from our Gaussian-hypoexponential model (blue dashed), instead agreeing with $(2/F^2-1)^{-1/2}$ (blue solid line), an expression obtained from a calculation of the state purity. This indicates that our model for the hypoexponential distribution [Eq.~\eqref{eq:Gaussian_hypoexponential_model}] should be refined in this regime. In addition, to illustrate the importance of circuit depth, for the open markers we indicate the purity at depths 24 and below. These deviate from our prediction due to a lack of anticoncentration in shallow circuits.}
    \label{fig:moments_of_noise_models}
\end{figure}

We next consider global coherent errors. Here, the weights follow a Gaussian distribution
\begin{align}
    \omega^\text{coh.}_i =  \exp(-i^2/(2\sigma_F^2))/\mathcal{Z}(\sigma_F)
\end{align}
with a width $\sigma_F$ that depends on the fidelity $F$. When $\sigma_F \gg 1$, the normalization factor is well-approximated by $\mathcal{Z}(\sigma_F)\approx \sqrt{2\pi \sigma_F^2}$. Performing the sum, we have 
\begin{equation}
 \sum_i \left(\omega_i^\text{coh.}\right)^2 \approx \frac{1}{4\pi \sigma_F^2} = \frac{F}{\sqrt{2}}.   
\end{equation}
Note the different scaling with $F$. In Fig.~\ref{fig:moments_of_noise_models}B, we verify these scaling behaviors for our two types of errors in an RUC simulation.

Note that the above prediction breaks down when $F\approx 1$. Using our Gaussian hypoexponential model above, the discrete Gaussian sums can be explicitly evaluated, giving $\sum_i \big(\omega_i^\text{coh.}\big)^2 = \theta_3(\exp[-1/\sigma_F^2])/\theta_3(\exp[-1/(2\sigma_F^2)])^2$,
where $\theta_3$ is a Jacobi theta function. Meanwhile, using methods developed in Ref.~\cite{Shaw2024BenchmarkingHighly}, one can perform an exact calculation of the state purity in this setting, which gives
 $\text{tr}(\hat{\rho}^2) = (2/F^2-1)^{-1/2}$. As seen in Fig.~\ref{fig:moments_of_noise_models}C, these expressions show deviations when $F\approx 1$, with numerical data from RUC simulations more consistent with the latter expression. 

In the presence of mixed noise, we further observe that since the weights factorize as $\omega^\text{mixed}_{ij} = \omega^\text{loc.}_i \omega^\text{coh.}_j$, so do the cumulants, in particular $\kappa_2 \approx F_\text{loc.}^2F_\text{coh.}/\sqrt{2}$. Coupled with knowledge of the fidelity $F = F_\text{loc.}F_\text{coh.}$, one could extract the fidelities of each noise type. In Fig.~\ref{fig:moments_of_noise_models}B, we show proof-of-principle numerical evidence for this behaviour in a RUC.

\begin{table*}[t!]
    
    \renewcommand\arraystretch{1.3}
    \setlength{\tabcolsep}{2pt}
    \centering
    \begin{tabular}{p{5cm}p{3cm}p{3cm}}
    \hline\hline
    \textbf{Noise channel} & 
    \textbf{$\kappa_2$} & \textbf{$\kappa_3$} \\
    \hline
    Global depolarization & $F^2$ & $2F^3$ \\
    Local incoherent & $F^2+(1-F)^2/k$ & $2\Big(F^3+(1-F)^3/k^2\Big)$ \\
    Global Gaussian coherent & $F/\sqrt{2}$ & $2F^2/\sqrt{3}$ \\
    
    \hline\hline
    \end{tabular}
    \caption{\textbf{Low order cumulants of PoPs for various selected noise channels.}}
    \label{ETab:moments}
\end{table*}

\subsection{Estimator of moments}

Remarkably, the moments can be efficiently estimated~\cite{Acharya2016EstimatingRenyi} with the following estimator:
\begin{equation}
    \hat{m}_k = D^{k-1} \sum_z \frac{(n_z)^{\underline{k}}}{M^{\underline{k}}} \approx D^{k-1} \sum_z p(z)^k,
    \label{eq:kth-moment-estimator}
\end{equation}
where $n^{\underline{k}}\equiv n(n-1)\cdots(n-k)$ is the $k$-th \textit{falling power} of $n$ and $n_z$ is the number of occurrences of the bitstring $z$ from $M$ measurements. This was shown to be an unbiased estimator of $m_k$, with optimal sample complexity scaling. Specifically, Ref.~\cite{Acharya2016EstimatingRenyi} showed that the $k$-th Renyi entropy (obtained from $m_k$) can be estimated to \textit{additive error} $\delta$ with $M = \Theta(D^{1-1/k}/\delta^2)$ samples,
where $D$ is the Hilbert-space dimension (more generally the number of possible measurement outcomes). In the same work, they showed that this scaling is optimal. Further, note that in the original reference~\cite{Acharya2016EstimatingRenyi}, 
the denominator of Eq.~\eqref{eq:kth-moment-estimator} is $M^k$ instead of $M^{\underline{k}}$. The former is the unbiased estimator in the Poisson sampling setting, while the latter is the unbiased estimator under realistic sampling. When $p_\text{avg}(z)$ is nontrivial, the moment estimator is $\hat{m}_k = \sum_z (n_z)^{\underline{k}}/(M^{\underline{k}} p_\text{avg}(z)^{k-1}) \approx \sum_z p(z)^{k-1}/p_\text{avg}(z)^{k-1}$.

We are interested in Eq.~\eqref{eq:kth-moment-estimator} because the low moments of the PoP can be estimated with far fewer samples (e.g.~$M=O(\sqrt{D})$ for the second moment) than the entire PoP, which requires $M = O(D)$. It is illustrative to consider the simplest nontrivial case, the second moment $m_2$. Our estimator $\hat{m}_2 = D\sum_z n_z(n_z-1)/(M(M-1))$ essentially counts the number of times a bitstring $z$ is seen more than once. Accordingly, this estimates the \textit{collision probability} $D\sum_z p(z)^2$, here rescaled to be typically an $O(1)$ quantity. The sample complexity is almost optimal: estimating the second moment to multiplicative accuracy $\delta$ only requires $\sqrt{D}/\delta^2$ measurements, a manifestation of the birthday paradox. For our purposes, it is also useful to estimate its sample requirements for \textit{additive} accuracy. Using properties of the multinomial distribution, we find that the expected uncertainty in the second moment $m_2$ is
\begin{align}
    \left(\Delta \hat{m}_2\right)^2 &= \left[\frac{2M^2-M}{M^{\underline{4}}} D m_2 + 4 \frac{M^{\underline{3}}}{M^{\underline{4}}} m_3 - 2 \frac{M^{\underline{2}}(2M-3)}{M^{\underline{4}}} m_2 \right]\\
    &\approx \frac{2Dm_2}{M^2} + \frac{4 m_3}{M} - \frac{4 m_2^2}{M}\,,  \label{eq:second_moment_sample_complexity}
\end{align}
where last term arises from anticorrelations between $n_z,n_{z'}$ which are not present in the Poisson sampling model. Since $m_2$ and $m_3$ are $O(1)$ quantities, this result indicates that for an additive error $|\hat{m}_2|=\delta$, the sample complexity behaves as $O(\text{max}(\sqrt{D}/\delta,1/\delta^2))$, where the crossover between the two scaling behaviors occurs at $M\approx D$.

To recover the result of Ref.~\cite{Acharya2016EstimatingRenyi} for the required number of samples to achieve a multiplicative error $\delta$, namely $\Theta(\sqrt{D}/\delta^2)$, we focus on the second term of Eq.~\eqref{eq:second_moment_sample_complexity}, then use the inequalities $m_3 = D^2 \sum_z p(z)^3 \leq D^2 \left(\sum_z p(z)^2\right)^{3/2} = \sqrt{D} m_2^{3/2}$, and $m_2 \geq 1$ to give the desired result that $(\Delta \hat{m}_2)^2 \sim  m_3/M \leq \sqrt{D} m_2^2/M$. The former inequality comes from the relation between $L_p$ norms, $L_p \leq L_q$ for $p\geq q$, relating $m_2$ and $m_3$ to the $L_2$ and $L_3$ norms of the probability vector $p(z)$, while the latter inequality is a straightforward application of the Cauchy-Schwarz inequality. Finally, note that this result is only valid in the asymptotic regime where the second term of Eq.~\eqref{eq:second_moment_sample_complexity} dominates the first term. This is valid at a crossover value of $M \approx D$, before which the sample complexity scales as $M \sim \sqrt{D}/\delta$ instead.

\section{Relationship between the hypoexponential weights and the channel Kraus operators}
\label{app:kraus}
In addition to their relative ease of measurement, the moments of the hypoexponential weight vector, $\vec{\omega}$, can also be theoretically analyzed and related to quantities that characterize the strength of the noise channel. Given our analysis, it is natural to ask how the weight vector $\vec{\omega}$ reflects the underlying quantum channel. We find that the properties of $\vec{\omega}$ are related to the strength of the channel, but there are additional terms that we expect to be subleading. We denote the channel $\Phi$, such that the output state is $\hat{\rho} \equiv \Phi(|\Psi_0\rangle\!\langle\Psi_0|) \equiv \sum_i K_i |\Psi_0\rangle\!\langle\Psi_0| K_i^\dagger$, where $K_i$ are Kraus operators that describe the channel, satisfying $\sum_i K_i^\dagger K_i = \hat{I}$. $\{K_i\}$ can be uniquely chosen to be orthogonal~\cite{Verstraete2003QuantumChannels}, in the sense that $\text{tr}(K_i K_j^\dagger) = D q_i \delta_{ij}$, such that $\sum_i q_i = 1$.

We first consider the second moment $\sum_i \omega_i^2$, which, using Eq.~\eqref{eq:moments_matching}, approximates the purity $\text{tr}(\Phi(|\Psi_0\rangle\!\langle\Psi_0|)^2)$. We approximate this purity with the Haar-averaged value:
\begin{align}
    \mathbb{E}_{\psi\sim\text{Haar}}[\text{tr}(\Phi(|\psi\rangle\!\langle\psi|)^2)] &= \frac{\sum_{i,j} \left(\text{tr}(K_i K_j^\dagger)\text{tr}(K_j K_i^\dagger)+\text{tr}(K_i K_i^\dagger K_j K_j^\dagger)\right)}{D(D+1)}\\
    &= \frac{D}{D+1} \left(\sum_i q_i^2 + \text{tr}\bigg[\Phi\left(\hat{I}/D\right)^2\bigg]\right)~,
\end{align}
where we have used the orthogonality conditions above. The first term is the ``purity" of the noise channel strengths $p_i$, while the second term is related to whether the channel is unital or not~\cite{Wallman2015EstimatingCoherence}. If the channel is unital, then $\Phi(\hat{I}/D) = \hat{I}/D$ so the second term has size $1/D$, which we expect to be exponentially smaller than the purity of the channel. Meanwhile, if the channel is non-unital -- in an extreme case consider the amplitude damping channel such that $\Phi(\hat{I}/D) = |0\rangle\!\langle 0|$ -- then the second term has size $1$, which is larger than $\sum_i q_i^2$. In other words, the purity of the state has two contributions, given by the decoherence strength of the channel and the purity of the steady state $\Phi(\hat{I}/D)$. Assuming that $\Phi(\hat{I}/D)$ is unital or close to maximally mixed, however, we conclude that the purity of a typical input state reflects how noisy the channel is.

Higher moments give additional terms which we may not simplify, such as $\sum_{hij} \text{tr}(K_h K_i^\dagger K_j) + \text{h.c.}$ for $k=3$. However, we may assume that such quantities are exponentially suppressed, since the orthogonal operators $K_i$ are obtained by the (reshaped) eigenvectors of the Choi matrix which describes the channel~\cite{Verstraete2003QuantumChannels}. If these eigenvectors have sufficiently complicated bipartite structure, we expect the contributions from such terms to be exponentially suppressed. We may repeatedly apply this argument to all orders $k$, and find that the (expected) dominant contribution is $\sum_i q_i^k$. This also bounds the typicality error, that is the difference between the moment $\text{tr}(\Phi(|\Psi\rangle\!\langle\Psi|)^k)$ of a single, typical random state $|\Psi\rangle$ and the ensemble average. Indeed, we also expect our conclusion to hold as long as the initial state $\ket{\Psi_0}$ has sufficiently small overlap with the eigenstates $\{|E\rangle\}$.
Therefore, we have 
\begin{equation}
    \frac{\kappa_k(p)}{(k-1)!} \approx \sum_i \omega_i^k \approx \text{tr}(\Phi(|\Psi_0\rangle\!\langle\Psi_0|)^k) \approx \sum_i q_i^k,
    \label{eq:approximate_chain}
\end{equation}
which therefore allows us to conclude that $\omega_i \approx q_i$, as long as the number of non-zero weights is less than the Hilbert-space dimension, i.e.~$\#\{\omega_i\} \ll D$.

\twocolumngrid
\bibliographystyle{adamref}
\bibliography{zotero.bib}

\end{document}